\documentclass[twocolumn]{aastex61}

\let\OldS\S
\renewcommand{\S}{\OldS{}}

\received{XX, 2017}
\revised{XX, 2017}
\accepted{\today}
\submitjournal{ApJ}

\shorttitle{Dark Galaxies with MUSE}
\shortauthors{Marino et al.}

\begin{document}
\bibliographystyle{plainnat}

\title{Dark Galaxy candidates at redshift $\sim$\,3.5 detected with MUSE\footnote{\scriptsize{Based on observations obtained at the Very Large Telescope (VLT) of the European Southern Observatory, Paranal, Chile (ESO Programme IDs 094.A-0396, 095.A-0708, 096.A-0345, 097.A-0251, 098.A-0678, 094.A-0131, 095.A-0200, 096.A-0222, 097.A-0089, 098.A-0216).}}}

\correspondingauthor{Raffaella Anna Marino}
\email{marinor@phys.ethz.ch}

\author{Raffaella Anna Marino}
\affil{Department of Physics, ETH Z$\ddot{u}$rich,Wolfgang$-$Pauli$-$Strasse\,27, 8093\,Z$\ddot{u}$rich, Switzerland}

\author{Sebastiano Cantalupo}
\affiliation{Department of Physics, ETH Z$\ddot{u}$rich,Wolfgang$-$Pauli$-$Strasse\,27, 8093\,Z$\ddot{u}$rich, Switzerland}

\author{Simon J$.$ Lilly}
\affiliation{Department of Physics, ETH Z$\ddot{u}$rich,Wolfgang$-$Pauli$-$Strasse\,27, 8093\,Z$\ddot{u}$rich, Switzerland}

\author{Sofia G$.$ Gallego}
\affiliation{Department of Physics, ETH Z$\ddot{u}$rich,Wolfgang$-$Pauli$-$Strasse\,27, 8093\,Z$\ddot{u}$rich, Switzerland}

\author{Lorrie A$.$ Straka}
\affiliation{Leiden Observatory, Leiden University, PO Box 9513, NL$-$2300 RA Leiden, the Netherlands}

\author{Elena Borisova}
\affiliation{Paul Scherrer Institute, WBBA/214, 5232 Villigen PSI, Switzerland}

\author{Roland Bacon}
\affiliation{Univ Lyon, Univ Lyon1, Ens de Lyon, CNRS, Centre de Recherche Astrophysique de Lyon UMR5574, F-69230, Saint-Genis-Laval, France}

\author{Jarle Brinchmann}
\affiliation{Leiden Observatory, Leiden University, PO Box 9513, NL$-$2300 RA Leiden, the Netherlands}
\affiliation{Instituto de Astrof\'{i}sica e Ci{\^e}ncias do Espa\c{c}o, Universidade do Porto, CAUP, Rua das Estrelas, PT4150-762 Porto, Portugal}

\author{C$.$ Marcella Carollo}
\affiliation{Department of Physics, ETH Z$\ddot{u}$rich,Wolfgang$-$Pauli$-$Strasse\,27, 8093\,Z$\ddot{u}$rich, Switzerland}

\author{Joseph Caruana}
\affiliation{Department of Physics, University of Malta, Msida MSD 2080, Malta}
\affiliation{Institute for Space Sciences and Astronomy, University of Malta, Msida MSD 2080, Malta}
 
\author{Simon Conseil}
\affiliation{Univ Lyon, Univ Lyon1, Ens de Lyon, CNRS, Centre de Recherche Astrophysique de Lyon UMR5574, F-69230, Saint-Genis-Laval, France}

\author{Thierry Contini}
\affiliation{Institut de Recherche en Astrophysique et Plan\'{e}tologie (IRAP), Universit\'{e} de Toulouse, CNRS, UPS, F-31400 Toulouse, France}

\author{Catrina Diener}
\affiliation{Institute of Astronomy, Madingley Road Cambridge, CB3 0HA, UK}

\author{Hayley Finley}
\affiliation{Universit\'{e} de Toulouse, UPS-OMP, 31400 Toulouse, France}
\affiliation{Institut de Recherche en Astrophysique et Plan\'{e}tologie (IRAP), Universit\'{e} de Toulouse, CNRS, UPS, F-31400 Toulouse, France}

\author{Hanae Inami}
\affiliation{Univ Lyon, Univ Lyon1, Ens de Lyon, CNRS, Centre de Recherche Astrophysique de Lyon UMR5574, F-69230, Saint-Genis-Laval, France}

\author{Floriane Leclercq}
\affiliation{Univ Lyon, Univ Lyon1, Ens de Lyon, CNRS, Centre de Recherche Astrophysique de Lyon UMR5574, F-69230, Saint-Genis-Laval, France}


\author{Sowgat Muzahid}
\affiliation{Leiden Observatory, Leiden University, PO Box 9513, NL$-$2300 RA Leiden, the Netherlands}

\author{Johan Richard}
\affiliation{Univ Lyon, Univ Lyon1, Ens de Lyon, CNRS, Centre de Recherche Astrophysique de Lyon UMR5574, F-69230, Saint-Genis-Laval, France}

\author{Joop Schaye}
\affiliation{Leiden Observatory, Leiden University, PO Box 9513, NL$-$2300 RA Leiden, the Netherlands}

\author{Martin Wendt}
\affiliation{Institut f\"{u}r Physik und Astronomie, Universit\"{a}t Potsdam,Karl-Liebknecht-Str. 24/25,  14476 Golm, Germany}
\affiliation{Leibniz-Institut f\"{u}r Astrophysik Potsdam (AIP), An der Sternwarte 16, 14482 Potsdam, Germany}

\author{Lutz Wisotzki}
\affiliation{Leibniz-Institut f\"{u}r Astrophysik Potsdam (AIP), An der Sternwarte 16, 14482 Potsdam, Germany}




\begin{abstract}
Recent theoretical models suggest that the early phase of galaxy formation could involve an epoch when galaxies are gas-rich but inefficient at forming stars: a ``dark galaxy" phase. Here, we report the results of our MUSE (Multi Unit Spectroscopic Explorer) survey for dark galaxies fluorescently illuminated by quasars at \textit{z}\,$>$\,3. Compared to previous studies which are based on deep narrow-band (NB) imaging, our integral field survey provides a nearly uniform sensitivity coverage over a large volume in redshift space around the quasars as well as full spectral information at each location. Thanks to these unique features, we are able to build control samples at large redshift distances from the quasars using the same data taken under the same conditions. By comparing the rest-frame equivalent width (EW$_{0}$) distributions of the Ly$\alpha$ sources detected in proximity to the quasars and in control samples, we detect a clear correlation between the locations of high EW$_{0}$ objects and the quasars. This correlation is not seen in other properties such as Ly$\alpha$ luminosities or volume overdensities, suggesting the possible fluorescent nature of at least some of these objects. Among these, we find 6 sources without continuum counterparts and EW$_{0}$ limits larger than 240\,$\mathrm{\AA}$ that are the best candidates for dark galaxies in our survey at $z>3.5$. The volume densities and properties, including inferred gas masses and star formation efficiencies, of these dark galaxy candidates are similar to previously detected candidates at $z\approx2.4$ in NB surveys. Moreover, if the most distant of these are fluorescently illuminated by the quasar, our results also provide a lower limit of \textit{t}=60 Myr on the quasar lifetime.

\end{abstract}




\def\nodata{ ~$\cdots$~}
\section{Introduction} \label{sec:intro}

Despite a great deal of progress in defining the demographics of galaxies at high redshift ($z\,>\,$3), our knowledge about the fuel for the formation of the first stars, i$.$e$.$ the cold gas (T\,$\leqslant$\,10$^{4}$ K) surrounding the galaxies, is still limited. In addition, due to small sample sizes and technical limitations of the current facilities \citep{2014ApJ...780...74F}, both how this gas forms the large$-$scale structure of the Universe, the Intergalactic Medium (IGM), and how it keeps star formation active over time are unclear processes \cite[hereafter C12]{2012MNRAS.425.1992C}.

It is well recognized that the densest and most filamentary parts of the IGM play a key role in the formation and evolution of galaxies (\citealt{2009RvMP...81.1405M} and references therein). Recent observations have raised our awareness of the nature of the IGM and CGM (Circum Galactic Medium), thanks to both the absorption \citep{2011ApJ...743...95G,2014MNRAS.445..794T} and emission  \citep[e$.$g$.$,][]{2016A&A...587A..98W,2016ApJ...831...39B} signatures of hydrogen at several scales and in different environments, from quasars (QSOs) to radio galaxies \citep[e$.$g$.$,][]{2014Natur.506...63C, 2015MNRAS.449.1298S,2017ASSL..430..195C}.

Theoretical models have suggested the existence of a primordial phase $-$ almost optically dark $-$ of galaxy formation in which there were gas$-$rich and residing in low$-$mass halos \citep[e$.$\,g$.$,][amongst others]{2009Natur.457..451D, 2012ApJ...753...16K, 2013ApJ...776...34K} with very low star formation efficiencies (SFEs\,$<$\,10$^{-11}$\,yr$^{-1}$). This less efficient star formation phase of the IGM gas at high redshift could be due to the metal$-$free gas present in the environment at that epoch, or to the H$_{2}$ self$-$regulation effect \citep{2012ApJ...749...36K} or even to a reduced CGM cooling rate \citep{2010MNRAS.403L..16C}.

Different approaches have been taken to further investigate this dark phase of galaxy formation in the literature. The different methods that have been used in the past to try to detect the ``starless" IGM gas, i$.$\,e$.$ just before a considerable star formation occurs, are:

\begin{enumerate}
\item [(i)] \textit{HI absorption systems} along the line$-$of$-$sight to bright background sources (QSOs) at high redshift \citep[e$.$g$.$,][among others]{2011Sci...334.1245F, 2013ApJ...762L..19P, 2013ApJ...776..136P,2014ApJ...795L..12L, 2015MNRAS.452.2553J} using one$-$dimensional data. This method does not help to discern between real isolated dark clouds or gas reservoirs within/around galaxies without the additional information on spatial extent that comes from the emission of the neutral gas.
\item [(ii)] \textit{HI$-$21 cm direct imaging} \citep[e$.$g$.$,][]{2005AJ....130.2598G, 2008A&A...482...43G}. This approach is observationally limited to the dark clouds detected in the local Universe because this line is too weak to be detected at high redshift using current ground$-$based telescopes.
\item [(iii)] \textit{Fluorescent emission induced by the cosmic ultraviolet background (UVB)}, as proposed by the pioneering works of  \cite{1987MNRAS.225P...1H} and \cite{1996ApJ...468..462G}. This radiation is produced by ionized gas that recombines and emits fluorescent HI Ly$\alpha$\footnote{HI Ly$\alpha$ line  $=$ atomic hydrogen de$-$excitation from the \textit{2$^{2}$P} to the \textit{1$^{2}$S} level that results in the emission of a single photon with energy 10.2 eV and $\lambda$ = 1215.67\,$\mathrm{\AA}$.} photons \citep{2005ApJ...628...61C}. The main drawback of this method is the intrinsic faintness of the UVB emission that would imply a Ly$\alpha$ surface brightness of SB\,$\sim$\,10$^{-20}$\,erg\,s$^{-1}$\,cm$^{-2}$\,arcsec$^{-2}$  \citep{2008ApJ...681..856R}, which makes the detection with current facilities very challenging.
\item [(iv)] \textit{QSO$-$induced fluorescent Ly$\alpha$ emission} can locally boost the signal from dense and otherwise dark gas clouds by orders of magnitude \citep[C12]{2001ApJ...556...87H,2005ApJ...628...61C, 2010ApJ...708.1048K} acting as a flashlight on its surroundings. Notwithstanding the complex interpretation of the physics behind the Ly$\alpha$ fluorescence \citep[e$.$g$.$,][among others]{2003A&A...407..147F, 2004MNRAS.353..301F, 2007ApJ...657..135C, 2008ApJ...681..856R, 2013ApJ...766...58H}, and thanks to the support of the 3D radiative transfer models, this seems to be the most promising observational approach and forms the basis of the present investigation using MUSE.
\end{enumerate}

Despite the predictions by several numerical simulations and observational efforts with 8$-$10 meter class telescopes, in most of the studies conducted so far, the proto$-$galactic phase preceding the first spark of SF has been poorly constrained. The most convincing observational evidence for this dark phase at high redshift are the objects presented in C12. Using the fluorescent emission induced by the QSO UM287 at redshift 2.4 they detected in a 20hr deep narrow$-$band (NB) image with VLT$-$FORS 12 dense and compact gas$-$rich emitters, named ``Dark Galaxies" (DG hereafter), with no detected continuum (stellar) counterpart. The rest$-$frame equivalent widths, EW$_{0}$, $>$ 240\,$\mathrm{\AA}$ of these DG cannot be easily explained by normal  star forming regions (Salpeter stellar initial mass function, \citealt{2002ApJ...565L..71M, 1993ApJ...415..580C}). There are several limitations in the methodology employed in C12. For instance, it required a custom$-$made NB filter centered on the QSO redshift. This implies that (\textsc{i}) the estimation of the QSO redshift must be very precise; (\textsc{ii}) the results have to take into account possible filter losses and (\textsc{iii}) the candidates need to be confirmed with spectroscopic data. Another limitation concerns the comparison of their results with previous works, because their control samples can be affected by the different observational strategies of the ``blank-field" surveys in the literature.

Therefore, the challenging question that we would like to consider here regarding the nature of the dark galaxies is: \\

\textit{Do DGs exist at higher (z\,$>$\,2.4) redshifts and what can be learned from their redshift evolution?} \\

In order to answer this question, we will use (1) an alternative approach of searching for the fluorescent Ly$\alpha$ emission produced by bright QSOs at \textit{z}\,$>$\,3 as well as (2) the advantages of an Integral Field Unit (IFU) like the MUSE instrument \citep{2010SPIE.7735E..08B} with the final aim at investigating how the IGM gas is converted into stars. The homogeneous data quality and large wavelength range, translating into a large cosmological volume, offered by the MUSE data presents an unparalleled opportunity for this kind of study, including an analysis of this process with bi$-$dimensional information. With the help of the third dimension, i$.$\,e$.$ the wavelength information missing from NB surveys, we have direct spectroscopic confirmation and also the possibility to explore the presence of other emission lines (e$.$\,g$.$, [C\,{\textsc{iv}}]\,$\lambda\lambda$\,1548,1550\, and [He\,{\textsc{ii}}]\,$\lambda$\,1640\,). 
More importantly, the use of Integral Field Spectroscopy (IFS) provides the ability to build control samples with essentially the same instrumental and observational conditions, as well as data reduction and analysis techniques, with respect to the main dataset. One drawback will be the relatively small MUSE field of view (MUSE FoV 1\,$\arcmin \times$\,1\,$\arcmin$) with respect to previous NB images by C12 (VLT$-$FORS FoV $\sim$\, 7\,$\arcmin \times$\,7\,$\arcmin$) in exploring the fluorescent volume around the QSO. Indeed, based on the C12 work, we expect to find only 1 or 2 DGs per MUSE field around each QSO. 
For this reason, in this paper we are combining medium-deep MUSE observations ($>$\,9 hours total exposure time per field) obtained on 6 different fields containing bright quasars. 

The QSOs photoionize the surrounding gas, boosting the faint Ly$\alpha$ fluorescent glow expected from the cold gas by a factor of 100$-$1000 (within a distance of about 10 comoving Mpc) with respect to fluorescence due to the UVB only. Uncertainties include the variable luminosities of the QSOs, the uncertain UV continuum \citep{2015MNRAS.449.4204}, the QSO opening angle \citep{2013ApJ...775L...3T} and further complexities related to the resonant nature of the Ly$\alpha$ line. 

Here, we present the MUSE detection of 11 high EW$_{0}$ ($>$ 240\,$\mathrm{\AA}$) objects within six medium-deep ($>$\,9 hours) fields at \textit{z}\,$>$\,3, of which 8 of these intriguing objects are possible DG candidates fluorescently illuminated by the QSOs. In addition, we present the discovery of a (control sample) population of $\sim$\, 200 Ly$\alpha$ emitters (LAEs) detected in the same fields. 

The paper is organized as follows. In \S{} 2 we describe the sample providing details of the MUSE observations, data reduction and post processing. In \S{} 3 we present the systematic analysis of both continuum detected and undetected Ly$\alpha$ emitters within the six MUSE fields. Our results are presented in \S{} 4 and we discuss our findings in \S{} 5. The summary and the conclusions are presented in \S{} 6. Finally, we publish the catalog of LAEs in the Appendix.\\

Throughout the paper we adopt a flat $\Lambda$CDM cosmology with Wilkinson Microwave Anisotropy Probe 9 cosmological parameters of $\Omega_{\Lambda}$\,=\,0.714, $\Omega_{M}$\,=\,0.286 and h\,=\,0.693 \citep{2013ApJS..208...19H}, corresponding to $\sim$\, 7.5 kpc/$\arcsec$ at redshift $\sim$\, 3. We use vacuum wavelengths for the spectral analysis and all magnitudes are in units of the AB system \citep{1983ApJ...266..713O}.\\

\begin{deluxetable*}{l|cccccc}[!]
\tablecaption{MUSE medium-deep fields major properties. \label{tab:fields}}
\tablehead{
\colhead{} & \colhead{Bulb} & \colhead{Hammerhead} & \colhead{Q0055-269} & \colhead{Q1317-0507} & \colhead{Q1621-0042} & \colhead{Q2000-330} 
}
\startdata
RA (J2000)		&	04:22:01.5		&		23:21:14.7   	&    00:57:58.1		&    13:20:30.0			&    16:21:16.9				&		20:03:24.0			\\
Dec (J2000)		&	-38:37:19		&		+01:35:54   	&    -26:43:14		&    -05:23:35			&    -00:42:50				&		-32:51:44			\\
Redshift\tablenotemark{a} 	&	3.094		    &		3.199			&    3.662			&    3.7			    &    3.7				    &		3.783			\\
{\it z}$_{\mathrm{ CIV}}$\tablenotemark{b} 	 & 3.110 & 3.202 & 3.634 & 3.701 & 3.689 &  3.759 \\
Exp. Time	(hr)	&	20   &		9 	&    10	&    9.75	&    8.75		&		10		\\
Class\tablenotemark{c}		&	Type II-AGN		        &		RQ-QSO				&    RQ-QSO 			    &    RQ-QSO 			        &    RQ-QSO 				        &		RL-QSO 			\\
V\tablenotemark{d} (AB mag) & 24.76 & 19.33 & 17.99 &  18.10 &  17.88 &  17.84\\ 
PSF\tablenotemark{e} ($\arcsec$)	    &	0.70		    &		0.76		  	&    0.84		&    0.74		    &    0.77				    &		0.84			\\
3$\sigma_{\mathrm{Cont}}$ @1$\arcsec$\tablenotemark{f}	(AB mag)	&	29.0	&		28.4	 		&    28.6			&    28.3				&    28.2				&		28.5				\\	 
Ly$\alpha$ Sensitivity @1$\arcsec$\tablenotemark{f} (AB mag)	&	30.8	&		30.2	 		&    30.5			&    30.6				&    30.3				&		30.7			\\	
\enddata
\tablenotemark{}{}\\
\tablenotemark{a}{ Redshift values from the catalog of \textcolor{blue}{V{\'e}ron-Cetty \& V{\'e}ron (2010)}. }\\
\tablenotemark{b}{ Computed from the luminosity$-$corrected \citep{2016ApJ...831....7S} C IV emission line measurement from the MUSE spectra.}\\
\tablenotemark{c}{ Class refers to the type of powering source in the field, i.e. AGN, Radio$-$quiet (RQ) QSO and Radio$-$loud (RL) QSO on the basis of the radio flux measurements (Flux[1.4 GHz] threshold  5mJy) presented in the \textcolor{blue}{V{\'e}ron-Cetty \& V{\'e}ron (2010)} catalog.}\\
\tablenotemark{d}{ Measured in a 3$\arcsec$ diameter aperture on the reconstructed MUSE$-$V image, i$.$e$.$ MUSE datacube convolved with the V-Johnson filter, without accounting for the foreground Galactic absorption.}\\
\tablenotemark{e}{ Mean FWHM of the Gaussian fit measured on different point sources in the final combined datacube at 7000\,$\mathrm{\AA}$ using both {\ttfamily SExtractor} and {\ttfamily QFitsView} tools.}\\
\tablenotemark{f}{ These values are computed within a 1$\arcsec$ diameter aperture.}\\
\end{deluxetable*}

\section{Sample and Observations} \label{sec:sample}

\begin{figure*}[ht!]
\vspace{1.4cm}
\includegraphics[scale=0.54]{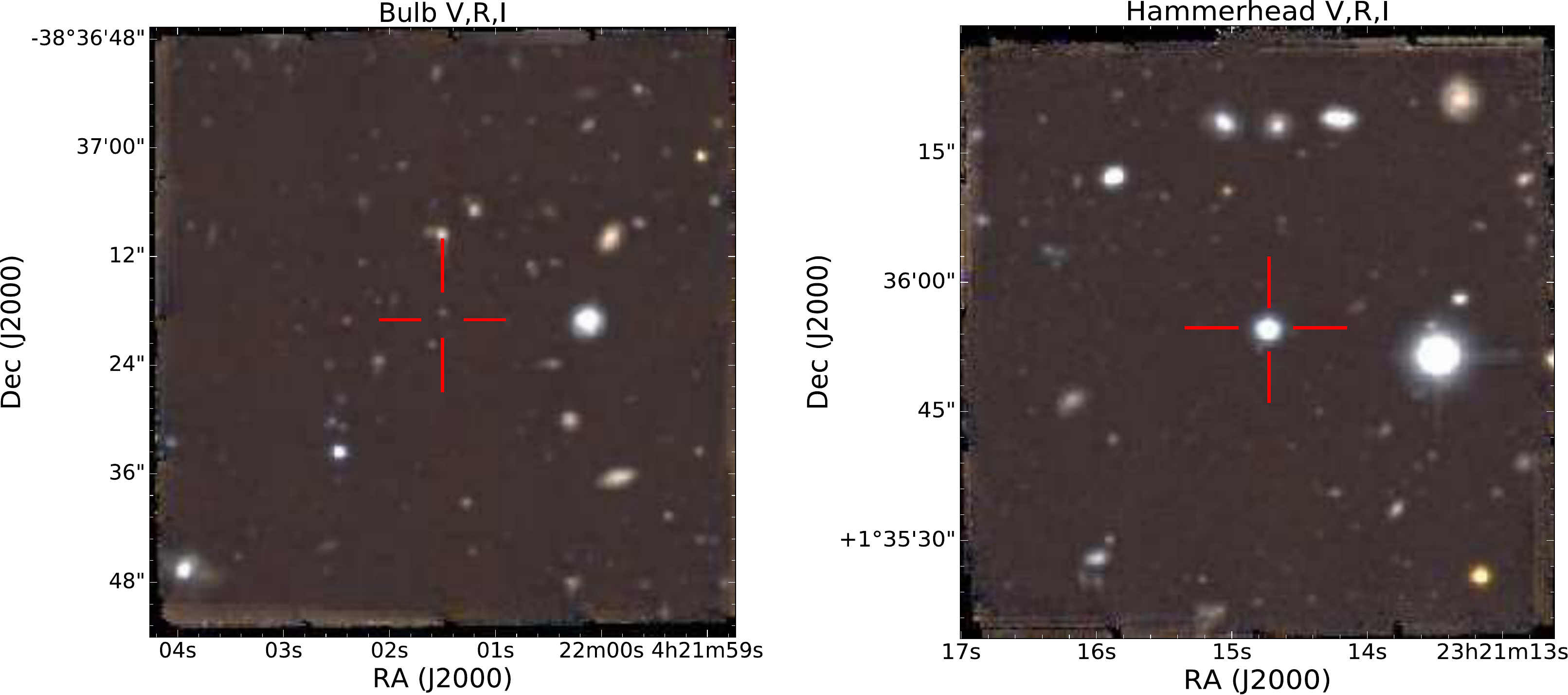}
\caption{Composite pseudo$-$color images of the low redshfit ($z\,<$\,3.2) MUSE fields. The RGB colors are assigned to V$-$, R$-$, and I$-$band images computed from the MUSE datacubes. Each image is 60 $\arcsec \times$ 60 $\arcsec$ and the red cross indicates the AGN and QSO location in the case of the Bulb and the Hammerhead fields, respectively. North is up and east to the left. \label{fig:RGB_lowz}}
\end{figure*}

Our observations were carried out with MUSE, the second generation IFU mounted on the Very Large Telescope (VLT) at the Nasmyth B focus of the Yepun (Unit Telescope 4) in Paranal, Chile. 
MUSE has uniquely powerful performance: relatively large FoV (in wide$-$field mode, WFM, 1\,$\arcmin \times$\,1\,$\arcmin$) combined with the excellent spatial sampling (0.2$\arcsec$) and spectral resolutions (R from $\sim$\,1750 to $\sim$\, 3500) over the wide optical wavelength window (from 4650\,$\mathrm{\AA}$ to 9300\,$\mathrm{\AA}$) and high throughput (35\,\% at 7500\,$\mathrm{\AA}$). \\  

\begin{figure*}[ht!]
\centering
\includegraphics[scale=0.56]{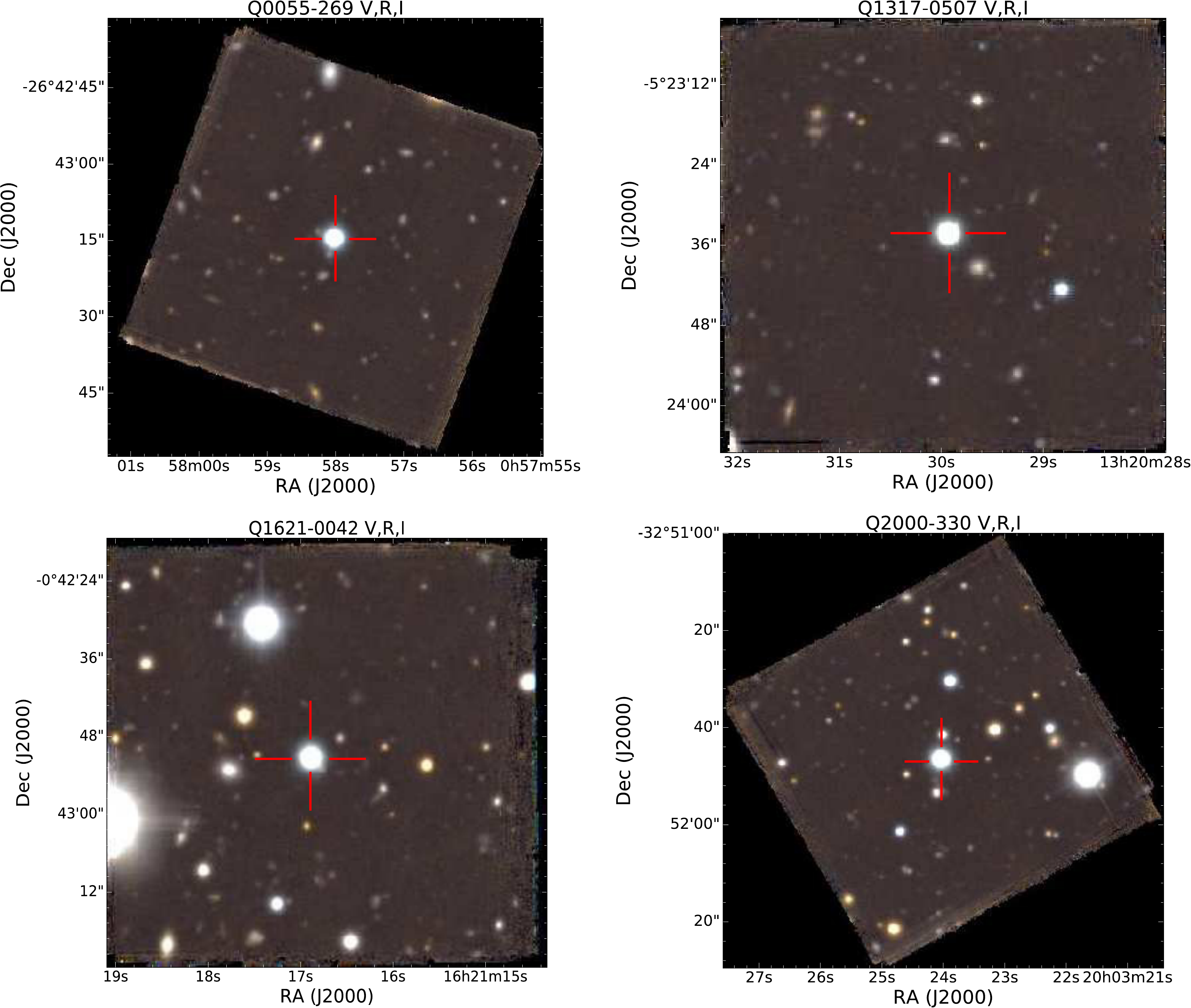}
\caption{Composite pseudo$-$color images of the MUSE QSO fields at \textit{z}\,$>$\,3.7. The RGB colors are assigned to V$-$, R$-$, and I$-$band images from the MUSE datacubes. The red cross indicates the QSO location. North is up and east is to the left.\label{fig:RGB_highz}}
\end{figure*}

\begin{deluxetable*}{ll|c|cc|ccc|c}
\tablecaption{Statistics of the detected emitters. For the six fields, we list the number of detection obtained in both the on$-$source datacube (the one centered on the QSO Ly$\alpha$ redshift) and the two control samples (Off$-$Blue and Off$-$Red). Each datacube has a width of 200\,$\mathrm{\AA}$ in the spectral direction, with the exception of the Bulb Off$-$Blue sample which, due to the (low) Ly$\alpha$ redshift of the AGN, has a width of 131\,$\mathrm{\AA}$. The last column (skylines layers) indicates the number of masked layers in the datacube due to presence of some residual skyline features. \label{tab:LAES_statistic}}
\tabletypesize{\small}
\tablehead{
\colhead{} &\colhead{} &  \colhead{Detected} & \colhead{LAEs} & \colhead{LAEs} & \colhead{[OII]} & \colhead{[OIII]} & \colhead{AGN/} &  \colhead{Skylines} \\
\colhead{} &\colhead{} &  \colhead{Emitters} & \colhead{w Continuum} & \colhead{w/o Continuum} & \colhead{Emitters} & \colhead{Emitters} & \colhead{Galaxies} &  \colhead{layers} 
}
\startdata
		     &	Off-Blue		     &		10   	&    2		&    4		& $-$ 	   &	1   &    3   & 	$-$  	\\
Bulb		&	On-Source		&		22   	&    7    &    11	&    2		&	 $-$    &	   2   & $-$  		\\
	         &	Off-Red		    &		14		&   1		&    9	    &    4     &     $-$   &	    $-$     & 14		\\
\hline	         
		     &	Off-Blue		     &		40   	&    9		&    28		&   2	   &	1   &    $-$    & 	5	\\
Hammerhead	&	On-Source		&		22   	&    3	    &    17	&    1	&	 1    &	 $-$     & 	 $-$	\\
	         &	Off-Red		    &		33	&   5		&    20	    &    2     &   6   &	  $-$       & 18		\\	
\hline	         
		     &	Off-Blue		     &		4   	&    $-$  	&    4		&    $-$	   &	 $-$ &    $-$    & 	 $-$	\\
Q0055-269	&	On-Source		&		13   	&    7	    &    5	&    1 	&	  $-$    &	 $-$     & 17		\\
	         &	Off-Red		    &		10	&   2		&    2    &    3     &    1   &	 2       & 21		\\	
\hline	         
		     &	Off-Blue		     &		8   	&    4		&    1		&   3	   &	 $-$   &  $-$      & 	20	\\
Q1317-0507 	&	On-Source		&		7   	&    3	    &    1	&    2		&	  $-$    &	   1   & 1		\\
	         &	Off-Red		    &		6	&   1		&    3    &    1     &   1    &	   $-$      & 27		\\	
\hline	         
		     &	Off-Blue		     &		8   	&    3		&    4		&   1	   &   $-$ &   $-$     & 	17	\\
Q1621-0042	&	On-Source	&		2   	&     $-$    &    2	&    	 $-$	&	  $-$   &	   $-$    & 	 $-$	\\
	         &	Off-Red		    &		8		&   3		&    3    &    1     &     1  &	    $-$     & 28		\\	
\hline	          
		     &	Off-Blue		     &		13   	&    6	&    6		&   1	   &	  $-$  &   $-$     & 	14	\\
Q2000-330 	&	On-Source		&		8   	&    3	    &    3	&    1	&	 1    &	 $-$      & 14	\\
	         &	Off-Red		    &		4		&   1		&    3	    &     $-$     &    $-$    &	     $-$    & 19		\\		
\hline
 &	 	TOTAL    &		232		&   60		&   126	    &     25    &     13  &	     8   &  $-$	\\	
\hline
\enddata
\end{deluxetable*}

\subsection{Sample} \label{subsec:sample}

The six medium-deep fields at \textit{z}\,$>$\,3 analyzed in this study were observed between September 2014 and April 2016. They form part of two MUSE Guaranteed Time of Observation (GTO) programs (094.A-0396, 095.A-0708, 096.A-0345, 097.A-0251, 098.A-0678 PI: S. Lilly; 094.A-0131, 095.A-0200, 096.A-0222, 097.A-0089, 098.A-0216 PI: J. Schaye). The observations comprise 270 exposures ($\approx$\, 65 hours) in total. Each MUSE datacube consists of 321\,$\times$\,328 spaxels with a sampling grid of 0.2$\arcsec \times$\,0.2$\arcsec \times$\,1.25\,$\mathrm{\AA}$ yielding $\sim$\,90,000 spectra per frame. We use {\ttfamily SExtractor} \citep{1996A&AS..117..393B} and {\ttfamily QFitsView}\footnote{{\ttfamily QFitsView} v3$.$1 is a FITS file viewer using the QT widget library and was developed at the Max Planck Institute for Extraterrestrial Physics by Thomas Ott.} on the NB images centered at 7000\,$\mathrm{\AA}$ to measure the seeing (mean full width half maximum, FWHM) on the final combined datacubes. We perform a Gaussian fit to the brightest point sources in each frame. From this, we obtain an average seeing across all frames better than 0.85\,$\mathrm{\AA}$. Most of the observations were carried out under clear or photometric conditions. From the quality assessment of the final combined MUSE datacubes, we obtain a mean (over the six fields) 3$\sigma$ flux continuum limit in a 1$\arcsec$ diameter aperture of 28.5 AB mag whereas 30.5 AB mag represents the mean sensitivity value for the Ly$\alpha$ flux detection (see Section \ref{subsec:limits} for details on how these sensitivities were computed). Table \ref{tab:fields} summarizes the measured properties for each field. Their short individual descriptions are provided in the next section. The composite pseudo$-$color images constructed from the MUSE datacube combining the broad V$-$, R$-$ and I$-$band images are shown in Figures \ref{fig:RGB_lowz} and \ref{fig:RGB_highz}. 
We decided to split our sample into two sub-samples by the redshifts of the targeted QSOs in the respective fields, since our observations target six fields with a difference in the QSO redshift of $\Delta\,z\,\approx\,$0.7 (maximum). Such difference can be important in terms of both cosmological surface brightness dimming \citep{1930PNAS...16..511T,1934rtc..book.....T} that scales as (1+$z$)$^{4}$, as well as in terms of the explored physical volume. Throughout the paper we will use the term ``lower redshift sample" to refer to the fields at \textit{z}\,$<$\,3.2 and ``high redshift" to refer to those at \textit{z}\,$>$\,3.7. \\

\begin{figure*}
\centering
\includegraphics[width=\columnwidth]{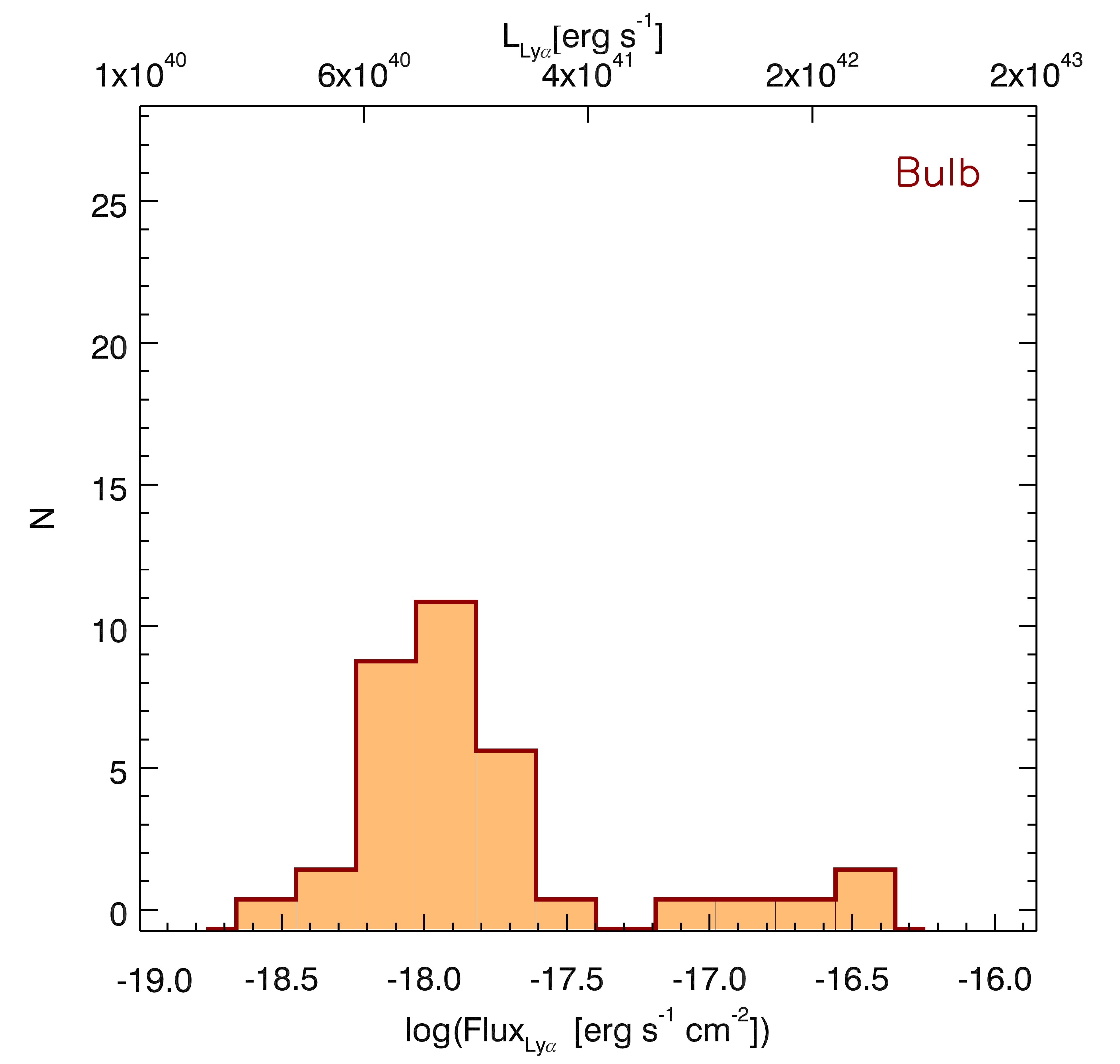}
\includegraphics[width=\columnwidth]{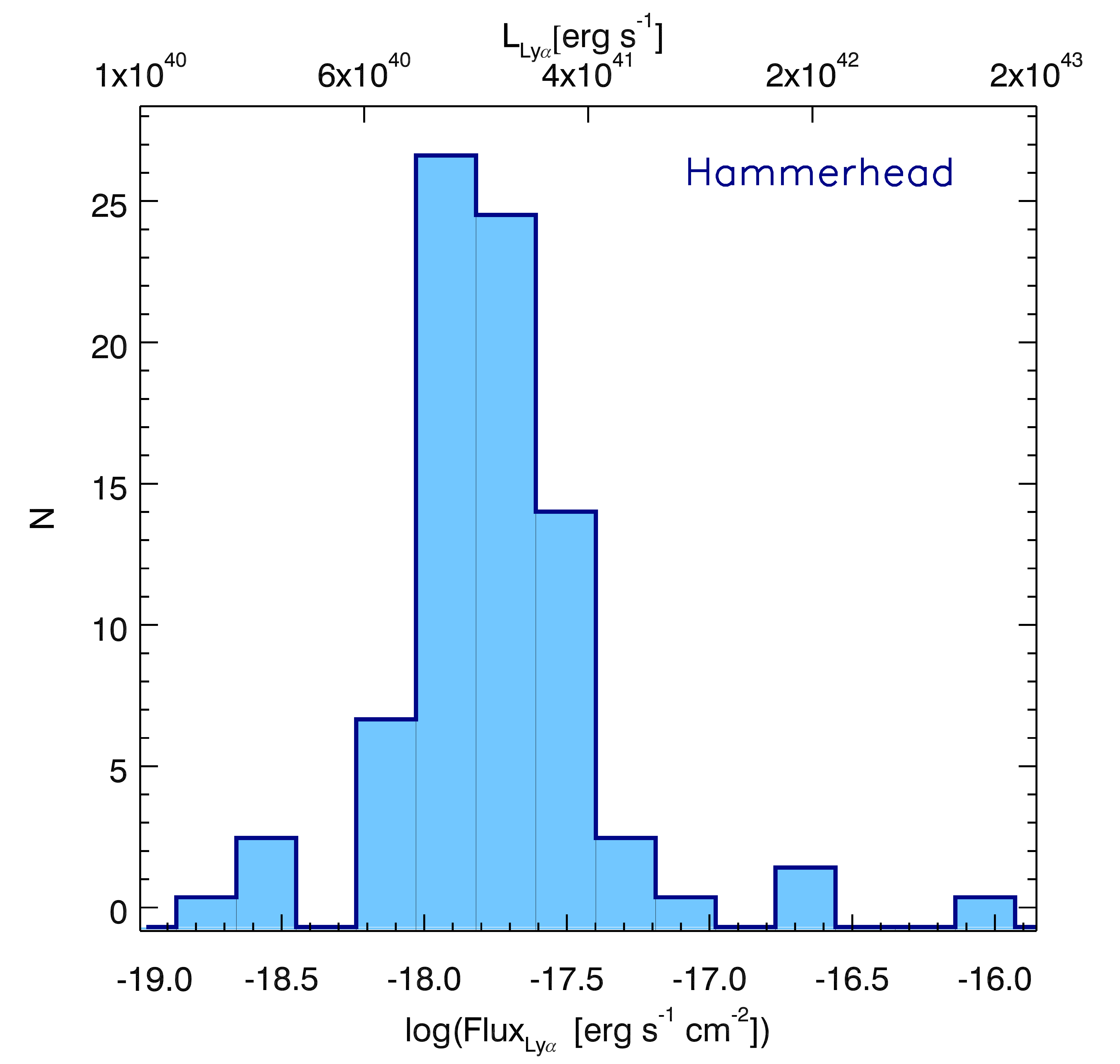}\\
\vspace{0.5cm}
\includegraphics[width=\columnwidth]{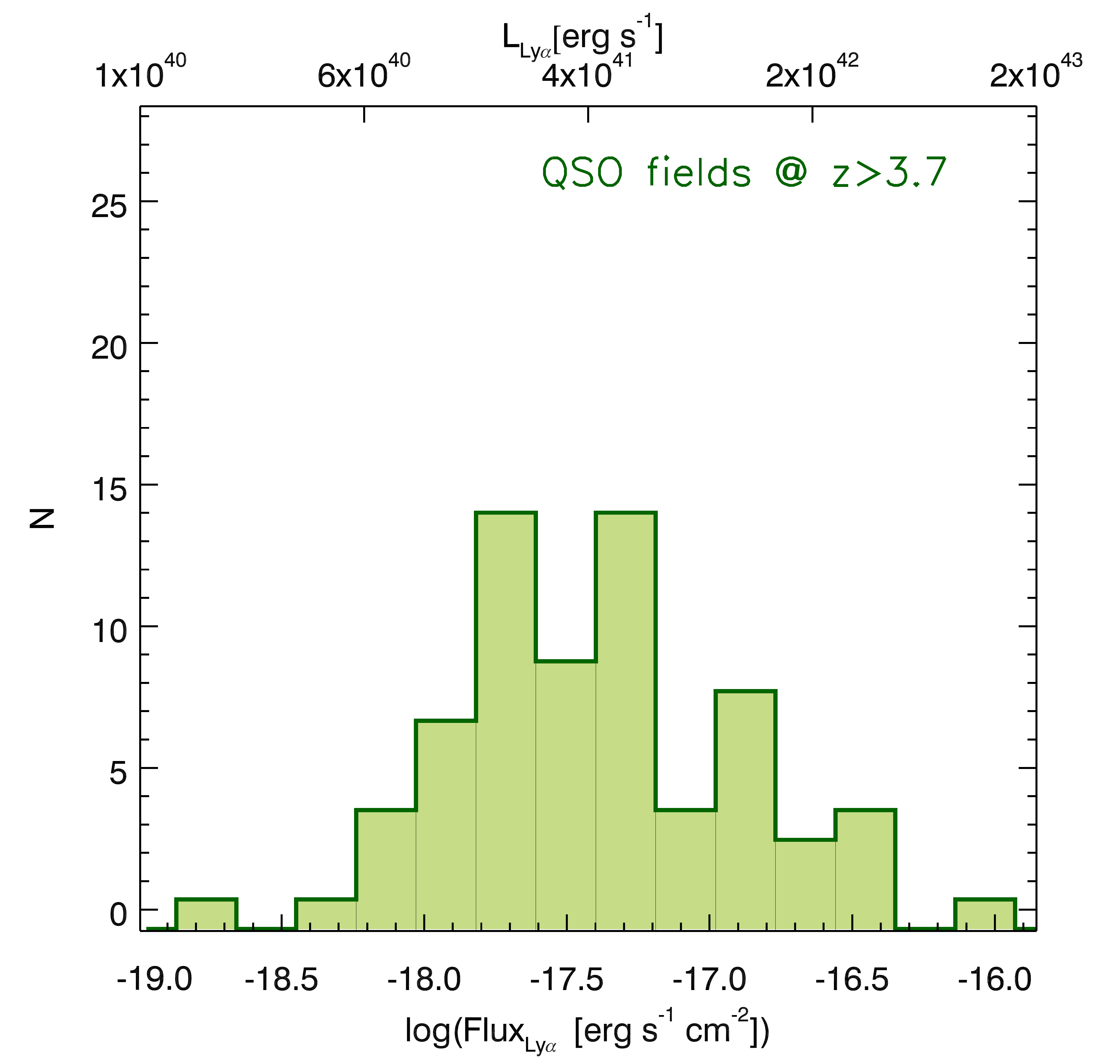}
\includegraphics[width=\columnwidth]{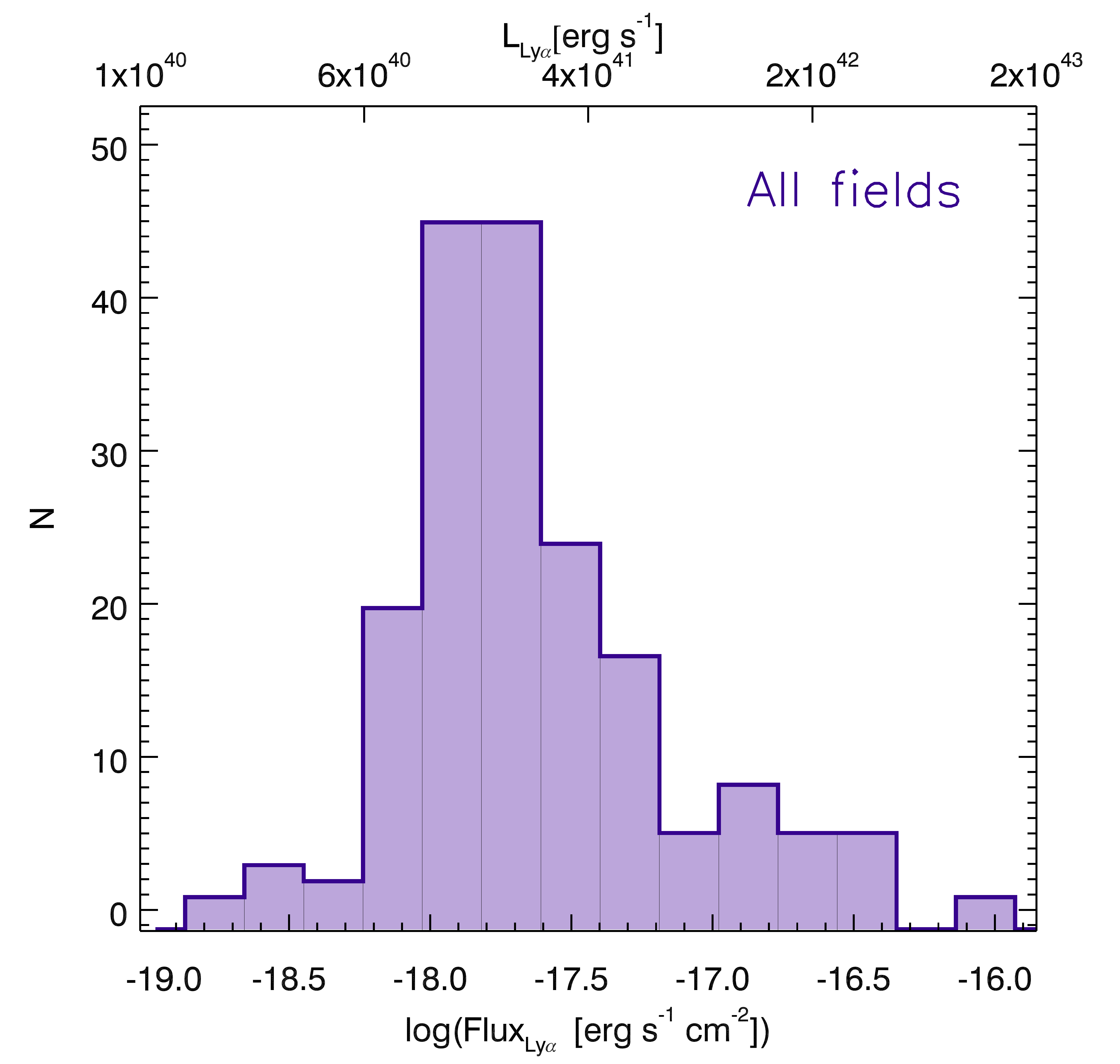}\\
\caption{Ly$\alpha$ flux (and luminosity, top x$-$axis) distribution of all the LAEs detected within our MUSE medium-deep fields. In the top row we plot the low redshift sample, with the Bulb LAEs on the left and the Hammerhead ones on the right. The high redshift sample is shown in the bottom left panel. The distribution of all the detected LAEs is presented (with purple histograms) in the bottom right panel. \label{fig:histo_flux}}
\end{figure*}

\vspace{1.cm}
\subsubsection{Notes on individual Fields} \label{subsec:fields_notes}
\hspace{2.cm}{ \it Low redshift sample}\\

{ \bf $-$ Q0422-3837 or Bulb Nebula:} This is the lowest redshift field,  \textit{z}\,$=$\,3.094, within our sample. Differently from other fields, this observation was targeting a known Ly$\alpha$ nebula around a galaxy that is $\sim$\,19 comoving Mpc (cMpc) from a bright QSO. It had been discovered through NB imaging \citep{2016ApJ...830..120B}. Our MUSE observations revealed a previously unknown Type II AGN at its center, $\alpha_{(J2000)}$=04\,:\,22\,:\,01.5 and $\delta_{(J2000)}$=-38\,:\,37\,:\,19 . It was observed during 20 hours with MUSE. The size of the Point Spread Function (PSF) measured on the final datacube at 7000\,$\mathrm{\AA}$ and based on different point sources is 0.7$\arcsec$ (the best seeing in our sample). This field is present in both the GALEX \citep{2012AAS...21934001S} and Spitzer \citep{2012AAS...21941001C} catalogs, but to our knowledge nothing remarkable about this field has been previously published. The name Bulb comes from the appearance of the Ly$\alpha$ nebula around this AGN in the NB survey that will be presented in a forthcoming paper (Cantalupo et al. in prep.). The RGB synthetic image is shown in the left panel of Fig. \ref{fig:RGB_lowz}, where the position of the AGN is marked with the red cross.\\

{ \bf $-$ Q2321+0135 or Hammerhead Nebula:} The second field in our low redshift sample is centered on a radio quiet (RQ)\footnote{This classification is taken from the \cite{2010A&A...518A..10V} catalog and is based on the radio flux measured at 1.4 GHz, which for a RQ QSO should be $<$\,5 mJy.} QSO at \textit{z}\,$=$\,3.199, $\alpha_{(J2000)}$=23\,:\,21\,:\,14.7 and $\delta_{(J2000)}$=+01\,:\,35\,:\,54, and it is presented in the right panel of Fig. \ref{fig:RGB_lowz}. This QSO was first spectroscopically discovered in Ly$\alpha$ emission by \cite{1987ApJ...316L...1S} and was also observed in the Sloan Digital Sky Survey \citep[SDSS,][]{2000AJ....120.1579Y} with a subsequent follow up by the Baryon Oscillation Spectroscopic Survey \citep[BOSS,][]{2012A&A...548A..66P}. These authors confirmed the detection of the C\,IV $\lambda$ 1550 line, which is likewise detected in the MUSE integrated spectra. The PSF measured at 7000\,$\mathrm{\AA}$ is 0.76$\arcsec$. Similarly to the Bulb case, a huge Ly$\alpha$ nebula around this QSO was discovered in NB imaging \citep{ElenaThesis}. More details on the Hammerhead will be provided in Marino et al. (in prep.). \\

\hspace{2.cm}{ \it High redshift sample}\\

{ \bf $-$ Q0055-269:} The RQ QSO Q0055-269, $\alpha_{(J2000)}$=  00\,:\,57\,:\,58.1 and $\delta_{(J2000)}$=-26\,:\,43\,:\,14, at  \textit{z}\,$=$\,3.662 is part of our high redshift sample. This interesting QSO presents several emission and absorption features also confirmed by previous UVES observations \citep{2013A&A...556A.141Z}, and it was the subject of many studies \citep[among others]{2002A&A...392..395C,2003ApJ...596..768S,2014MNRAS.441.1916B}. The PSF measured on the 10hr MUSE datacube is 0.84$\arcsec$, and it was observed with a position angle (PA) of 70$^{\circ}$ as plotted in the top$-$left panel of Fig. \ref{fig:RGB_highz}.\\

{ \bf $-$ Q1317-0507:} Q1317-0507 is a RQ QSO at $\alpha_{(J2000)}$=13\,:\,20\,:\,30.0 and $\delta_{(J2000)}$=-05\,:\,23\,:\,35, at  \textit{z}\,$=$\,3.7. Despite the poor photometric data available in the literature, this QSO has good spectral coverage with UVES. The original time exposure was 10 hours but due to a satellite passing by during one observation, we simply rejected one exposure (15 minutes). The RGB image of this field is shown in the top$-$right panel of Fig. \ref{fig:RGB_highz} and the PSF measured is 0.74$\arcsec$. \\

{ \bf $-$ Q1621-0042:} This RQ QSO, $\alpha_{(J2000)}$=16\,:\,21\,:\,16.9 and $\delta_{(J2000)}$=-00\,:\,42\,:\,50, with \textit{z}\,$=$\,3.7, is part of the SDSS-DR7 quasar catalog by \cite{2010AJ....139.2360S}. Due to the availability of panchromatic photometric observations together with UVES spectra, this is one of the metal rich QSO used to probe the time evolution of the C\,IV absorbers \citep{2013ApJ...763...37C}. The PSF for the 35 combined exposures (i$.$\,e$.$ 8.45 hours, we had to exclude one problematic exposure due to its offset shifts) is 0.77$\arcsec$. \\

{ \bf $-$ Q2000-330:} The highest redshift field and the only radio loud (RL) QSO within our sample is located at $\alpha_{(J2000)}$=20\,:\,03\,:\,24.0 and $\delta_{(J2000)}$=  -32\,:\,51\,:\,44 with \textit{z}\,$=$\,3.783. The high resolution spectrum of this QSO was taken with the HIRES \citep[High Resolution Echelle Spectrometer,][]{1994SPIE.2198..362V} instrument and it is part of the KODIAQ survey \citep{2015AJ....150..111O} along with several other investigations mainly focused on characterizing the CGM. It was observed with MUSE during 10 hours with a PA of 30$^{\circ}$. The PSF in the final datacube has a Gaussian FWHM of 0.84$\arcsec$ at $\lambda=7000\,\mathrm{\AA}$. \\

\vspace{1.cm}

\subsection{Data Reduction and Post-Processing} \label{subsec:reduction}
The reduction of all 65\,hr MUSE data was performed using some of the standard recipes from the latest version of the ESO MUSE Data Reduction Software \citep[DRS, pipeline version 1.6,][]{2015scop.confE..53W}, complemented with the {\ttfamily CubExtractor} package ({\ttfamily CubEx} hereafter, version 1.6; Cantalupo, in prep.) developed to optimally improve the flat$-$fielding correction and the sky subtraction steps for our specific science case. After retrieving the raw data for each night, we first created the master calibration files using the MUSE pipeline, i$.$e$.$ the master$-$bias, the master$-$flat, the twilight and illumination correction, and wavelength calibration files. Using the DRS routine {\itshape MUSE scibasic} we then processed each individual science exposure, both standard stars and QSO fields, applying the master calibration correction with the recommended parameters. For the illumination correction step we always used the lamp flat$-$field and the twilight frames closest in time to each individual observation. All these instrumental signatures are removed for each IFU (24 in total), and as output this recipe gives the pre$-$reduced pixel tables for every IFU exposure. Next we use the {\itshape MUSE scipost} routine to create the individual datacubes, by merging the pixel tables from all IFUs of each exposure. During this step, we also performed the flux calibration using the response curve and telluric absorption correction from one spectrophotometric standard observed during the same night. In addition, {\itshape scipost} applies the geometry and astrometry tables available for each run to the science frames and performs a re$-$sampling (drizzle algorithm that maximizes the pixel fraction used) onto a 3D grid in order to construct the final datacube.
Due to the fact that our observing strategy for each field included a 4\,$\times$\,90$^{\circ}$ rotation pattern with small ($<$\,1$\arcsec$) offsets, the automatic correction for the absolute astrometry obtained with the pipeline is a source of some uncertainty. For this reason, a double check of the pipeline astrometry correction was required and in the case of clear residual offsets we followed a more classic {\ttfamily SExtractor} approach in order to correct these offsets.

Once the {\itshape pipeline$-$level} datacubes were registered, we performed the post$-$processing using the routines {\ttfamily CubeFix}, {\ttfamily CubeAdd2Mask}, {\ttfamily CubeSharp} and {\ttfamily CubeCombine} within the {\ttfamily CubEx} package (Cantalupo, in prep.), since we are interested in reaching very faint surface brightness levels.
In particular, using {\ttfamily CubeFix} we were able to remove the typical checker$-$board pattern that is seen after the standard data reduction with the pipeline. This is achieved because we self$-$calibrate each individual exposure at the level of the IFU, slice$-$by$-$slice and vertical stacks using the sky$-$continuum and the sky$-$lines as ``flat sources" together with an iterative masking of any possible continuum sources. Thanks to the {\ttfamily CubeFix} flat$-$fielding correction, we were able to reduce the residuals to less than 0.1\% of the sky level.
Afterwards, we visually inspected the white$-$light (WL) images created from each {\itshape CubeFixed}$-$datacube. In those cases where the edges of the individual IFU slices were still visible, or if there is a bright satellite trail or even a problematic channel, we performed a manual masking using {\ttfamily CubeAdd2Mask}. 

Then, we performed a local and flux$-$conserving sky subtraction on the {\itshape CubeFixed$-$CubeMasked$-$} datacube using the {\ttfamily CubeSharp} routine. This empirical correction takes into account the sky line spread function (LSF) shifts and the variation across the MUSE FoV, conserving the flux and minimizing the residuals.
Both {\ttfamily CubeFix} and {\ttfamily CubeSharp} were performed twice in order to minimize the contamination from possible unmasked sources when the illumination correction was applied. 

Lastly, the {\itshape CubeFixed-CubeSharped} datacubes were combined with a 3$\sigma$ clipping using both mean and median statistics with the {\ttfamily CubeCombine} routine. In the case of our analysis, we use the mean$-$combined datacubes. We also refer the reader to \cite{2016ApJ...831...39B, 2016MNRAS.462.1978F, 2017MNRAS.467.4802F,2017arXiv170505728N} for further details and additional applications of these reduction procedures. \\

\begin{figure*}[ht!]
\centering
\includegraphics[scale=0.34]{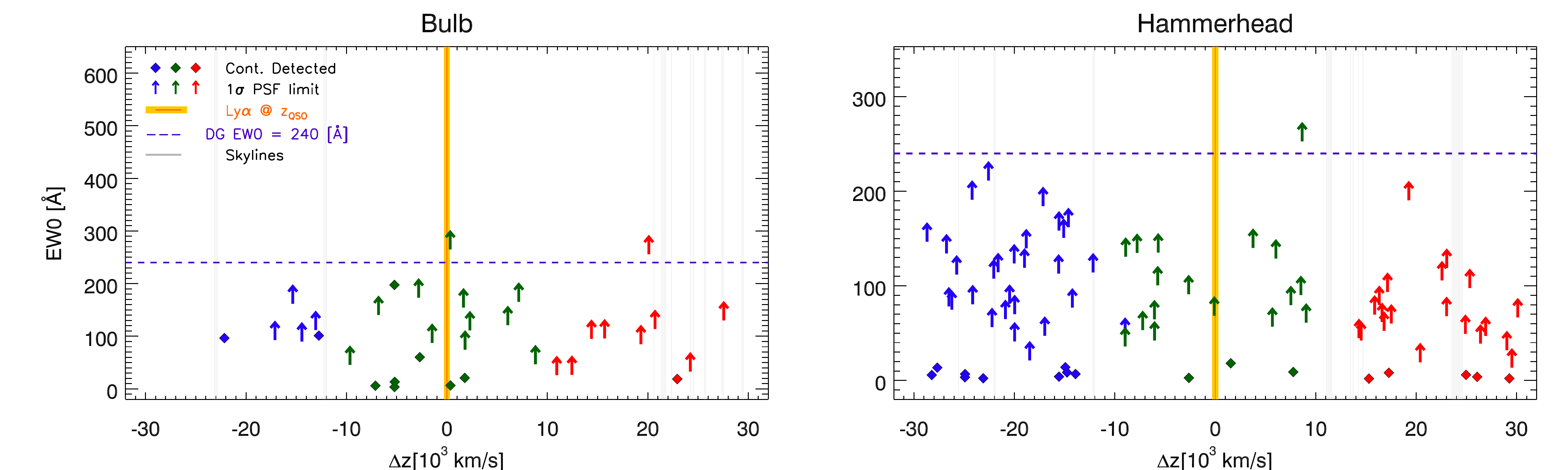}
\caption{Rest-frame Ly$\alpha$ equivalent width (EW$_{0}$) values versus the spectral distance (velocity) to the QSO of the detected Ly$\alpha$ emitters for MUSE \textit{z}\,$<$\,3.2 sample. Blue and red symbols indicate those LAEs detected in the control samples, while green symbols show the LAEs closer to the QSO. Diamonds symbolize those LAEs with continuum counterparts, and the arrows show the lower limit (at 1$\sigma$) EW$_{0}$ values for continuum undetected LAEs. The QSO velocity (plus the 1$\sigma$ error) associated with the systemic redshift calibration (415 km s$^{-1}$) is marked with the shaded yellow area and it was computed from the Ly$\alpha$ wavelength. The vertical grey shaded lines denote the masked OH skylines. The horizontal dashed line indicates the EW$_{0}$ threshold (240\,$\mathrm{\AA}$) for the dark galaxy candidates. \label{fig:EW0_multi_low_z}}
\end{figure*}

\begin{figure*}[]
\centering
\includegraphics[scale=0.34]{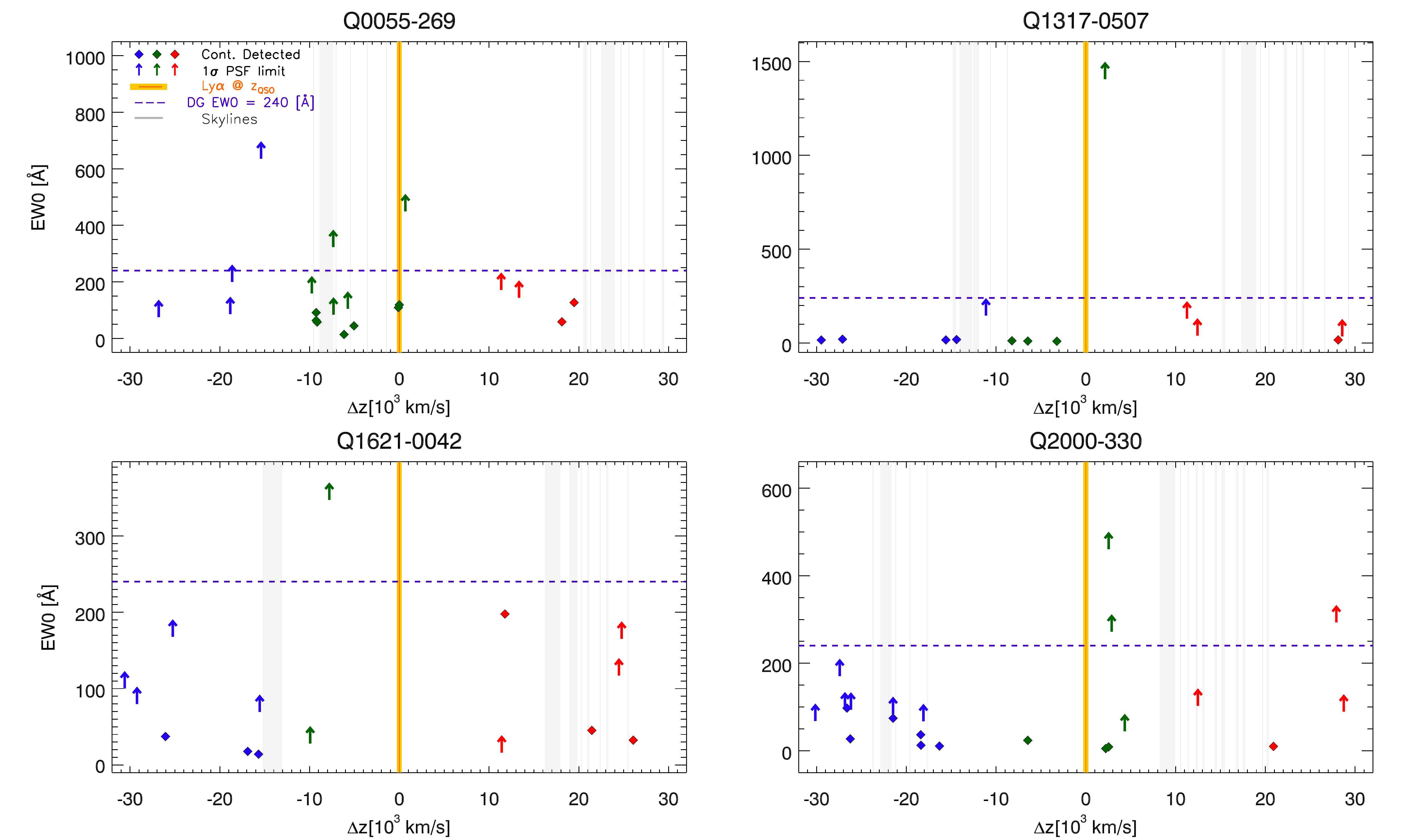}
\caption{Rest$-$frame Ly$\alpha$ equivalent width (EW$_{0}$) values versus the spectral distance (velocity) to the QSO of the detected Ly$\alpha$ emitters for the MUSE \textit{z}\,$>$\,3.7 sample. Symbols and colors are the same as for Fig. \ref{fig:EW0_multi_low_z}.  \label{fig:EW0_multi_high_z}}
\end{figure*}

\section{Analysis} \label{sec:analysis}
The goal of our analysis is to detect Ly$\alpha$ emitters within our sample. In this section, we describe our systematic search and classification of the Ly$\alpha$ emission candidates detected in the MUSE datacubes. In order to perform a consistent comparison between the 6 MUSE fields, we emphasize that the same methodology has been applied to all the MUSE fields as detailed below.

\subsection{PSF and Continuum Subtraction} \label{subsec:PSF_Cont}
Ideally, since we are interested in emission line objects around the QSO (in both spectral and spatial directions) and not in QSO Ly$\alpha$ nebula, removing the nuclear contribution of the quasar should not be necessary for the detection of faint and compact targets. Nonetheless, we decided to perform a PSF subtraction to ensure minimum contamination from the QSO PSF in our LAEs detection by making use of the empirical PSF subtraction of the {\ttfamily CubePSFSub} routine (part of the {\ttfamily CubeExtractor} package). Using an averaged$-$sigma$-$clipping algorithm, {\ttfamily CubePSFSub} constructs and rescales the QSO PSF using the NB images created for each wavelength layer, giving excellent results on large scales around the QSO \citep{2016ApJ...831...39B}. The next step was the subtraction of the brightest foreground continuum sources within our fields that were carefully removed using {\ttfamily CubeBKGSub}. This routine estimates the continuum voxel$-$by$-$voxel\footnote{The volumetric (3D spatial and spectral) pixel element in IFU datacubes).} on the basis of a median$-$filtering performed on the spectrum, which is integrated in 50\,$\mathrm{\AA}$ bins and smoothed with a median filter radius of 3 pixels. This allows us to avoid any prominent line features and also to reduce the computational time. Some residuals are still visible in the output datacube, but this has a minimal impact on the extraction procedure of our LAEs considering that we are masking all the bright continuum sources detected from the WL image.

\subsection{Detection and extraction of Ly$\alpha$ emitters} \label{subsec:Lya_det}
One of the most important advantages of the IFS is that we can explore the same spatial area over a wide spectral range. To exploit the full capabilities of our MUSE data, our strategy to detect Ly$\alpha$ emitters within our sample was to build three different sub$-$cubes from each datacube with the same spectral width 200\,$\mathrm{\AA}$ (or 160 spectral pixels). The on$-$source datacube is centered on the QSO Ly$\alpha$ wavelength. Two control sample sub$-$cubes adjacent to the on$-$source datacube were extracted on the blue and red sides. For practical reasons, they have the same spectral width as the on$-$source sub$-$cube. This choice of the spectral width is justified in terms of the maximum volume \citep[10 cMpc,][]{2013ApJ...775L...3T} where the signature of the fluorescent emission can be detected. 

\begin{figure*}[!]
\centering
\includegraphics[width=\linewidth]{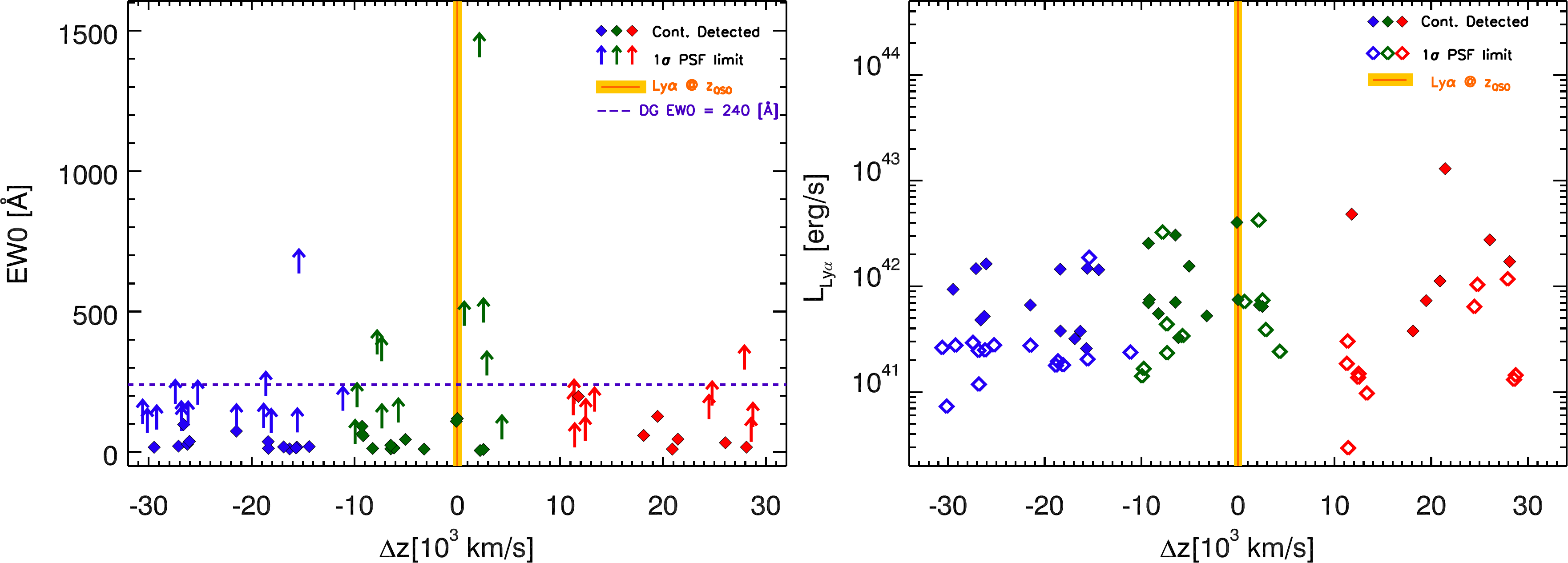}
\caption{ Stacked EW$_{0}$(Ly$\alpha$) values (left) and Ly$\alpha$ luminosities (right) versus the spectral distance (velocity) from the QSO for the fields at \textit{z}\,$>$\,3.7 (Q1317, Q0055, Q1621, Q2000). Symbols and colors are the same as for Fig. \ref{fig:EW0_multi_low_z}, except in the case of the luminosities distribution where we use empty diamonds (instead of arrows) to plot the continuum undetected (CU) LAEs. \label{fig:high_z}}
\end{figure*}

In total we extracted 6 on$-$source datacubes. These are represented with green symbols in Figures \ref{fig:EW0_multi_low_z}, \ref{fig:EW0_multi_high_z} and \ref{fig:high_z}. We also extracted a total of 12 control sample datacubes represented with blue and red colors in the same figures.
As mentioned above, the difference in redshifts between our fields corresponds to slightly different analyzed volumes along the spectral direction, because of the constant area coverage. These distances span a range from 36 physical Mpc (pMpc) at redshift\,$<$\,3.2 to 27 pMpc at redshift\,$>$\,3.7. 
We blindly implemented three$-$dimensional source detection on the 18 reduced and post$-$processed datacubes using {\ttfamily CubExtractor} with the same threshold parameters. 

Aside from the routines described above, the main purpose of the {\ttfamily CubExtractor} software is the 3D automatic extraction of sources based on a novel approach used in computer science vision to detect connected regions in binary digital images (see Cantalupo in prep.). The algorithm uses subsets of connected components uniquely labeled on a user$-$defined property basis, i$.$\,e$.$ connected$-$labelling$-$component \citep{Shastock}. Specifically, we first smooth (with a radius of 0.4$\arcsec$) both the science and variance datacubes only in the spatial directions for each wavelength layer. Then we require that all detected objects fulfill three conditions: a minimum of 40 connected voxels above a signal$-$to$-$noise ratio (SNR) threshold of 3.5 (after the re$-$scaling factor accounting for the propagated variance is applied) along with a SNR measured on the 1D extracted spectrum above 4.5. 

Since the extraction process is based on the noise, estimating the noise correctly is a crucial ingredient of our selection criteria. Since the MUSE pipeline variance tends to be an underestimate of this noise (see Sect$.$3 in \citealt{2015A&A...575A..75B}), we use the propagated variance datacube computed by {\ttfamily CubEx} that takes the noise sources introduced by both the MUSE pipeline and the {\ttfamily CubEx} post-processing steps into account. The propagated variance is used to calculate the re$-$scaling factor applied to each wavelength layer, which in the most extreme case is $\approx$\,1.95. We also carefully mask the brightest and extended continuum sources detected in the WL image of each datacube\footnote{In order to select the  brightest and extended continuum sources, we run {\ttfamily CubEx} on the datacubes using as detection threshold a SNR of 10 and we also required that each object have a minimum of 100 connected voxels.}, as well as possible skyline residuals to minimize possible artificial detections.

\begin{figure*}
\centering
\includegraphics[width=\columnwidth]{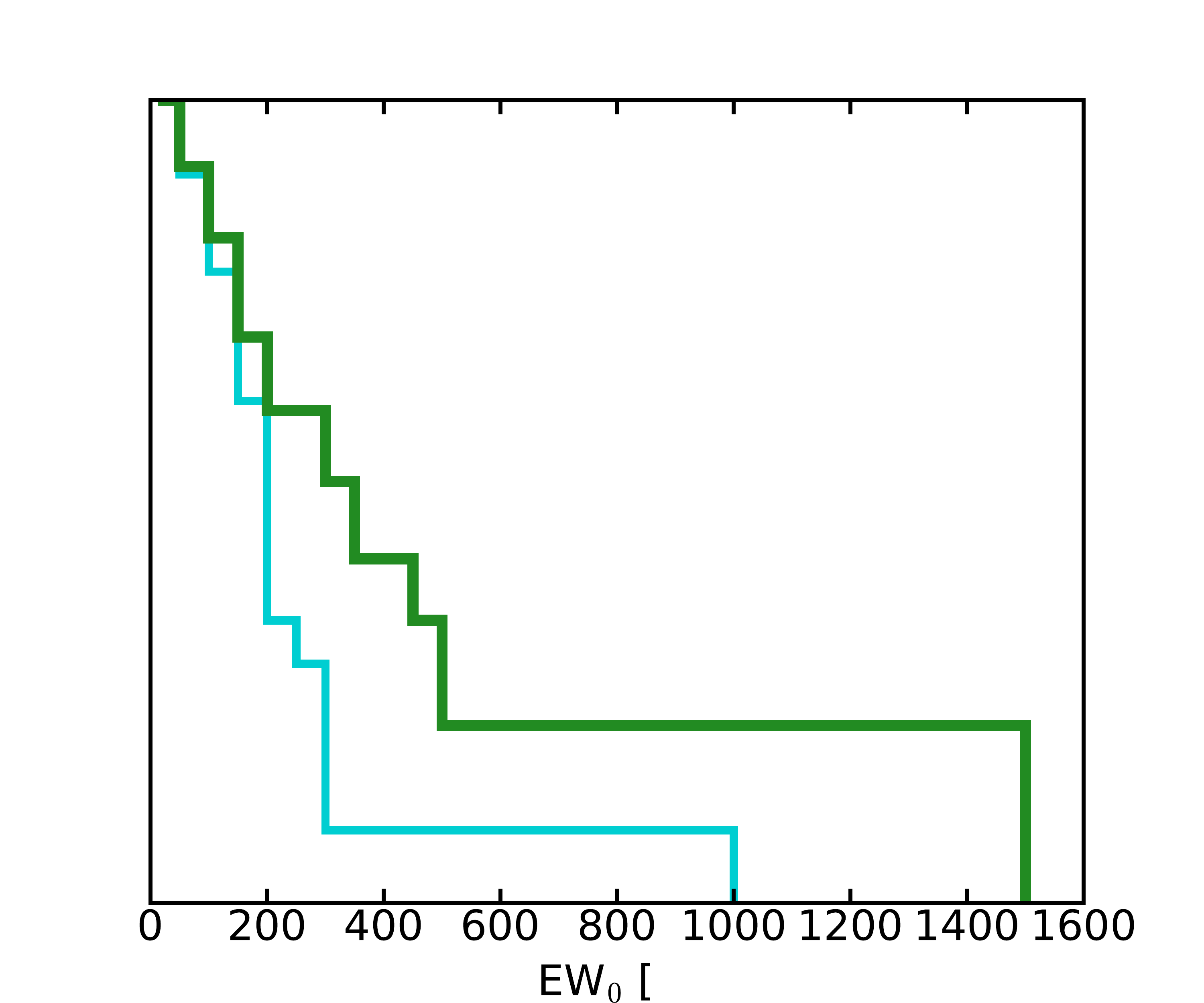}
\includegraphics[width=\columnwidth]{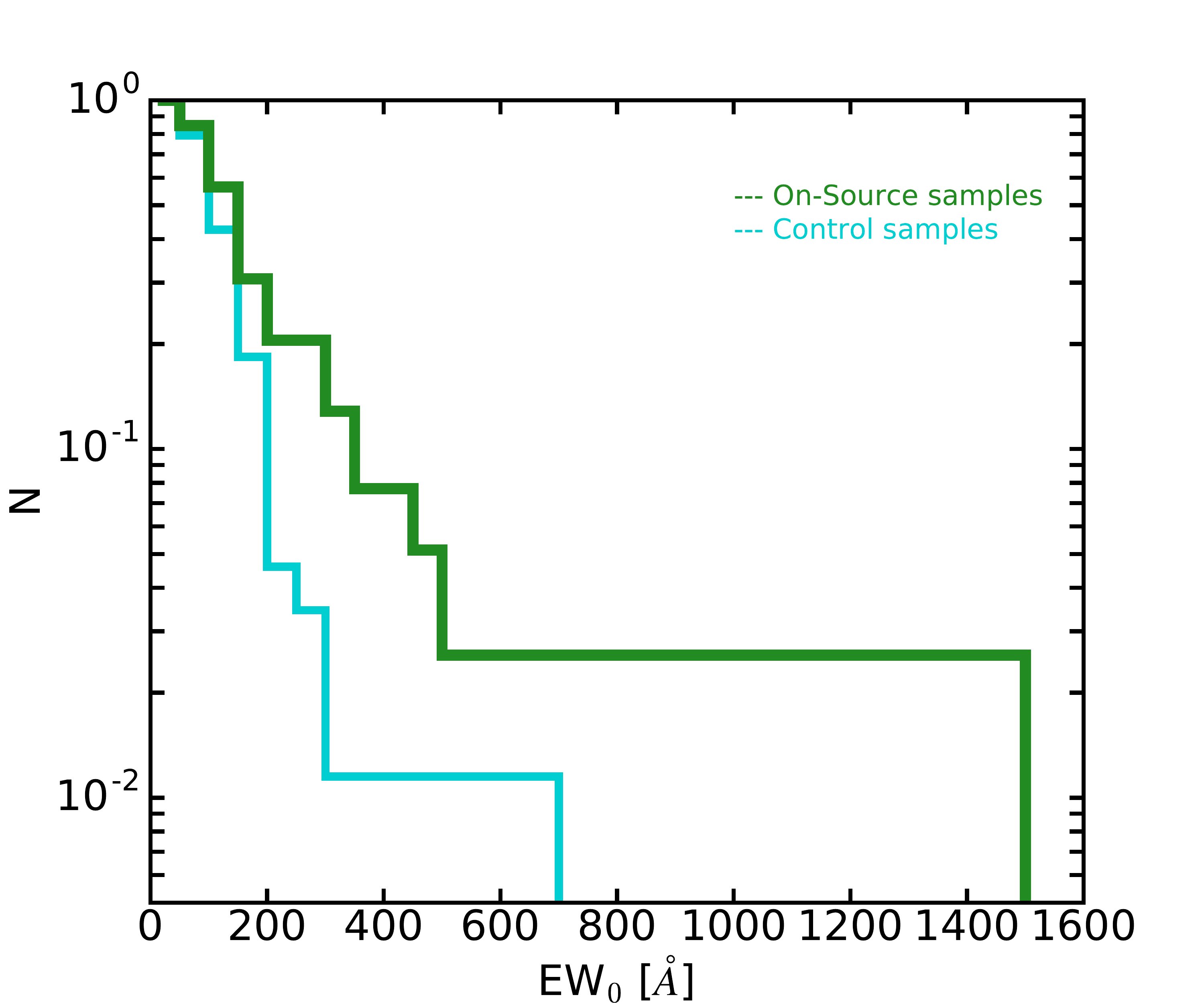}
\caption{Cumulative rest-frame equivalent width (EW$_{0}$) distribution of all LAEs (left panel; all points in Figs. \ref{fig:EW0_multi_low_z} and \ref{fig:EW0_multi_high_z}) and the undetected continuum LAEs (right panel; only arrow symbols in Figs. \ref{fig:EW0_multi_low_z} and \ref{fig:EW0_multi_high_z}). Cyan solid line represents the control sample distribution while the green line marks the fluorecently illuminated QSO LAEs. \label{fig:histo_cum}}
\end{figure*}

As a result of the three$-$dimensional segmentation map, we obtain a full catalog of all the line emitters automatically detected in each MUSE field for the on$-$source and control sample datacubes.

\subsection{Classification of the Ly$\alpha$ emitters} \label{subsec:lya_class}

Although we extensively tested our selection criteria, visual inspection is necessary to remove possible spurious detections of LAEs, such as possible contaminants from [O\,{\textsc{ii}}]\,$\lambda\lambda$\,3726,3729, [O\,{\textsc{iii}}]\,$\lambda$5007 and AGN emitters that were able to pass through the previous masking. Therefore, for each object in our catalog, we tabulated both spectral and photometric information. Specifically, we visually checked the extracted 1D spectrum, where, accounting for different redshift solutions, we were able to distinguish pure Ly$\alpha$ and other emitters by identifying the most prominent emission and absorption line features. 

Regarding the photometric properties, from the MUSE datacubes we produce (1) the optimally$-$extracted (OE), (2) the classical pseudo-NB, (3) continuum and (4) WL images centered on each candidate with a typical size of\,  30$\arcsec \times$\, 30$\arcsec$.
The OE images are constructed by combining all voxels  along the wavelength direction that are within the corresponding 3D mask of each detected object from the PSF$-$ and continuum$-$subtracted MUSE datacubes. This image can be interpreted as a pseudo$-$NB with a spectral width optimized for the SNR of the candidate (see also Appendix A in \citealt{2016ApJ...831...39B} for a detailed comparison of the OE with the pseudo$-$NB images). 

As we will discuss in the next section, the choice of the continuum image is very critical, especially because based on this image we define a line emitter to be continuum (or not) detected. The ideal case would be the availability of Hubble Space Telescope (HST) images, but these are not available for these fields. We can however take advantage of our IFU datacubes and build the broad$-$band continuum image. Hence, our approach was to create three continuum images by coadding different spectral ranges. For each field, we considered the spectral layers redward of their QSO Ly$\alpha$ emission and the continuum images were created combining 800 (1000\,$\mathrm{\AA}$), 1600 (2000\,$\mathrm{\AA}$) and all ($\sim$\,3000\,$\mathrm{\AA}$) the wavelength layers in the red part of the datacube. Finally, due to the limited coverage on the blue side of the QSO Ly$\alpha$ emission, we conservatively assumed a continuum slope of $\beta$=-2 ($f_{\lambda}  \propto  \lambda^{\beta}$ in wavelength space, \citealt{1999ApJ...521...64M}) to take in account the shape of the continuum. Then we performed a global statistic of these continuum images while masking the sources in each field. Our final selection of the best continuum image was the deepest one of the three. From the tests performed on our data, the 2000\,$\mathrm{\AA}$ continuum image turns out to be the deepest, because its width represents the best spectral compromise able to minimize the contribution of the sky$-$residual layers. For the sake of completeness, we also checked the classical pseudo$-$NB and white light images. 
The results of our classification are summarized in Table \ref{tab:LAES_statistic}, where we provide the full statistics of the detected line emitters. In a total volume of $\sim$\,90 physical Mpc$^{3}$, we find 186 LAEs, 25 [O\,{\textsc{ii}}], 13 [O\,{\textsc{iii}}] emitters and 8 AGN candidates.

\begin{deluxetable*}{l|c|c|c|c|c|c|c|c|c|c|c}
\tablecaption{Derived properties of the Dark Galaxy candidates.  \label{tab:dg_table}}
\tabletypesize{\tiny}
\tablewidth{0pt}
\tablehead{
\colhead{Field} & \colhead{ID}  & \colhead{RA} & \colhead{Dec} & \colhead{Area} & \colhead{$\lambda_{\mathrm{detected}}$}  &  \colhead{Redshift}  &  \colhead{Flux(Ly$\alpha$)\tablenotemark{a}} & \colhead{L(Ly$\alpha$)} & \colhead{Flux(Cont$_{\mathrm{PSF}}$)\tablenotemark{b}} &\colhead{ EW$_{0}$(Ly$\alpha$)\tablenotemark{c}} & \colhead{M$_{\mathrm{gas}}$\tablenotemark{d}}  \\
\colhead{}    & \colhead{}    & \colhead{(J2000)} & \colhead{(J2000)} & \colhead{(pixels$^{2}$)} & \colhead{(\AA)} & \colhead{} & \colhead{(10$^{-17}$\,erg\,s\,$^{-1}$\,cm$^{-2}$)} &   \colhead{(10$^{41}$\,erg\,s\,$^{-1}$)} &  \colhead{(10$^{-20}$\,erg\,s\,$^{-1}$\,cm$^{-2}$\,\AA$^{-1}$)} &  \colhead{(\AA)} &  \colhead{(10$^{9}$ M$_{\sun}$)} 
}
\startdata
Bulb &       24  & 04:22:02.904 &  -38:37:43.71 &   41 & 4984.50  & 3.102  &   0.16 $\pm$ 0.01 & 1.35 &   -0.02 $\pm$ 0.14 &   $>$265 & 0.2\\
Hammerhead & 78 & 23:21:14.776 &  01:36:02.12 &   49 & 5175.52 &  3.259   &   0.29 $\pm$ 0.02 & 2.82 &   0.10 $\pm$ 0.27 &   $>$253  & 0.4 \\
Q0055-269 &  9  & 00:58:00.108 & -26:43:26.42 &   98 & 5585.45   &  3.596   &   0.35 $\pm$ 0.02 & 4.40 &   0.05 $\pm$ 0.24 &   $>$323  & 0.6 \\
Q0055-269 &  39 & 00:57:57.721 & -26:42:57.52 &  121 & 5665.77   &  3.662   &   0.55 $\pm$ 0.02 & 7.14 &   0.08 $\pm$ 0.26 &   $>$450  & 1.0\\
Q1317-0507 & 14 & 13:20:29.317 & -05:23:52.02 &  314 & 5732.48   &  3.717   &   3.12 $\pm$ 0.06 & 42.1 &   0.26 $\pm$ 0.47 &   $>$1406 & 5.9 \\
Q1621-0042 & 2  & 16:21:14.791 &  00:42:26.18 &   71 & 5644.02   &  3.644   &   2.52 $\pm$ 0.05 & 32.4 &   1.12 $\pm$ 1.56\tablenotemark{e} &   $>$347  & 4.5 \\
Q2000-330 & 18 & 20:03:24.882 & -32:51:46.95 &   81 & 5825.74   &  3.794   &   0.52 $\pm$ 0.02 & 7.38 &   0.04 $\pm$ 0.24 &   $>$461& 1.0  \\
Q2000-330 & 20 & 20:03:25.213 & -32:52:04.57 &   55 & 5829.11   &  3.797   &   0.27 $\pm$ 0.02 & 3.87 &   0.01 $\pm$ 0.21 &   $>$272 & 0.5  \\
\enddata
\tablenotemark{}{}

\tablenotemark{a}{The Ly$\alpha$ flux is computed from the curve$-$of$-$growth analysis detailed in Sec$.$4.3.}

\tablenotemark{b}{The continuum flux is computed as the maximum between the fluxes measured in 9 adjacent PSF size apertures, i$.$e$.$ it will be always positive.}

\tablenotemark{c}{The rest$-$frame EWs were determined using the PSF$-$ aperture approach, see Eq$.$ 4. }

\tablenotemark{d}{The gas masses are computed using Eq$.$ 8 in C12.}

\tablenotemark{e}{This measurement is relatively high due to the position of this target in the edge of the FoV.}

\end{deluxetable*}

\subsection{Estimation of our detection limits} \label{subsec:limits}

In order to compute the minimum flux for which we would not be able to detect any candidates, we determine our detection limits for both the continuum and Ly$\alpha$ emission line using the standard deviation (\textit{std}) of 100 random locations for each field in our sample. The \textit{std} is calculated on the continuum and pseudo$-$NB Ly$\alpha$ images where we mask out all the bright sources with special attention to the scattered light and halos of bright foreground stars. We explored successively larger apertures, with radii from 0.2$\arcsec$ to 2$\arcsec$ (including the PSF radius) and a 3$\sigma$ clipping algorithm. We also compare these values with the results from pixel$-$by$-$pixel statistics, i$.$e$.$ the theoretical photon count noise variance, to measure the level of systematics resulting from the sky and continuum subtraction. The typical surface brightness values obtained in a 10hr datacube within an aperture of 1$\arcsec$ in diameter are of the order of 10$^{-20}$\,erg\,s\,$^{-1}$\,cm$^{-2}\, \mathrm{\AA}^{-1}$ arcsec$^{-2}$ in the case of the continuum and 10$^{-19}$\,erg\,s\,$^{-1}$\,cm$^{-2}$ arcsec$^{-2}$ for the Ly$\alpha$ emission. Fig.\,\ref{fig:histo_flux} shows the distribution of the Ly$\alpha$ fluxes and luminosities of the selected LAE candidates.

\section{Results} \label{sec:results}

In this section we present our sample of $\sim$\,200 LAEs detected in our MUSE datacubes in proximity of quasars and in the control regions within a total volume of $\sim$\,90 physical Mpc$^{3}$. In particular, we will focus on 
 the LAE Ly$\alpha$ luminosities and equivalent width and their distribution in function of distance from the quasars. The overall properties of the sample is presented in Table \ref{tab:all_LAEs_table} in Appendix A. \\

\vspace{0.5cm}

\subsection{Ly$\alpha$ flux estimations}
Given the recent findings of the extended and diffuse nature of the Ly$\alpha$ emission from LAEs \citep[][\textcolor{blue}{Leclercq et al. 2017, submitted}]{2016A&A...587A..98W}, measuring reliable Ly$\alpha$ fluxes is not a trivial task because it might depend on both the methodology and available data.

In our analysis, the Ly$\alpha$ fluxes were accurately computed from the curve$-$of$-$growth analysis \citep[C$.$o$.$G$.$ following][]{2016arXiv160902920D, 2016A&A...587A..98W} performed on the pseudo NB image centered on the QSO Ly$\alpha$ wavelength with a width of 200\,$\mathrm{\AA}$.
 By collapsing the corresponding spectral channels of the on$-$source datacube and assuming the {\ttfamily CubeEx} coordinates for each target, the Ly$\alpha$ C$.$o$.$G$.$ was computed using the fluxes extracted from concentric circular annuli of increasing radii (in steps of 0.2$\arcsec$) up to 4$\arcsec$. This results in a reasonable value for the characterization of compact objects and their possible extended emission. The total Ly$\alpha$ flux of each object was then determined from the integrated value out to the radius where the surface brightness within a 0.2$\arcsec$ annulus is equal to or less than zero. Using the C$.$o$.$G$.$ approach, we are able to recover LAEs as faint as 10$^{-19}$\,erg\,s\,$^{-1}$\,cm$^{-2}$. Fig. \ref{fig:histo_flux} shows the distribution of the Ly$\alpha$ fluxes and luminosities for each low redshift field (Bulb in orange and Hammerhead in blue), for the high redshift fields (in green) and for the full sample (in purple). 
Although there are definitely uncertainties and limitation in our calculations of Ly$\alpha$ fluxes, we stress that we have used exactly the same method for both the main and the control sample.

\begin{deluxetable*}{l|c|c|c|c|c|c|c|c|c|c}
\tablecaption{Derived properties of Lyman $\alpha$ candidates with EW$_{0}$ $>$ 240\,$\mathrm{\AA}$ detected in the control samples.\label{tab:Lae_high_EW_table}}
\tabletypesize{\tiny}
\tablewidth{0pt}
\tablehead{
\colhead{Field} & \colhead{ID}  & \colhead{RA} & \colhead{Dec} & \colhead{Area} & \colhead{$\lambda_{\mathrm{detected}}$}  &  \colhead{Redshift}  &  \colhead{Flux(Ly$\alpha$)} & \colhead{L(Ly$\alpha$)} & \colhead{Flux(Cont$_{\mathrm{PSF}}$)} &\colhead{ EW$_{0}$(Ly$\alpha$)} \\
\colhead{}    & \colhead{}    & \colhead{(J2000)} & \colhead{(J2000)} & \colhead{(pixels$^{2}$)} & \colhead{(\AA)} & \colhead{} & \colhead{(10$^{-17}$\,erg\,s\,$^{-1}$\,cm$^{-2}$)} &   \colhead{(10$^{41}$\,erg\,s\,$^{-1}$)} &   \colhead{(10$^{-20}$\,erg\,s\,$^{-1}$\,cm$^{-2}$\,\AA$^{-1}$)} &  \colhead{(\AA)} 
}
\startdata
Bulb & 22 &	04:21:59.656 & -38:37:39.14 & 105 & 5182.19 & 3.264 & 0.28 $\pm$ 0.01 & 2.77 &  0.03 $\pm$ 0.17 & $>$370 \\  
Q0055-269  &   6 & 00:57:59.131 &  -26:43:10.75 & 149 & 5504.99 & 3.530 & 1.57 $\pm$ 0.04 & 18.7 & 0.36  $\pm$ 0.55 & $>$636 \\
Q2000-330 &  7 &  20:03:23.891 & -32:51:58.87 & 132 & 6079.46 & 4.003 & 0.73 $\pm$ 0.03 & 11.7 & 0.30  $\pm$ 0.50 & $>$293 \\
\enddata
\end{deluxetable*}

\begin{figure*}
\centering
\includegraphics[width=0.98\columnwidth, angle=180]{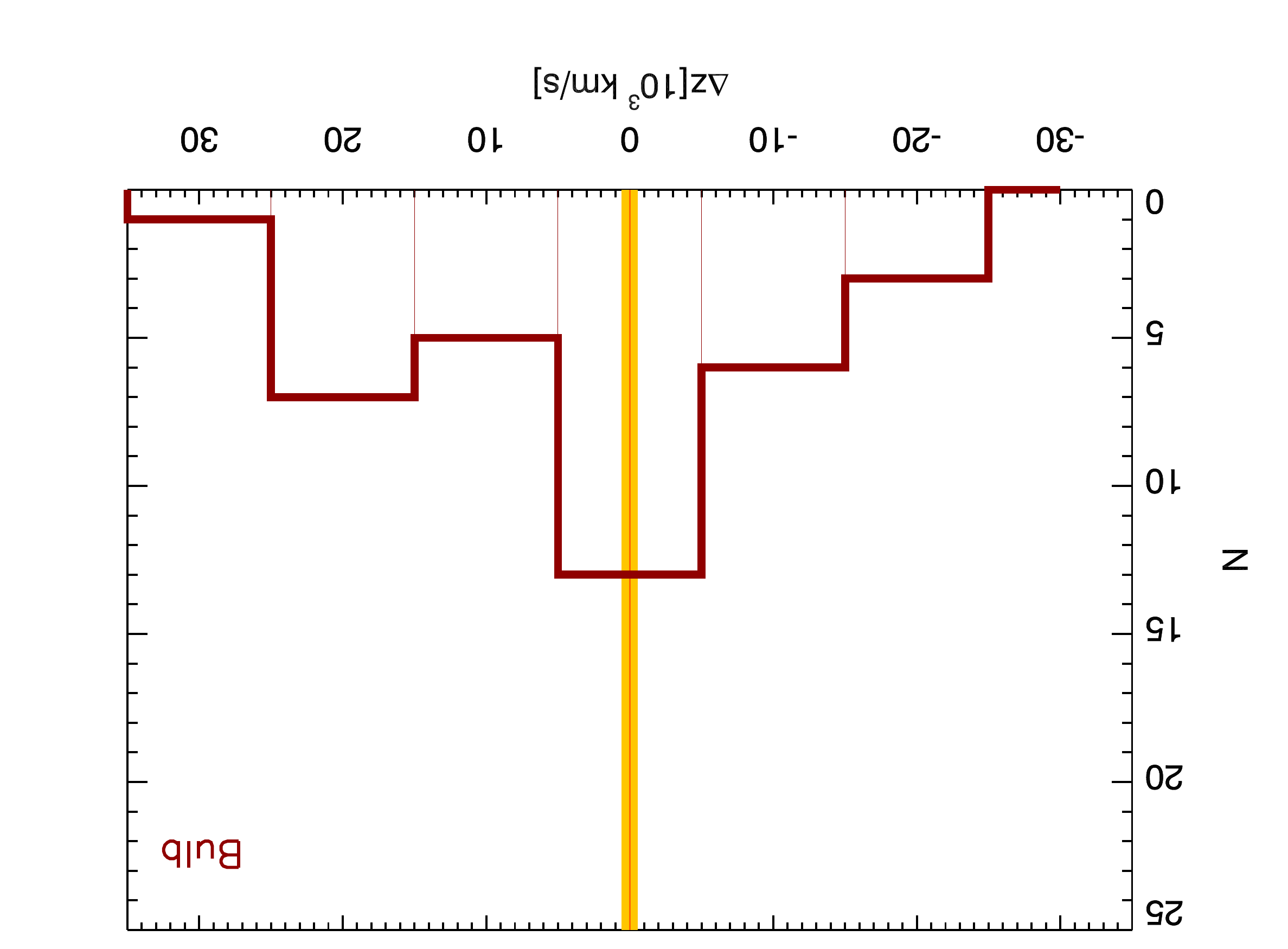}
\includegraphics[width=0.98\columnwidth, angle=180]{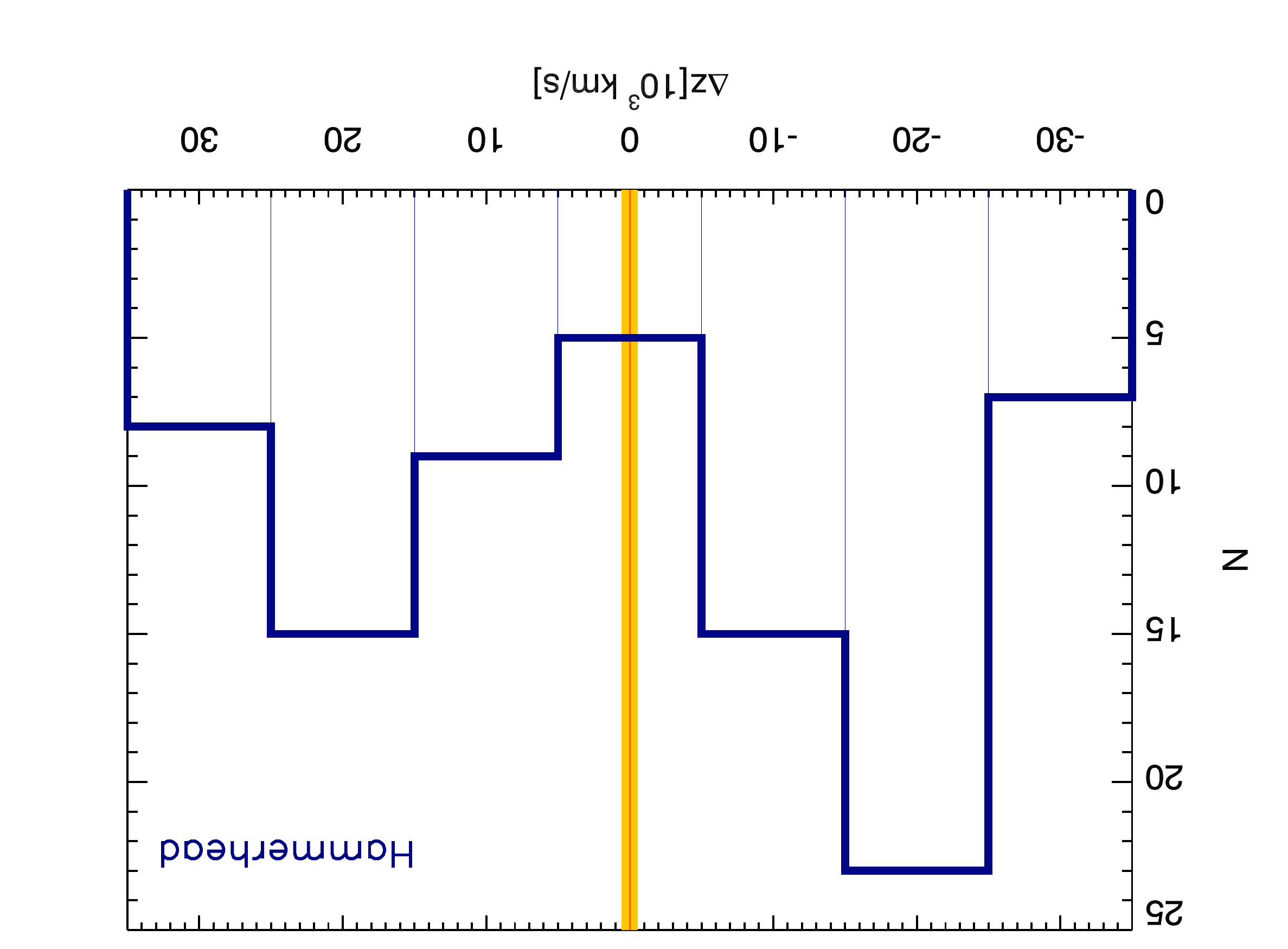}\\
\includegraphics[width=0.98\columnwidth, angle=180]{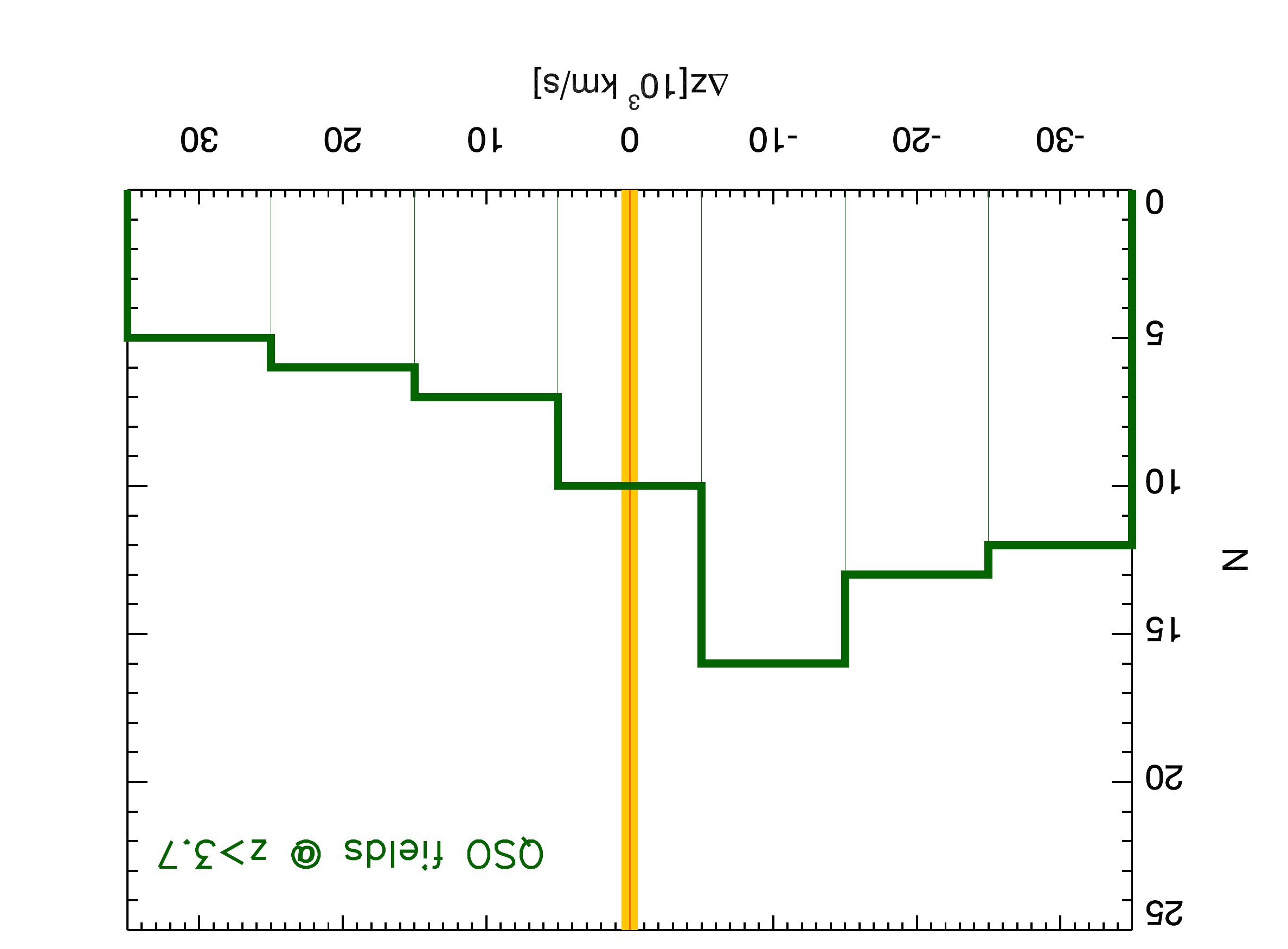}
\includegraphics[width=0.98\columnwidth, angle=180]{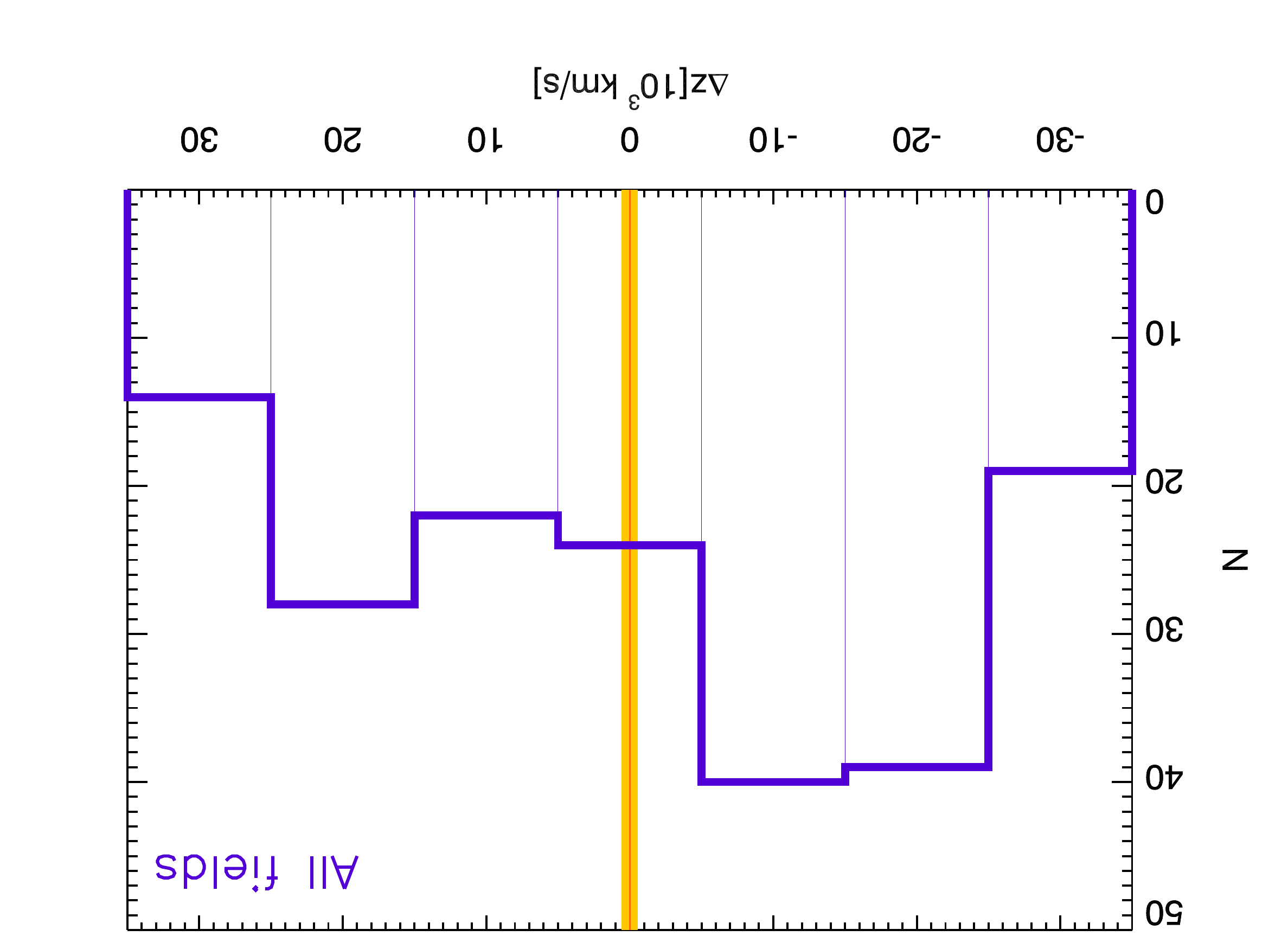}
\caption{Ly$\alpha$ emitters distribution as a function of the velocity separation with the QSO. The top panels show the Bulb (on the left) and Hammerhead (on the right) number densities while in the bottom left panel the results for the MUSE \textit{z}\,$>$\,3.7 sample are shown. The LAEs distribution of all the MUSE fields is shown in the bottom right panel.  \label{fig:histo_distance}}
\end{figure*}

\subsection{The distribution of the Ly$\alpha$ equivalent width} \label{subsec:laes_EW}

The equivalent width (EW) is a quantitative way of describing the strength of spectral features, both in emission and absorption, compared to the continuum emission. Physically, EWs depend on the initial mass function (IMF) and of the gas metallicity from which stars form, as well as being a useful diagnostic to understand what kind of mechanisms are triggering and sustaining the star formation \citep[e$.$g$.$,][]{2002A&A...382...28S,2003A&A...397..527S}. 
Similar to the Ly$\alpha$ fluxes, the Ly$\alpha$ EW estimation is not unique, and it is very sensitive to the methodology used as well as to the data available for the EW measurements.

In general, we compute the EW as the following ratio:
\begin{equation}
\mathrm{EW}(\mathrm{Ly}\alpha)=\frac{\mathrm{Flux}_{\,\mathrm{Ly}\alpha}}{\mathrm{Flux\,Density}_{\mathrm{\,Continuum}}} 
\end{equation}

where the numerator corresponds to the Ly$\alpha$ flux. Flux$_{\mathrm{Ly}\alpha}$ is computed from the C$.$o$.$G$.$ analysis and it is in units of erg\,s\,$^{-1}$\,cm$^{-2}$. The denominator is the continuum flux density measured in the MUSE continuum image (centered at $\lambda\,\sim$\,6000\,$\mathrm{\AA}$) and extrapolated to the wavelength of the line, assuming that the monochromatic fluxes f$_{\nu}$ of all objects are flat in the frequency space. The unit in this case is erg\,s\,$^{-1}$\,cm$^{-2} \mathrm{\AA}^{-1}$. However, as explained below, we will use different estimates of these fluxes depending on the nature of the analyzed object.
The rest$-$frame EW(Ly$\alpha$), EW$_{0}$(Ly$\alpha$), is:

\begin{equation}
\mathrm{EW}_{0}(\mathrm{Ly}\alpha)=\frac{\mathrm{EW}(\mathrm{Ly}\alpha)}{(1+z)}.
\end{equation}

The redshift used in the above equation is defined as the flux centroid of the three$-$dimensional segmentation mask associated with each detected object. Stellar population synthesis models predict that in the case of continuously star$-$forming galaxies, the EW$_{0}$(Ly$\alpha$) produced by Population II stars (hereafter PopII stars) cannot be higher than 240\,$\mathrm{\AA}$ except in very extreme cases \citep{1993ApJ...415..580C,2002A&A...382...28S}. EW$_{0}$(Ly$\alpha$) values above this value may in principle be expected for metal$-$free PopIII stellar systems \citep{2003A&A...397..527S,2010A&A...523A..64R} and/or \textit{Dark Galaxies} (C12).\\

In order to compute the EW$_{0}$(Ly$\alpha$) of our targets, we decide to follow two different approaches depending on the detection (or not) of our LAE in the continuum image. First, in order to establish if our LAE is detected in the continuum, we measure the continuum flux of our target as the maximum value obtained from the measured continuum flux in 9 different and contiguous positions around the central coordinates of the targets within an aperture with radius equal to the PSF size. This method takes into account possible offsets between the spatial peak of the Ly$\alpha$ emission and the stellar continuum (note that the PSF values, listed in Table \ref{tab:fields}, are all larger than the offsets proposed in \citealt{2014ApJ...785...64S}). Second, if the continuum flux of the target within the PSF size aperture, F$_{\mathrm{Cont}\,@\,\mathrm{PSF}}$, is higher than 3 times the standard deviation, \textit{std}, of the continuum image (3$\sigma_{\mathrm{Cont}}$, i$.$e$.$ the local noise, see Section \ref{subsec:limits} for a detailed explanation on how we computed this value) the LAE is considered detected in the continuum. In the case of F$_{\mathrm{Cont}\,@\,\mathrm{PSF}}\,<\,$3$\,\sigma_{\mathrm{Cont}}$ our LAE is considered continuum undetected. Of the 186 LAEs selected in our sample, 54\% were undetected in the continuum. In the 4th and 5th columns of Table \ref{tab:LAES_statistic} this statistic is provided for each field. 

In the case of the continuum detected (CD) LAEs, we used the matched$-$aperture approach as in C12 and the EW$_{0}$(Ly$\alpha$) is computed as follows:

\vspace{-0.1cm}
\begin{equation}
\scriptsize
\mathrm{EW}_{0}(\mathrm{Ly}\alpha)|_{\mathrm{CD}}=\frac{\mathrm{Flux}_{\mathrm{Ly}\alpha}(\mathrm{R})}{\mathrm{Flux}_{\mathrm{Cont}}(\mathrm{R})+1\sigma(\mathrm{R})} \times \frac{1}{(1+z)}.
\end{equation}

where Flux$_{\mathrm{Ly}\alpha}$(R) is the Ly$\alpha$ flux within the radius R derived from the C$.$o$.$G$.$ analysis, $\sigma$(R) is the \textit{std} of the continuum scaled to the same R apertures and Flux$_{\mathrm{Cont}}$(R) is the continuum flux measured in the same aperture as the Ly$\alpha$ flux. We also masked the contribution of the visible bright continuum objects that were contaminating the measurements extracted from the target aperture, as well as possible contamination from fainter foreground objects inside the aperture.

\begin{figure*}[!]
\centering
\includegraphics[scale=0.85,width=\linewidth]{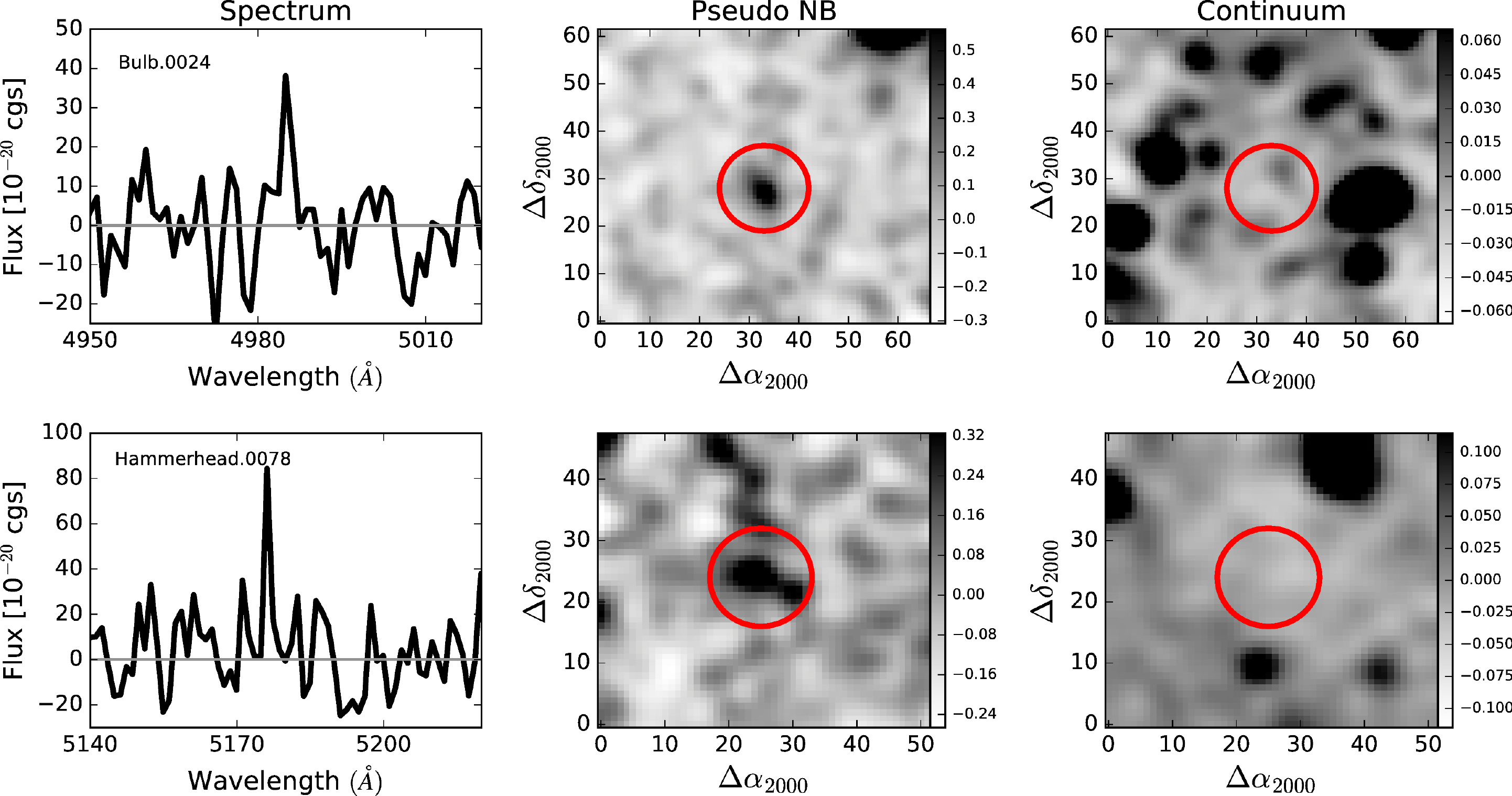}
\caption{Dark Galaxy candidates detected in the MUSE \textit{z}\,$<$\,3.2 fields. $-$ Left: the MUSE spectrum within a wavelength range highlighting the observed Ly$\alpha$ emission. The spectrum has been smoothed with a 2 pixel gaussian filter. $-$ Middle: The MUSE Ly$\alpha$ pseudo narrow$-$band image is shown. The position of the candidate is marked by the red circle. The image was smoothed using a 2 pixels gaussian kernel and the Ly$\alpha$ flux is shown in z$-$scale. $-$ Right: Continuum broad$-$band image obtained from the MUSE datacube. We applied a gaussian smoothing with a 2 pixels radius. The continuum flux is plotted with a z$-$scale stretch between $\pm$ 5 $\sigma$. In each panel North is up and East is left. Plate scale is 0.2$\arcsec$/pix. \label{fig:DG2_low_z}}
\end{figure*}

For those LAEs undetected in the continuum image (CU), we used the PSF$-$aperture approach and EW$_{0}$(Ly$\alpha$) is obtained via:

\vspace{-0.1cm}
\begin{equation}
\scriptsize
\mathrm{EW}_{0}(\mathrm{Ly}\alpha)|_{\mathrm{CU}}=\frac{\mathrm{Flux}_{\mathrm{Ly}\alpha}(\mathrm{R})}{max[1\sigma_{\mathrm{Cont}},\mathrm{Flux}_{\mathrm{Cont}}(\mathrm{R}_{\mathrm{PSF}})+1\sigma_{\mathrm{Cont}}]} \times \frac{1}{(1+z)}.
\end{equation}

where the Flux$_{\mathrm{Ly}\alpha}$(R) is derived as in the case of the CD LAEs and here the continuum flux is computed using the one in the PSF aperture plus 1$\sigma$. This method proposed by \cite{1998PhRvD..57.3873F} ensures an upper limit for the continuum estimation, if the flux in the PSF aperture is positive, otherwise the continuum flux is at least 1$\sigma$. This upper limit in the continuum will yield a lower limit in the estimation of the EW$_{0}$. 

Despite the complexity and the limitations in estimating the EW$_{0}$, we would like to stress here that we are more interested in the relative distribution of the EW$_{0}$ values around the QSOs rather than their absolute values. Similarly to any other measured properties of the Ly$\alpha$ emitters in our sample, we have used exactly the same methods
to estimate the EW$_{0}$ independent of the position of the object relative to the quasar redshift, both in the main and in the control sample. 

In Figures \ref{fig:EW0_multi_low_z} and \ref{fig:EW0_multi_high_z}, we present the EW$_{0}$(Ly$\alpha$) distribution as a function of the redshift difference (spectral distance) from the QSO for the low and the high redshift samples, respectively. The vertical yellow shaded area represents the position of the QSO, while the grey lines indicate the masked position of OH skylines. The CD LAEs are plotted with diamond symbols while the arrows symbolize the lower limit EW$_{0}$(Ly$\alpha$) estimations for the CU LAEs. Green colors represent the LAEs detected in the on$-$source (QSO) samples, while the blue and the red ones indicate the control samples. The horizontal dashed line at 240\,$\mathrm{\AA}$ denotes the EW$_{0}$(Ly$\alpha$) limit expected for ``normal" star$-$forming galaxies.

In all MUSE high$-$\textit{z} fields we found a higher occurrence of objects with EW$_{0}$(Ly$\alpha$)$>240\mathrm{\AA}$ closer to the QSOs rather than in the control samples.  
In order to quantify the observed overdensity of high EW$_{0}$(Ly$\alpha$) objects around the QSO, i$.$\,e$.$ in the on$-$source sample with respect to the control samples, we looked at the EW$_{0}$(Ly$\alpha$) cumulative distribution. In the left hand panel of Figure \ref{fig:histo_cum}, the green line indicates the EW$_{0}$(Ly$\alpha$) cumulative distribution of all (CD and CU) LAEs detected around the QSO. The cyan line denotes the detections in the control samples. In the right hand panel, we plot the same but for the continuum undetected LAEs. It is clear in both cases that for EW$_{0}$(Ly$\alpha$) $>$ 240\,$\mathrm{\AA}$ the number of LAEs in the on$-$source samples is higher. 
We quantified the probability that the on$-$source and the control samples are drawn from the same parent population using two non$-$parametric statistical tests; the Anderson$-$Darling (AD) test, which is more sensitive to the tails of the distribution, and the Kolmogorov$-$Smirnov (KS) test, which is more sensitive to the center of the distribution. We decided to use both tests due to our moderate sample size and the fact that the difference between the two samples is more prominent for EW$_{0}$(Ly$\alpha$) $>$ 240\,$\mathrm{\AA}$. The resulting \textit{p-values} are lower than 0.007 in both tests. Specifically, in the case of CD and CU LAEs (left panel of Figure \ref{fig:histo_cum}), we obtained $p_{KS}\,=\,$0.007 and $p_{AD}\,=\,$0.005, whereas in the case of the CU LAEs alone (right panel of Figure \ref{fig:histo_cum}) the \textit{p-values} are 0.001 in both KS and AD tests. Such low \textit{p-values} allow us to reject the null hypothesis that the two samples belong to the same population, hence the on$-$source and the control samples are statistically different.

In the left panel of Figure  \ref{fig:high_z}, we combined the EW$_{0}$(Ly$\alpha$) distribution of the high redshift fields in order to highlight that most of the high EW detections are located closer in redshift to the QSO in the on$-$source sample. This difference is not due to luminosity effects: If we analyze the Ly$\alpha$ luminosities of these LAEs, plotted in the right panel of Figure \ref{fig:high_z} as empty diamonds, the distribution of our data does not suggest any significant difference in the luminosities of the LAEs in the on$-$source with respect to the ones in the control samples.

Similarly, this excess of high EW objects is not connected to an apparent enhancement in the number density of LAE in proximity of the quasars with respect to the control fields, as shown in Fig$.$ \ref{fig:histo_distance} where we plot the distribution of the LAEs as a function of the distance from the central ionizing source (AGN in the case of the Bulb field and QSOs for the others). With the exception of the Bulb field, which hosts a lower luminosity AGN, we do not find evidence for an overdensity of LAEs around any of the MUSE QSOs, although the statistical sample is small. Our result is in agreement with the recent findings of \cite{2017arXiv170406050U} using a sample of $\sim$\,150 QSOs and of \cite{2017arXiv170504753K} using $\sim$\,300 LAEs in different environments. 

We will discuss in Section \ref{sec:discussion} the implication of these results in light of our search for dark galaxies candidates fluorescently illuminated by the quasars.

\begin{figure*}[]
\centering
\includegraphics[scale=0.5]{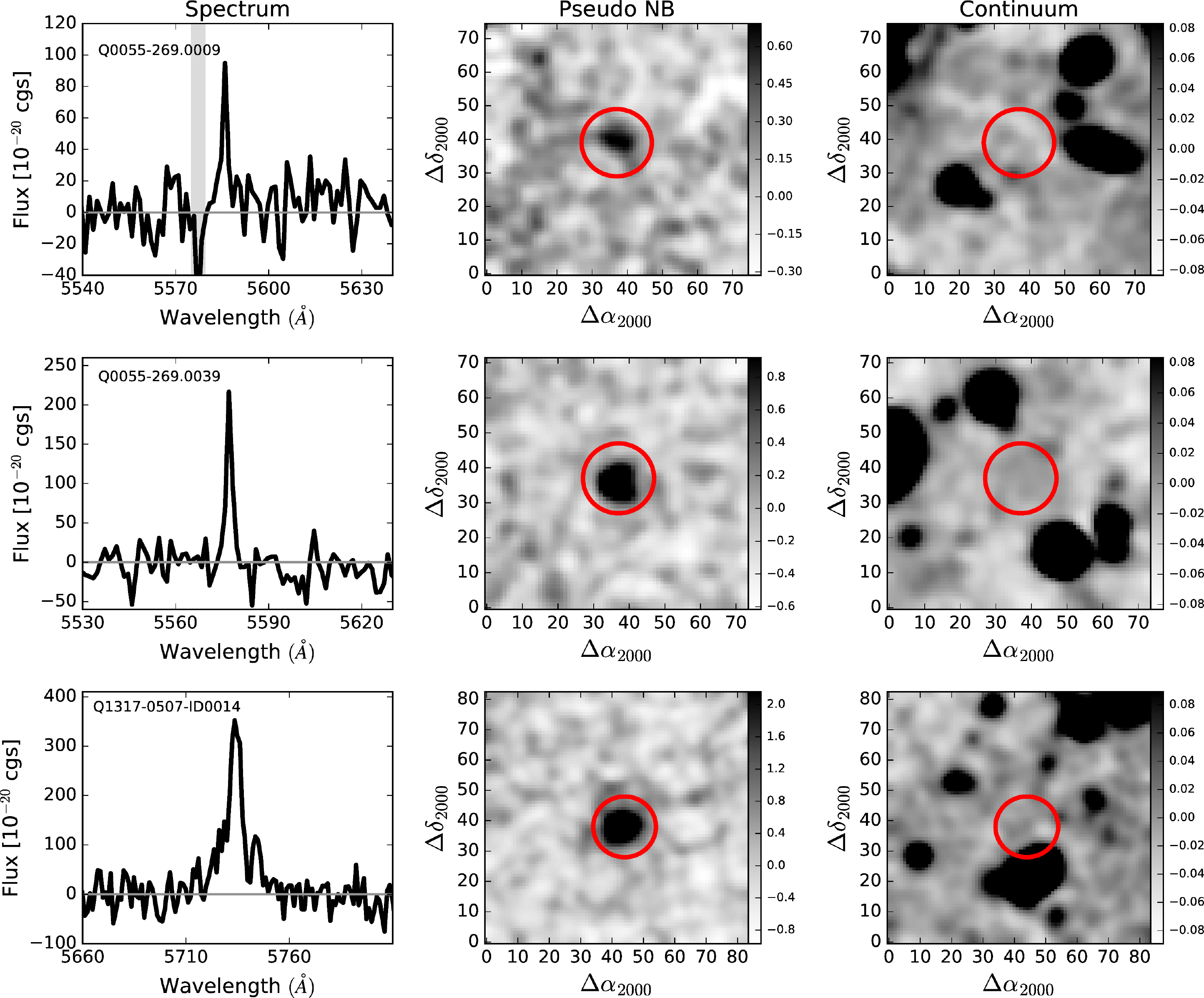}\\
\vspace{0.2cm}
\includegraphics[scale=0.5]{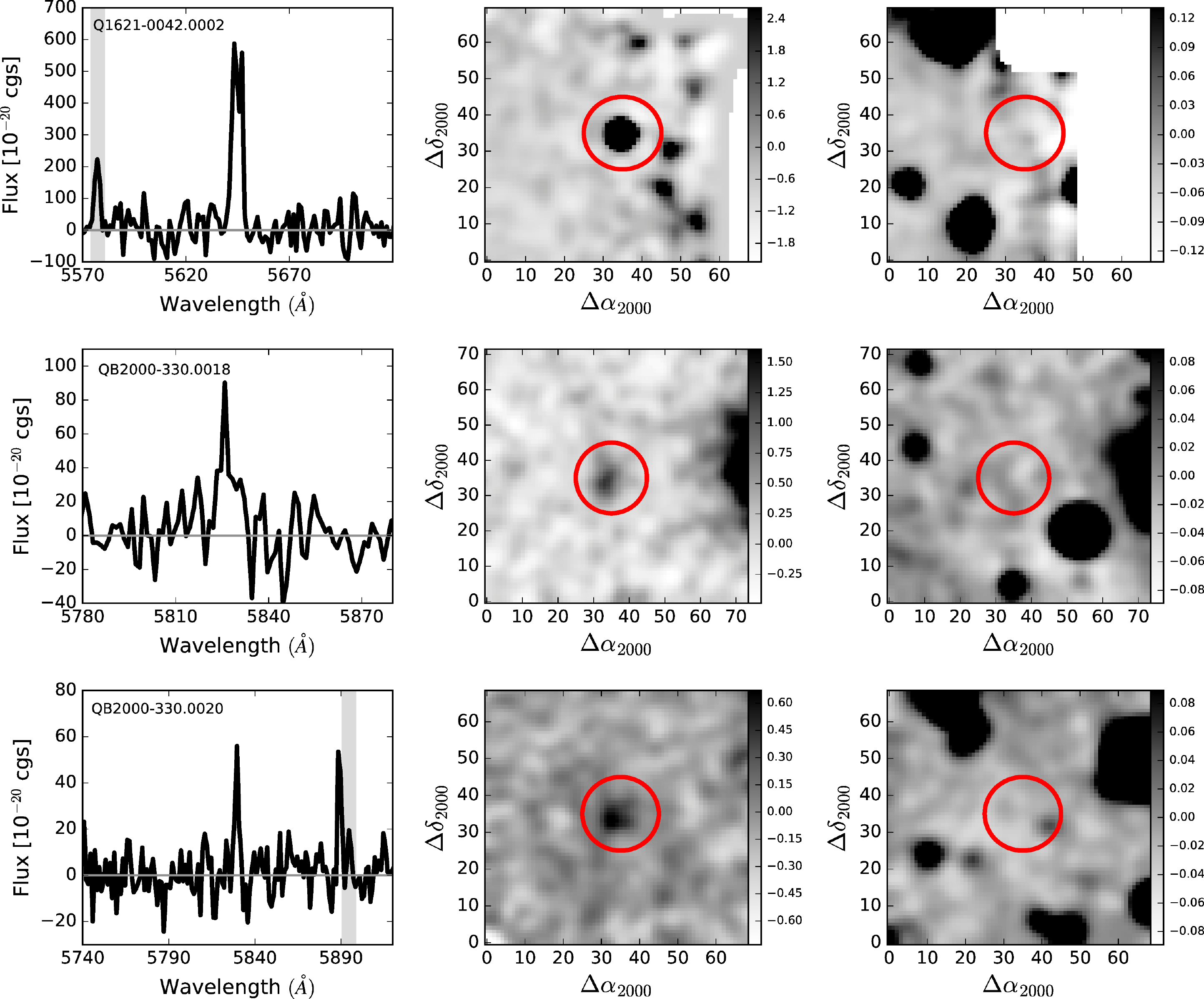}
\caption{Dark Galaxy candidates detected in the MUSE QSO \textit{z}\,$>$\,3.7 fields. Panels have the same meaning as in Fig.\ref{fig:DG2_low_z}. \label{fig:DG2_high_z}}
\end{figure*}

\begin{figure}[!]
\includegraphics[width=\linewidth]{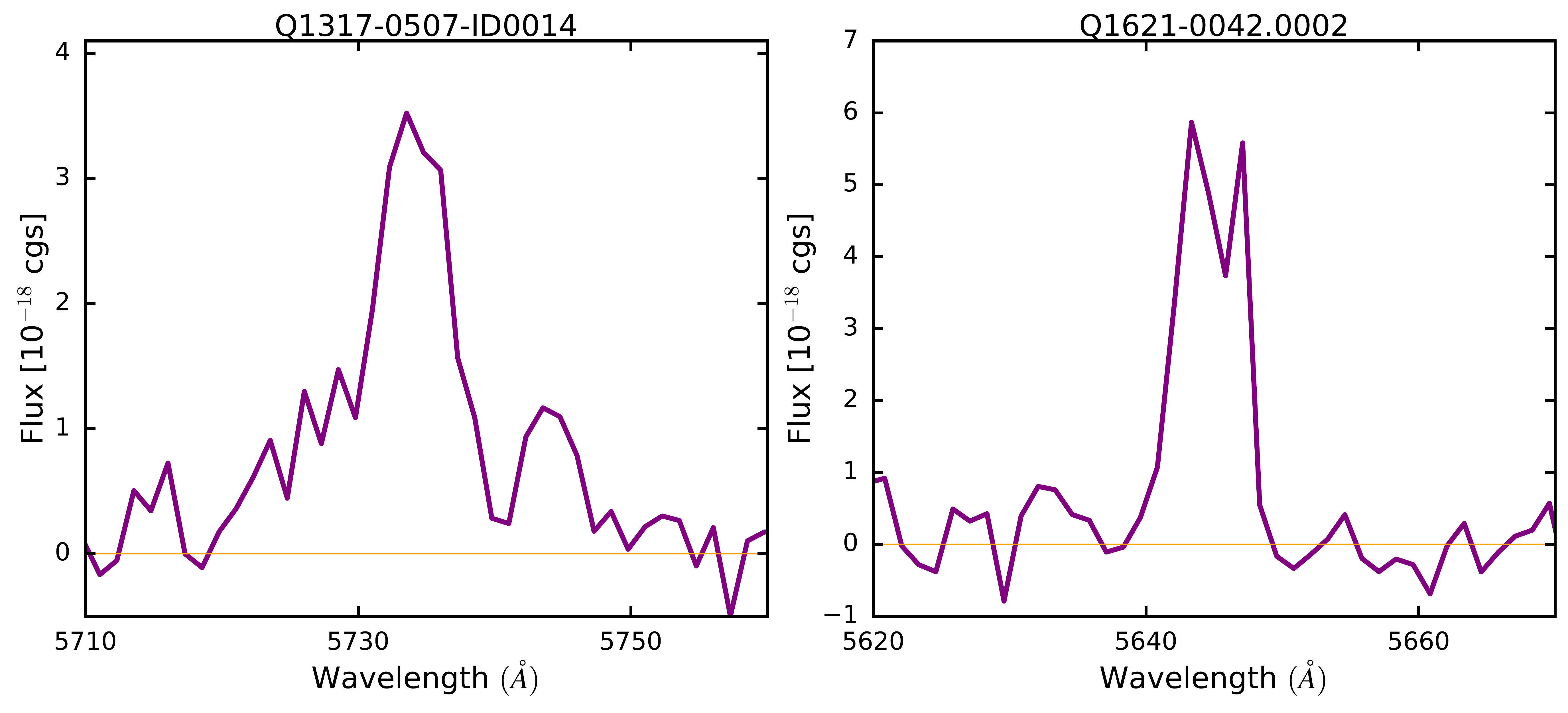}
\caption{Zoomed-in portion of the Ly$\alpha$ line profile for the double peaked Dark Galaxy candidates. Fluxes are given in units of 10$^{-18}$\,erg\,s\,$^{-1}$\,cm$^{-2}$
.\label{fig:DG2i}}
\end{figure}

\subsection{\texorpdfstring{High EW$_{0}$ sources }%
                               {High EW0 sources }}

As shown in the previous section, 11 of the 200 LAEs in the total volume explored in this study, including the control samples, present a lower limit on their EW$_{0}$(Ly$\alpha$) larger than 240\,$\mathrm{\AA}$ (arrows in Figs. \ref{fig:EW0_multi_low_z} and \ref{fig:EW0_multi_high_z} above the purple horizontal dashed line). 
We have demonstrated that these high EW$_{0}$ objects tend to be more frequent in proximity of the quasars and in our high redshift sample. In particular, 6 of these are detected
in our on$-$source sub$-$cubes around the 4 high redsfhit quasars, representing about 25\% of the total detected LAEs (24) in this volume. This value is significantly larger than the corresponding fraction in the control samples for the high redshift quasars (about 4\%) and for the two fields at low redshift. 

In total, 8 high EW$_{0}$ objects are present in the on$-$source samples, i.e. within 10\,$^3$ km/s from the quasars (AGN in the case of the Bulb). 
In Figs$.$ \ref{fig:DG2_low_z} and \ref{fig:DG2_high_z}, we show the spectra and postage stamps of these 8 high EW$_{0}$ objects detected in the low and high redshift samples, respectively.
In particular, for each target, the left panel illustrates a zoom$-$in of the MUSE spectrum around the detected Ly$\alpha$ emission line while the central and right panels specifically show the Ly$\alpha$ pseudo NB and continuum images obtained from the MUSE datacubes. The position of each object is indicated with a red circle.
Their Ly$\alpha$ emission appear compact, similarly to their analogues detected at $z\approx2.4$ by C12. Coordinates, derived photometric and spectral properties, as well as EW$_{0}$ lower limits are reported in Table \ref{tab:dg_table}.

The Ly$\alpha$ line profiles of these sources is typically asymmetric and in two cases, highlighted in Fig$.$ \ref{fig:DG2i}, the emission appears double$-$peaked. Since the shape of the Ly$\alpha$ profile may be sensitive to the gas kinematics, HI geometry and dust content, our plan is to further investigate these two double$-$peaked high EW$_{0}$ sources as well as the $\sim$\,60 double$-$peaked LAEs in our total sample with the help of radiative transfer models in a separate paper.

The main properties of the 3 high EW$_{0}$ sources in our control samples are summarized in Table \ref{tab:Lae_high_EW_table} and their postage stamps are shown in Fig. \ref{fig:highEW} (in the Appendix). We note that these objects do not show any other prominent lines in their spectra. When we considered the 3D extension,  i$.$\,e$.$ spatial and spectral pixels detected above a threshold, we do not find any significant difference between the 8 objects near to the AGN/QSO and these three high EW$_{0}$ objects.

\section{Discussion} \label{sec:discussion}

The most prominent and characteristic feature of quasar fluorescent illumination is a boost in the EW$_{0}$(Ly$\alpha$) of LAEs, leading to:
(\textsc{i}) a higher frequency of objects without continuum counterparts and (\textsc{ii}) EW$_{0}$ limits above 240\,$\mathrm{\AA}$ with respect to ``blank$-$fields" \citep[e$.$g$.$,][C12]{2005ApJ...628...61C,2007ApJ...657..135C}. 
Because the measurement of EWs$_{0}$ relies on different methodologies in the literature and because of the different observational techniques and instruments, a proper comparison between the EW$_{0}$ of LAEs detected in ``quasar$-$fields" and ``blank$-$fields" has been difficult in previous surveys.

Thanks to the new MUSE Integral Field Spectrograph, we were able to obtain a homogeneous sample of Ly$\alpha$ emitting sources around 6 AGN/QSO at \textit{z}\,$>$\,3.2 and we were able to build control samples using the same data, the same data reduction and analysis techniques. 
As expected in the case of fluorescent illumination, we detected an overall excess of high EW$_{0}$ sources in proximity of the quasars with respect to the control samples (Figs. \ref{fig:EW0_multi_low_z} and \ref{fig:EW0_multi_high_z}). 
We stress again that, despite the uncertainties and limitation on the measurement of absolute values or limits for the EW$_{0}$, we have used exactly the same methods for our estimates for each source independent of its distance from the quasar. 
The excess of high EW$_{0}$ sources is more prominent in the four quasar fields at $z\sim3.7$. 
The field$-$to$-$field variations could be possibly due to the relatively small MUSE FoV and limited volume probed around each individual quasar. However, they could also suggest intrinsic differences in the quasar properties, such as, e.g., opening angle or age.  
In any case, as demonstrated in Section \ref{subsec:laes_EW}, the EW$_{0}$ distribution in the combined sample around the quasars (on$-$source) is statistically different than the EW$_{0}$ in the control samples at a high significance level.

Is there any other mechanism intrinsic to the sources that would enhance the EW$_{0}$(Ly$\alpha$) in proximity of quasars without the need for fluorescent ``illumination"?
High values of EW$_{0}$, if intrinsic, may be due to younger stellar population, different IMF or lower metallicities \citep[see, e$.$g$.$,][]{1993ApJ...415..580C, 2002ApJ...565L..71M, 2002A&A...382...28S, 2012ApJ...753...16K, 2012MNRAS.425...87O}. 
In order for these processes to produce an excess of high EW$_{0}$ sources in proximity of the quasar, a relation between the quasar environment and intrinsic galaxy properties would be required. We have explored if the Ly$\alpha$ luminosity and the number density of galaxies are different in proximity of the quasar, possibly indicating a different ``environment" but we have found no statistically different results between the on$-$source and the control samples with respect to these quantities. Moreover, the compact Ly$\alpha$ morphology and the isolated nature of our high EW$_{0}$ objects do not suggest any possible effects due to merger activities, although our spatial resolution and the lack of HST imaging would not allow us to detect interactions below scales of a few kpc.  While we cannot categorically rule out such a possibility, we see no reason to favour it.

In contrast, the high luminosities of our quasars, the demonstrated existence of ``quasar proximity effect" in absorption \citep[at least along our line-of-sight][]{1982MNRAS.198...91C, 2008A&A...491..465D, 2011MNRAS.412.2543C}, and the detection of bright Ly$\alpha$ nebulae around these quasars \citep[][\textcolor{blue}{Marino et al., in prep.}]{2016ApJ...831...39B} showing that quasars are illuminating their surroundings, all suggest that quasar fluorescence is the most likely explanation for the excess of the compact high EW$_{0}$ sources correlated with the quasar redshift in our survey. In this case, the 8 high EW$_{0}$ sources without detectable continuum counterparts and EW$_{0}$ limits larger than 240\,$\mathrm{\AA}$ are the best candidates for Dark Galaxies fluorescently illuminated by the quasars in our survey. The number densities, luminosities and morphologies of these sources are very similar to their 12 analogues detected by C12 at $z\approx2.4$ using NB imaging around a single bright quasars. 

How many of these sources have intrinsically high EW$_{0}$ without the need of fluorescent ``illumination"? Let us consider the fraction of high EW$_{0}$ sources in our on$-$source and control sample at different redshifts. The combined high redshift sample has 25\% high EW object on$-$source and only about 5\% in the control sample suggesting that about 1 to 2 of the 6 high redshift LAE with EW$_{0}$ limits above 240\,$\mathrm{\AA}$ could be objects with intrinsically high EW$_{0}$. Our fraction of 5\% of high EW$_{0}$ away from quasars at $z\sim3.6$ is consistent with other studies, despite the different methodologies to measure the EW$_{0}$. For instance, \textcolor{blue}{Hashimoto et al. (2017, submitted)} measured a fraction of about 3\% of high EW$_{0}$ object in the Hubble Ultra Deep Field (HUDF) using deep MUSE Ly$\alpha$ datacubes and the deepest HST continuum measurements available to date. 
Despite the small number statistics, this suggest that a significant fraction of the 8 sources with EW$_{0}$ limits above 240\,$\mathrm{\AA}$ and without continuum counterparts in our survey are strong candidates for Dark Galaxies detected at $z>3$. 

From the luminosities of these sources and following the approach of C12, we can estimate their total gas masses and star formation efficiencies. In particular, using equation 8 in C12, we estimate gas masses spanning a range between M$_{\mathrm{gas}} \sim$ 0.2 and 6 $\times$ 10$^{9}$\, M$_{\sun}$ similarly to the DG candidates in C12.

To estimate their limit on the star formation rate (SFR), we use the limit on their continuum magnitude (28.8 AB mag extracted from a 1$\arcsec$ diameter aperture from the stacked continuum image, see Fig$.$ \ref{fig:stack}, re$-$centered at the position of the DG candidates) and convert this value into a SFR using \cite{2010A&A...511A..61O} and assuming: \textsc{i}) a Salpeter initial mass function (IMF), \textsc{ii}) a color excess E(B$-$V)$=$0 and, \textsc{iii}) an extended burst of 250 Myr. 
The constraint achieved for the SFR is  0.02 M$_{\sun}$ yr$^{-1}$, which yields a star formation efficiency SFE (=SFR/M$_{\mathrm{gas}}$) of 2.13 $\times$ 10$^{-11}$ yr$^{-1}$ indicating that, similar to their analogues at $z\approx2.4$ (C12), our DG are very inefficient at forming stars. 

Finally, the distribution of boosted fluorescent Ly$\alpha$ emitters can be also used to constraint the QSO lifetimes \citep[e$.$g$.$,][]{2007ApJ...657..135C,2012ApJ...752...39T, 2016ApJ...831...39B}. Assuming that our DG candidates are fluorescently illuminated by the QSO, we used a simple geometrical model presented in \citealt{2016ApJ...830..120B} in order to constrain how long the QSO was shining on these proto$-$clouds of neutral gas, i$.$\,e$.$ the quasar life time $t_{Q}$. Considering the most distant DG candidate within our sample and taking the mean error in the systemic redshift into account, we obtain a distance of 8.7 physical Mpc that corresponds to $t_{Q} \sim$ 60 Myr. Our estimate is compatible with the results obtained for different QSOs at redshift $\sim$\, 3 analyzed in previous studies \citep{2016ApJ...830..120B}. \\

\section{Summary and Conclusions} \label{sec:conclusions}

Making use of medium$-$deep ($\sim$\,10 hr) MUSE IFU GTO observations around five bright QSO and one Type-II AGN, we have searched for fluorescently illuminated dark galaxies at \textit{z} $>$ 3.2 among Ly$\alpha$ emitters in proximity of the quasars.
Differently than previous surveys based on narrow-band imaging (e.g., C12) and therefore restricted to a fixed volume, we have been able to build control samples at large distances from the quasars using the same data, the same data reduction and analysis techniques thanks to the new capability of the MUSE instrument.

Within a volume of 90 physical Mpc$^3$, including the control sample regions, we have identified $\sim$\,200 line emitters using the automatic source extraction software CubExtractor (\textcolor{blue}{Cantalupo, in prep.}) complemented with visual analysis. After inspecting their spectral properties in the large wavelength range provided by MUSE, we have found that 186 of these sources are Ly$\alpha$ emitters (LAEs) between redshifts 3.1 and 4.0. 
We estimated their EW$_{0}$(Ly$\alpha$) in a homogenous way among the main and the control samples using two different approaches depending whether the sources are detected or not in the continuum. 
Among all LAEs, we found 11 objects with EW$_{0}$(Ly$\alpha$) lower limits larger than 240 $\mathrm{\AA}$ - the theoretical limit for galaxies with PopII stellar population \citep{2002ApJ...565L..71M, 1993ApJ...415..580C}. 
The analysis of the EW$_{0}$(Ly$\alpha$) distribution reveals that these high EW$_{0}$ LAEs tend to preferentially reside within $\sim10^4$ km/s from the quasar systemic redshift.
In particular, 6 of the 8 LAEs with EW$_{0}$(Ly$\alpha$)$>240\,\mathrm{\AA}$ in our high redshift sample lie in close proximity of the quasars.
These sources represent about 25\% of all LAEs detected within a velocity distance of $\sim10^4$ km/s from the high redshift quasars in our sample. This fraction is significantly higher than the corresponding value in the control samples (4\%). 

This excess of high EW$_{0}$ sources correlated with distance from the quasar is totally consistent with the expectations from quasar fluorescent illumination  \citep[e$.$g$.$,][C12]{2007ApJ...657..135C, 2012ApJ...752...39T, 2016ApJ...831...39B}.
Alternative scenarios would require a tight link between distance from the quasars and intrinsic galaxy properties. However, the lack of any correlation between number density of detected LAEs and their luminosities with the distance from the quasars (consistently with other surveys) do not offer currently any support to these alternative scenarios. 

In the fluorescent case, the 8 LAEs with EW$_{0}$(Ly$\alpha$)$>240\,\mathrm{\AA}$ and without continuum counterpart located in close proximity of the quasars represent the best candidates so far for Dark Galaxies at $z>3$. Their properties, such as their number densities, compact morphology, luminosities, derived gas masses ($\sim$\,10$^{9}$ M$_{\sun}$) and star formation efficiencies (SFE $<$ 2.13 $\times$ 10$^{-11}$ yr$^{-1}$) are remarkably similar to their analogues detected at $z\approx2.4$ with NB imaging by C12. 

Although our current sample is limited, this study demonstrate the potential of MUSE observations for the robust detection and characterisation of Dark Galaxies candidates fluorescently illuminated by quasars at $z>3$. 
Compared to NB imaging, the main limitation given by the relatively small MUSE FoV is compensated by the large wavelength range (offering the opportunity to build robust control samples), the immediate spectroscopic confirmation, and the lack of filter (and slit) losses. Every quasar field observed with MUSE will therefore offer the potential to discover new Dark Galaxy candidates and provide crucial information on the early and dark phases of galaxy formation.

\acknowledgments{
 {\it Acknowledgments.} 
This work is based on observations taken at ESO/VLT in Paranal and we would like to thank the ESO staff for their assistance and support during the MUSE GTO campaigns. This work was supported by the Swiss National Science Foundation. 
We thank J. Woo and M. Maseda for their comments on the draft. SC gratefully acknowledges support from Swiss National Science Foundation grant PP00P2\_163824.
JR acknowledges support from the ERC starting grant 336736-CALENDS. JB acknowledges support by Funda{\c c}{\~a}o para a Ci{\^e}ncia e a Tecnologia (FCT) through national funds (UID/FIS/04434/2013) and Investigador FCT contract IF/01654/2014/CP1215/CT0003., and by FEDER through COMPETE2020 (POCI-01-0145-FEDER-007672). TC acknowledges support of the ANR FOGHAR (ANR-13-BS05-0010-02), the OCEVU Labex (ANR-11- LABX- 0060) and the A*MIDEX project (ANR-11-IDEX-0001-02) funded by the ``Investissements d’avenir" French government program managed by the ANR. JS and SM acknowledge support from European Research Council (ERC), grant agreement 278594-GasAroundGalaxies. JS acknowledges support from the Netherlands Organisation for Scientific Research (NWO) VICI grant 639.043.409.
This research made use of \textit{Astropy}, a community$-$developed core PYTHON package for  astronomy \citep{2013A&A...558A..33A}, \textit{NumPy} and \textit{SciPy} (Oliphant 2007), \textit{Matplotlib} (Hunter 2007), \textit{IPython} (Perez \& Granger 2007), and of NASAs Astrophysics Data System Bibliographic Services.
}

          
\clearpage

\appendix

\section{Appendix information}


\begin{center}
\begin{longtable}{l|c|c|c|c|c|c|c}
\caption{Physical properties of the Ly$\alpha$ candidates. The columns are: (1) Field; (2) Object ID; (3) Object area; (4) Redshift; (5) Ly$\alpha$ flux; (6) Ly$\alpha$ luminosity; (7) Continum flux; (8) Rest$-$frame equivalent width. \label{tab:all_LAEs_table}}\\
\hline\hline
Field & ID & Area & Redshift & Flux(Ly$\alpha$) & L(Ly$\alpha$) & Flux(Cont$_{\mathrm{PSF}}$) & EW$_{0}$(Ly$\alpha$)\\
 &  & (pixels$^{2}$) &  & (10$^{-17}$\,erg\,s\,$^{-1}$\,cm$^{-2}$) & (10$^{41}$\,erg\,s\,$^{-1}$) & (10$^{-20}$\,erg\,s\,$^{-1}$\,cm$^{-2}$) & $\mathrm{\AA}$ \\
\hline\hline
Bulb & 15b & 225 & 2.9 & 1.53 $\pm$ 0.04 & 11.5 &  1.83 $\pm$ 0.60 & 96 $\pm$ 32 \\
Bulb & 30b & 39  & 3.0 & 0.11 $\pm$ 0.01 & 0.88 &  0.17 $\pm$ 0.31 & $>$93 \\ 
Bulb & 34b & 88  & 3.0 & 0.09 $\pm$ 0.01 & 0.72 & -0.07 $\pm$ 0.14 & $>$162 \\ 
Bulb & 36b & 50  & 3.0 & 0.13 $\pm$ 0.01 & 0.99 &  0.21 $\pm$ 0.35 & $>$90 \\ 
Bulb & 40b & 49  & 3.0 & 0.12 $\pm$ 0.01 & 0.94 &  0.12 $\pm$ 0.26 & $>$112 \\ 
Bulb & 41b & 303 & 3.0 & 3.37 $\pm$ 0.06 & 27.0 &  2.09 $\pm$ 0.70 & 101 $\pm$ 34 \\
Bulb &  2q & 47  & 3.0 & 0.09 $\pm$ 0.01 & 0.73 &  0.35 $\pm$ 0.49 & $>$46 \\    
Bulb &  4q & 45  & 3.0 & 0.09 $\pm$ 0.01 & 0.79 &  3.91 $\pm$ 0.16 & 6 $\pm$ 1 \\
Bulb &  6q & 45  & 3.0 & 0.11 $\pm$ 0.01 & 0.89 &  0.04 $\pm$ 0.19 & $>$140 \\    
Bulb &  8q & 121 & 3.1 & 0.91 $\pm$ 0.03 & 7.64 &  1.29 $\pm$ 0.45 & 13 $\pm$ 5 \\
Bulb &  9q & 306 & 3.1 & 2.89 $\pm$ 0.05 & 24.3 &  3.61 $\pm$ 0.72 & 198 $\pm$ 39 \\
Bulb & 14q & 36  & 3.1 & 0.06 $\pm$ 0.01 & 0.53 &  3.61 $\pm$ 0.18 & 5 $\pm$ 1 \\
Bulb & 15q & 55  & 3.1 & 0.20 $\pm$ 0.01 & 1.74 &  0.14 $\pm$ 0.29 & $>$173 \\    
Bulb & 19q & 52  & 3.1 & 0.12 $\pm$ 0.01 & 1.01 &  0.18 $\pm$ 0.33 & $>$87 \\    
Bulb & 22q & 248 & 3.1 & 2.41 $\pm$ 0.05 & 20.6 &  3.08 $\pm$ 0.59 & 60 $\pm$ 12 \\
Bulb & 25q & 74  & 3.1 & 0.15 $\pm$ 0.01 & 1.30 &  0.09 $\pm$ 0.23 & $>$155 \\    
Bulb & 26q & 39  & 3.1 & 0.12 $\pm$ 0.01 & 1.04 &  4.08 $\pm$ 0.26 & 6 $\pm$ 1 \\
Bulb & 28q & 44  & 3.1 & 0.10 $\pm$ 0.01 & 0.88 &  0.18 $\pm$ 0.33 & $>$74 \\    
Bulb & 29q & 40  & 3.1 & 0.14 $\pm$ 0.01 & 1.22 &  0.16 $\pm$ 0.30 & $>$111 \\    
Bulb & 32q & 39  & 3.1 & 0.18 $\pm$ 0.01 & 1.59 &  0.48 $\pm$ 0.27 & 21 $\pm$ 12 \\
Bulb & 34q & 43  & 3.1 & 0.08 $\pm$ 0.01 & 0.74 &  0.02 $\pm$ 0.16 & $>$121 \\    
Bulb & 36q & 174 & 3.2 & 0.39 $\pm$ 0.02 & 3.56 &  0.43 $\pm$ 0.57 & $>$165 \\    
Bulb & 38q & 53  & 3.2 & 0.08 $\pm$ 0.01 & 0.78 &  0.29 $\pm$ 0.44 & $>$47 \\    
Bulb &  3r  & 41 & 3.2 & 0.05 $\pm$ 0.01 & 0.49 &  0.34 $\pm$ 0.48 & $>$26 \\  
Bulb &  6r  & 41 & 3.2 & 0.04 $\pm$ 0.01 & 0.36 &  0.20 $\pm$ 0.34 & $>$27 \\  
Bulb &  9r  & 35 & 3.2 & 0.08 $\pm$ 0.01 & 0.73 &  0.05 $\pm$ 0.19 & $>$95 \\  
Bulb & 11r  & 69 & 3.2 & 0.11 $\pm$ 0.01 & 1.03 &  0.12 $\pm$ 0.27 & $>$96 \\  
Bulb & 21r  & 46 & 3.3 & 0.13 $\pm$ 0.01 & 1.24 &  0.21 $\pm$ 0.35 & $>$85 \\  
Bulb & 28r  & 44 & 3.3 & 0.07 $\pm$ 0.01 & 0.69 & -0.20 $\pm$ 0.14 & $>$114 \\  
Bulb & 39r  & 34 & 3.3 & 0.22 $\pm$ 0.01 & 2.19 &  0.53 $\pm$ 0.23 & 19 $\pm$ 8 \\
Bulb & 82r  & 38 & 3.3 & 0.07 $\pm$ 0.01 & 0.75 &  0.38 $\pm$ 0.53 & $>$33 \\  
Bulb & 83r  & 57 & 3.3 & 0.22 $\pm$ 0.01 & 2.22 &  0.24 $\pm$ 0.38 & $>$130 \\  
Hammerhead &  3b  &  37 & 2.9 & 0.37 $\pm$ 0.02 & 2.87 &  1.03 $\pm$ 0.64 & $>$147 \\   
Hammerhead &  6b  &  41 & 3.0 & 0.09 $\pm$ 0.01 & 0.72 &  1.75 $\pm$ 0.46 & 6 $\pm$ 2 \\ 
Hammerhead &  8b  &  48 & 3.0 & 0.35 $\pm$ 0.02 & 2.72 &  3.94 $\pm$ 0.50 & 14 $\pm$ 2 \\ 
Hammerhead &  11b &  44 & 3.0 & 0.21 $\pm$ 0.02 & 1.65 &  0.42 $\pm$ 0.40 & $>$134 \\   
Hammerhead &  12b &  33 & 3.0 & 0.25 $\pm$ 0.02 & 1.93 &  1.42 $\pm$ 0.79 & $>$78 \\   
Hammerhead &  14b &  43 & 3.0 & 0.13 $\pm$ 0.01 & 1.01 &  0.52 $\pm$ 0.43 & $>$75 \\   
Hammerhead &  15b &  56 & 3.0 & 0.18 $\pm$ 0.01 & 1.39 &  0.42 $\pm$ 0.40 & $>$112 \\   
Hammerhead &  18b &  39 & 3.0 & 0.17 $\pm$ 0.01 & 1.38 &  5.90 $\pm$ 0.45 & 4 $\pm$ 1 \\ 
Hammerhead &  19b &  37 & 3.0 & 0.17 $\pm$ 0.01 & 1.36 &  4.42 $\pm$ 0.44 & 7 $\pm$ 1 \\ 
Hammerhead &  22b &  41 & 3.0 & 0.14 $\pm$ 0.01 & 1.09 &  0.49 $\pm$ 0.42 & $>$80 \\   
Hammerhead &  23b &  38 & 3.0 & 0.18 $\pm$ 0.01 & 1.45 &  0.03 $\pm$ 0.24 & $>$191 \\   
Hammerhead &  28b &  53 & 3.0 & 0.07 $\pm$ 0.01 & 0.55 &  4.90 $\pm$ 0.53 & 3 $\pm$ 1 \\ 
Hammerhead &  30b &  79 & 3.0 & 0.31 $\pm$ 0.02 & 2.51 &  0.35 $\pm$ 0.37 & $>$211 \\   
Hammerhead &  34b &  47 & 3.0 & 0.08 $\pm$ 0.01 & 0.68 &  0.37 $\pm$ 0.37 & $>$56 \\   
Hammerhead &  36b &  47 & 3.0 & 0.13 $\pm$ 0.01 & 1.02 &  0.17 $\pm$ 0.29 & $>$108 \\   
Hammerhead &  42b &  46 & 3.0 & 0.25 $\pm$ 0.02 & 2.00 &  0.78 $\pm$ 0.54 & $>$114 \\   
Hammerhead &  45b &  29 & 3.0 & 0.16 $\pm$ 0.01 & 1.29 &  0.95 $\pm$ 0.61 & $>$65 \\   
Hammerhead &  46b &  42 & 3.0 & 0.15 $\pm$ 0.01 & 1.20 &  0.56 $\pm$ 0.45 & $>$81 \\   
Hammerhead &  47b &  51 & 3.0 & 0.31 $\pm$ 0.02 & 2.51 &  0.97 $\pm$ 0.62 & $>$124 \\   
Hammerhead &  48b &  43 & 3.0 & 0.16 $\pm$ 0.01 & 1.30 &  0.84 $\pm$ 0.56 & $>$70 \\   
Hammerhead &  49b &  44 & 3.0 & 0.11 $\pm$ 0.01 & 0.93 &  1.14 $\pm$ 0.68 & $>$41 \\   
Hammerhead &  53b &  40 & 3.0 & 0.28 $\pm$ 0.02 & 2.31 &  0.89 $\pm$ 0.58 & $>$119 \\   
Hammerhead &  56b &  50 & 3.0 & 0.32 $\pm$ 0.02 & 2.63 &  0.85 $\pm$ 0.57 & $>$140 \\   
Hammerhead &  60b &  84 & 3.0 & 0.03 $\pm$ 0.01 & 0.27 &  0.37 $\pm$ 0.38 & $>$21 \\   
Hammerhead &  64b &  73 & 3.0 & 0.30 $\pm$ 0.02 & 2.50 &  0.43 $\pm$ 0.40 & $>$184 \\   
Hammerhead &  66b &  67 & 3.0 & 0.11 $\pm$ 0.01 & 0.93 &  0.90 $\pm$ 0.58 & $>$47 \\   
Hammerhead &  72b &  80 & 3.1 & 0.43 $\pm$ 0.02 & 3.63 &  1.11 $\pm$ 0.67 & $>$158 \\   
Hammerhead &  74b &  75 & 3.1 & 0.23 $\pm$ 0.02 & 1.93 &  0.69 $\pm$ 0.50 & $>$113 \\   
Hammerhead &  78b &  43 & 3.1 & 0.02 $\pm$ 0.01 & 0.17 &  2.02 $\pm$ 0.47 & 4 $\pm$ 1 \\ 
Hammerhead &  81b &  33 & 3.1 & 0.50 $\pm$ 0.02 & 4.19 &  1.46 $\pm$ 0.81 & $>$151 \\   
Hammerhead &  83b & 224 & 3.1 & 1.76 $\pm$ 0.04 & 14.9 &  7.77 $\pm$ 1.06 & 14 $\pm$ 2 \\ 
Hammerhead &  84b & 221 & 3.1 & 1.99 $\pm$ 0.04 & 16.8 &  16.2 $\pm$ 1.1  & 8 $\pm$ 1 \\
Hammerhead &  87b &  36 & 3.1 & 0.27 $\pm$ 0.02 & 2.30 &  0.46 $\pm$ 0.41 & $>$162 \\   
Hammerhead &  89b &  41 & 3.1 & 0.24 $\pm$ 0.02 & 2.03 &  1.33 $\pm$ 0.76 & $>$77 \\   
Hammerhead &  91b &  37 & 3.1 & 0.17 $\pm$ 0.01 & 1.44 &  2.25 $\pm$ 0.44 & 7 $\pm$ 1 \\ 
Hammerhead &  95b &  45 & 3.1 & 0.18 $\pm$ 0.01 & 1.58 &  0.41 $\pm$ 0.39 & $>$114 \\   
Hammerhead & 118b &  50 & 3.1 & 0.13 $\pm$ 0.01 & 1.17 &  1.19 $\pm$ 0.70 & $>$46 \\   
Hammerhead &   1q &  65 & 3.1 & 0.36 $\pm$ 0.02 & 3.17 &  1.11 $\pm$ 0.67 & $>$131 \\  
Hammerhead &   2q &  50 & 3.1 & 0.10 $\pm$ 0.01 & 0.91 &  1.19 $\pm$ 0.70 & $>$36 \\  
Hammerhead &   7q &  34 & 3.1 & 0.13 $\pm$ 0.01 & 1.17 &  0.03 $\pm$ 0.24 & $>$135 \\  
Hammerhead &  11q &  47 & 3.1 & 0.12 $\pm$ 0.01 & 1.03 &  0.74 $\pm$ 0.53 & $>$53 \\  
Hammerhead &  14q &  59 & 3.1 & 0.13 $\pm$ 0.01 & 1.16 &  1.28 $\pm$ 0.74 & $>$42 \\  
Hammerhead &  15q &  35 & 3.1 & 0.20 $\pm$ 0.01 & 1.78 &  1.29 $\pm$ 0.74 & $>$65 \\  
Hammerhead &  17q &  46 & 3.1 & 0.19 $\pm$ 0.01 & 1.68 &  0.56 $\pm$ 0.45 & $>$100 \\  
Hammerhead &  18q &  59 & 3.1 & 0.19 $\pm$ 0.01 & 1.68 &  0.27 $\pm$ 0.33 & $>$135 \\  
Hammerhead &  28q &  63 & 3.2 & 0.15 $\pm$ 0.01 & 1.35 &  0.41 $\pm$ 0.39 & $>$91 \\  
Hammerhead &  30q &  39 & 3.2 & 0.22 $\pm$ 0.01 & 1.98 &  2.05 $\pm$ 0.45 & 3 $\pm$ 1\\
Hammerhead &  38q &  42 & 3.2 & 0.12 $\pm$ 0.01 & 1.10 &  0.47 $\pm$ 0.41 & $>$68 \\  
Hammerhead &  42q & 432 & 3.2 & 6.68 $\pm$ 0.08 & 62.8 &  44.8 $\pm$ 1.48 & 18 $\pm$ 1\\
Hammerhead &  59q &  42 & 3.2 & 0.17 $\pm$ 0.01 & 1.59 &  0.14 $\pm$ 0.28 & $>$140 \\  
Hammerhead &  63q &  44 & 3.2 & 0.11 $\pm$ 0.01 & 1.10 &  0.62 $\pm$ 0.47 & $>$57 \\  
Hammerhead &  64q &  75 & 3.2 & 0.28 $\pm$ 0.02 & 2.70 &  0.71 $\pm$ 0.51 & $>$129 \\  
Hammerhead &  71q &  49 & 3.2 & 0.28 $\pm$ 0.02 & 2.71 &  1.49 $\pm$ 0.82 & $>$80 \\  
Hammerhead &  72q &  90 & 3.2 & 0.24 $\pm$ 0.02 & 2.39 &  2.11 $\pm$ 0.64 & 9 $\pm$ 3\\
Hammerhead &  77q &  49 & 3.2 & 0.21 $\pm$ 0.01 & 2.02 &  0.77 $\pm$ 0.54 & $>$90 \\    
Hammerhead &  81q & 118 & 3.2 & 0.10 $\pm$ 0.01 & 1.03 &  0.44 $\pm$ 0.40 & $>$61 \\ 
Hammerhead &  12r &  41 & 3.3 & 0.14 $\pm$ 0.01 & 1.43 &  1.25 $\pm$ 0.73 & $>$45 \\
Hammerhead &  15r &  41 & 3.3 & 0.07 $\pm$ 0.01 & 0.71 &  0.39 $\pm$ 0.38 & $>$42 \\
Hammerhead &  19r &  39 & 3.3 & 0.14 $\pm$ 0.01 & 1.44 & 2.74 $\pm$ 0.45 & 3$\pm$1 \\
Hammerhead &  22r &  51 & 3.3 & 0.14 $\pm$ 0.01 & 1.48 &  0.64 $\pm$ 0.48 & $>$70 \\
Hammerhead &  23r &  47 & 3.3 & 0.21 $\pm$ 0.01 & 2.15 &  0.95 $\pm$ 0.61 & $>$80 \\
Hammerhead &  27r &  38 & 3.3 & 0.14 $\pm$ 0.01 & 1.48 &  0.78 $\pm$ 0.54 & $>$62 \\
Hammerhead &  29r &  46 & 3.3 & 0.14 $\pm$ 0.01 & 1.44 &  0.97 $\pm$ 0.61 & $>$52 \\
Hammerhead &  32r &  97 & 3.3 & 0.30 $\pm$ 0.02 & 3.08 &  1.27 $\pm$ 0.74 & $>$94 \\
Hammerhead &  33r &  57 & 3.3 & 0.11 $\pm$ 0.01 & 1.19 & 2.31 $\pm$ 0.55 & 8$\pm$2 \\
Hammerhead &  34r &  45 & 3.3 & 0.18 $\pm$ 0.01 & 1.91 &  1.20 $\pm$ 0.71 & $>$60 \\
Hammerhead &  38r & 191 & 3.3 & 0.74 $\pm$ 0.03 & 7.74 &  1.67 $\pm$ 0.89 & $>$190 \\
Hammerhead &  43r &  47 & 3.4 & 0.07 $\pm$ 0.01 & 0.73 &  1.50 $\pm$ 0.83 & $>$19 \\
Hammerhead &  55r &  42 & 3.4 & 0.37 $\pm$ 0.02 & 3.99 &  1.46 $\pm$ 0.81 & $>$106 \\
Hammerhead &  60r &  52 & 3.4 & 0.07 $\pm$ 0.01 & 0.72 & -0.02 $\pm$ 0.23 & $>$68 \\
Hammerhead &  61r &  46 & 3.4 & 0.22 $\pm$ 0.02 & 2.35 &  0.49 $\pm$ 0.42 & $>$119 \\
Hammerhead &  63r &  58 & 3.4 & 0.58 $\pm$ 0.02 & 6.32 & 11.0 $\pm$ 0.55 & 7$\pm$1 \\
Hammerhead &  65r &  41 & 3.4 & 0.18 $\pm$ 0.01 & 1.92 &  1.49 $\pm$ 0.82 & $>$49 \\
Hammerhead &  67r &  31 & 3.4 & 0.22 $\pm$ 0.02 & 2.37 &  0.70 $\pm$ 0.51 & $>$98 \\
Hammerhead &  68r &  59 & 3.4 & 0.18 $\pm$ 0.01 & 1.97 & 6.02 $\pm$ 0.56 & 5$\pm$1 \\
Hammerhead &  70r &  41 & 3.4 & 0.04 $\pm$ 0.01 & 0.43 &  0.00 $\pm$ 0.23 & $>$39 \\
Hammerhead &  71r &  45 & 3.4 & 0.11 $\pm$ 0.01 & 1.18 &  0.73 $\pm$ 0.52 & $>$47 \\
Hammerhead &  87r &  44 & 3.4 & 0.10 $\pm$ 0.01 & 1.13 &  1.23 $\pm$ 0.72 & $>$32 \\
Hammerhead &  89r &  41 & 3.4 & 0.07 $\pm$ 0.01 & 0.77 & 2.81 $\pm$ 0.46 & 3$\pm$1 \\
Hammerhead &  90r &  58 & 3.4 & 0.03 $\pm$ 0.01 & 0.35 &  0.74 $\pm$ 0.52 & $>$14 \\
Hammerhead &  91r &  55 & 3.4 & 0.15 $\pm$ 0.01 & 1.67 &  0.70 $\pm$ 0.51 & $>$67 \\
Q0055-269  &   3b & 39  & 3.4 & 0.11 $\pm$ 0.01 & 1.18 & 0.13  $\pm$ 0.32 & $>$75 \\
Q0055-269  &   4b & 46  & 3.5 & 0.15 $\pm$ 0.01 & 1.78 & 0.21  $\pm$ 0.40 & $>$86 \\
Q0055-269  &   5b & 42  & 3.5 & 0.17 $\pm$ 0.01 & 1.96 & -0.13 $\pm$ 0.19 & $>$200 \\
Q0055-269  &   1q & 50  & 3.6 & 0.14 $\pm$ 0.01 & 1.66 & -0.05 $\pm$ 0.19 & $>$159 \\
Q0055-269  &   2q & 253 & 3.6 & 2.08 $\pm$ 0.05 & 25.6 & 1.67  $\pm$  1.00 & 64  $\pm$ 38 \\
Q0055-269  &   5q & 63  & 3.6 & 0.57 $\pm$ 0.02 & 6.99 & 1.67  $\pm$  0.54 & 91  $\pm$ 30 \\
Q0055-269  &   7q & 49  & 3.6 & 0.19 $\pm$ 0.01 & 2.34 & 0.30  $\pm$ 0.49 & $>$84 \\
Q0055-269  &  17q & 118 & 3.6 & 0.61 $\pm$ 0.02 & 7.49 & 0.68  $\pm$  0.65 & 58  $\pm$ 55 \\
Q0055-269  &  19q & 33  & 3.6 & 0.27 $\pm$ 0.02 & 3.39 & 0.38  $\pm$ 0.56 & $>$105 \\
Q0055-269  &  20q & 45  & 3.6 & 0.26 $\pm$ 0.02 & 3.27 & 1.60  $\pm$  0.37 & 14  $\pm$ 3 \\
Q0055-269  &  31q & 156 & 3.6 & 1.23 $\pm$ 0.04 & 15.5 & 1.66  $\pm$  0.83 & 44  $\pm$ 22 \\
Q0055-269  &  32q & 52  & 3.7 & 0.58 $\pm$ 0.02 & 7.49 & 0.96  $\pm$  0.27 & 119 $\pm$ 34 \\
Q0055-269  &  38q & 190 & 3.7 & 3.11 $\pm$ 0.06 & 40.3 & 2.78  $\pm$  0.74 & 109 $\pm$ 29 \\
Q0055-269  &  1r & 61 & 3.7 & 0.22 $\pm$ 0.02 & 3.02 & 0.17  $\pm$ 0.27 & $>$171 \\
Q0055-269  &  3r & 33 & 3.8 & 0.07 $\pm$ 0.01 & 0.97 & -0.09 $\pm$ 0.10 & $>$143 \\
Q0055-269  & 12r & 42 & 3.8 & 0.27 $\pm$ 0.02 & 3.78 & 1.68 $\pm$ 0.36 & 59  $\pm$ 13 \\
Q0055-269  & 14r & 97 & 3.8 & 0.51 $\pm$ 0.02 & 7.33 & 0.72 $\pm$ 0.55 & 127 $\pm$ 96 \\
Q1317-0507 &   1b & 113 & 3.5 & 0.83 $\pm$ 0.03 &  9.36 & 1.24 $\pm$ 0.68 & 16 $\pm$ 9 \\ 
Q1317-0507 &   3b & 125 & 3.5 & 1.28 $\pm$ 0.04 &  14.7 & 0.90 $\pm$ 0.81 & 21 $\pm$ 19 \\
Q1317-0507 &   9b & 174 & 3.6 & 1.21 $\pm$ 0.03 &  14.8 & 1.01 $\pm$ 0.96 & 17 $\pm$ 16 \\
Q1317-0507 &  12b & 72  & 3.6 & 1.17 $\pm$ 0.03 &  14.4 & 1.14 $\pm$ 0.62 & 19 $\pm$ 10 \\
Q1317-0507 &  18b & 36  & 3.6 & 0.19 $\pm$ 0.01 &  2.37 & 0.07 $\pm$ 0.28 & $>$147 \\
Q1317-0507 &   3q & 46 & 3.6 & 0.43 $\pm$ 0.02 & 5.53 & 4.88 $\pm$ 0.29 & 12 $\pm$ 1 \\
Q1317-0507 &   4q & 77 & 3.6 & 0.55 $\pm$ 0.02 & 7.08 & 0.66 $\pm$ 0.64 & 11 $\pm$ 11 \\
Q1317-0507 &   7q & 74 & 3.7 & 0.40 $\pm$ 0.02 & 5.26 & 1.82 $\pm$ 0.63 & 10 $\pm$ 3 \\
Q1317-0507 &   3r & 26 & 3.8 & 0.13 $\pm$ 0.01 & 1.85 & -0.03 $\pm$  0.21 & $>$130 \\
Q1317-0507 &   6r & 44 & 3.8 & 0.10 $\pm$ 0.01 & 1.37 & 0.30  $\pm$ 0.51 & $>$40 \\
Q1317-0507 &  16r & 99 & 3.9 & 1.12 $\pm$ 0.03 & 17.2 & 8.80 $\pm$ 0.56 & 17 $\pm$ 1 \\
Q1317-0507 &  17r & 51 & 3.9 & 0.09 $\pm$ 0.01 & 1.31 & 0.27  $\pm$ 0.48 & $>$36 \\
Q1621-0042 &   1b & 38  & 3.5 & 0.23 $\pm$ 0.02 & 2.62 & 0.25 $\pm$ 0.52 & $>$100 \\
Q1621-0042 &   2b & 30  & 3.5 & 0.24 $\pm$ 0.02 & 2.77 & 0.41 $\pm$ 0.68 & $>$80 \\
Q1621-0042 &   5b & 33  & 3.5 & 0.24 $\pm$ 0.02 & 2.77 & 0.05 $\pm$ 0.32 & $>$168 \\
Q1621-0042 &   4b & 149 & 3.5 & 1.41 $\pm$ 0.04 & 16.3 & 1.28 $\pm$ 1.12 & 37 $\pm$ 33 \\
Q1621-0042 &   9b & 28  & 3.6 & 0.26 $\pm$ 0.02 & 3.19 & 1.54 $\pm$ 0.42 & 18 $\pm$ 5 \\
Q1621-0042 &  10b & 50  & 3.6 & 0.21 $\pm$ 0.01 & 2.59 & 1.38 $\pm$ 0.65 & 14 $\pm$ 7 \\
Q1621-0042 &  11b & 43  & 3.6 & 0.17 $\pm$ 0.01 & 2.04 & 0.25 $\pm$ 0.52 & $>$69 \\
Q1621-0042 &   1q & 41  & 3.6 & 0.11 $\pm$ 0.01 & 1.41 & 0.59 $\pm$ 0.86 & $>$28 \\
Q1621-0042 &   2r & 42  & 3.8 & 0.02 $\pm$ 0.01 & 0.30 & -0.23 $\pm$  0.27 & $>$16 \\
Q1621-0042 &   3r & 136 & 3.8 & 3.39 $\pm$ 0.06 & 48.2  & 0.45 $\pm$ 1.04 & 198 $\pm$ 460 \\
Q1621-0042 &   7r & 137 & 3.9 & 8.69 $\pm$ 0.09 & 130.0 & 0.45 $\pm$ 0.95 & 45  $\pm$ 97 \\
Q1621-0042 &   9r & 61  & 3.9 & 0.42 $\pm$ 0.02 & 6.41 & 0.47  $\pm$ 0.73 & $>$117 \\
Q1621-0042 &  10r & 72  & 3.9 & 0.68 $\pm$ 0.03 & 10.4 & 0.57  $\pm$ 0.84 & $>$165 \\
Q1621-0042 &  11r & 114 & 3.9 & 1.80 $\pm$ 0.04 & 27.6  & 6.72 $\pm$ 0.74 & 32  $\pm$ 4 \\
Q2000-330 &   2b & 45  & 3.5 & 0.06 $\pm$ 0.01 & 0.73 & -0.05 $\pm$ 0.20 & $>$68 \\ 
Q2000-330 &   4b & 36  & 3.5 & 0.24 $\pm$ 0.02 & 2.93 & 0.11  $\pm$ 0.32 & $>$170 \\
Q2000-330 &   5b & 44  & 3.5 & 0.20 $\pm$ 0.01 & 2.46 & 0.27  $\pm$ 0.47 & $>$94 \\
Q2000-330 &   6b & 47  & 3.5 & 0.40 $\pm$ 0.02 & 4.81 & 2.80 $\pm$ 0.37 & 97 $\pm$ 14 \\
Q2000-330 &   7b & 76  & 3.6 & 0.43 $\pm$ 0.02 & 5.21 & 1.23 $\pm$ 0.52 & 27 $\pm$ 11 \\
Q2000-330 &   8b & 47  & 3.6 & 0.21 $\pm$ 0.01 & 2.49 & 0.28  $\pm$ 0.48 & $>$94 \\
Q2000-330 &  14b & 102 & 3.6 & 0.22 $\pm$ 0.02 & 2.75 & 0.37  $\pm$ 0.57 & $>$84 \\
Q2000-330 &  16b & 87  & 3.6 & 0.54 $\pm$ 0.02 & 6.66 & 2.04 $\pm$ 0.57 & 74 $\pm$ 21 \\
Q2000-330 &  25b & 176 & 3.6 & 1.15 $\pm$ 0.03 & 14.5 & 0.77 $\pm$ 0.98 & 37 $\pm$ 46 \\
Q2000-330 &  26b & 48  & 3.6 & 0.30 $\pm$ 0.02 & 3.78 & 2.43 $\pm$ 0.41 & 12 $\pm$ 2 \\
Q2000-330 &  27b & 33  & 3.6 & 0.14 $\pm$ 0.01 & 1.80 & 0.26  $\pm$ 0.46 & $>$67 \\
Q2000-330 &  29b & 65  & 3.6 & 0.29 $\pm$ 0.02 & 3.76 & 0.67 $\pm$ 0.61 & 11 $\pm$ 10 \\
Q2000-330 &   1q & 93 & 3.7 & 2.27 $\pm$ 0.05 & 30.6 & 1.54 $\pm$ 0.70 & 24 $\pm$ 11 \\
Q2000-330 &  17q & 42 & 3.8 & 0.47 $\pm$ 0.02 & 6.64 & 3.09 $\pm$ 0.30 & 5 $\pm$ 1 \\
Q2000-330 &  19q & 43 & 3.8 & 0.46 $\pm$ 0.02 & 6.45 & 0.65 $\pm$ 0.45 & 9 $\pm$ 6 \\
Q2000-330 &  23q & 38 & 3.8 & 0.17 $\pm$ 0.01 & 2.41 & 0.59 $\pm$ 0.79 & $>$45 \\
Q2000-330 &  1r & 42 & 3.9 & 0.10 $\pm$ 0.01 & 1.49 & -0.20 $\pm$  0.20 & $>$103 \\
Q2000-330 &  5r & 59 & 3.9 & 0.72 $\pm$ 0.03 & 11.2 & 3.37 $\pm$ 0.29 & 10 $\pm$ 1 \\
Q2000-330 &  8r & 46 & 4.0 & 0.09 $\pm$ 0.01 & 1.44 & -0.30 $\pm$  0.20 & $>$89 \\
\hline\hline
\end{longtable}
\end{center}

\newpage

\begin{figure}[!h]
\includegraphics[scale=0.85,width=\linewidth]{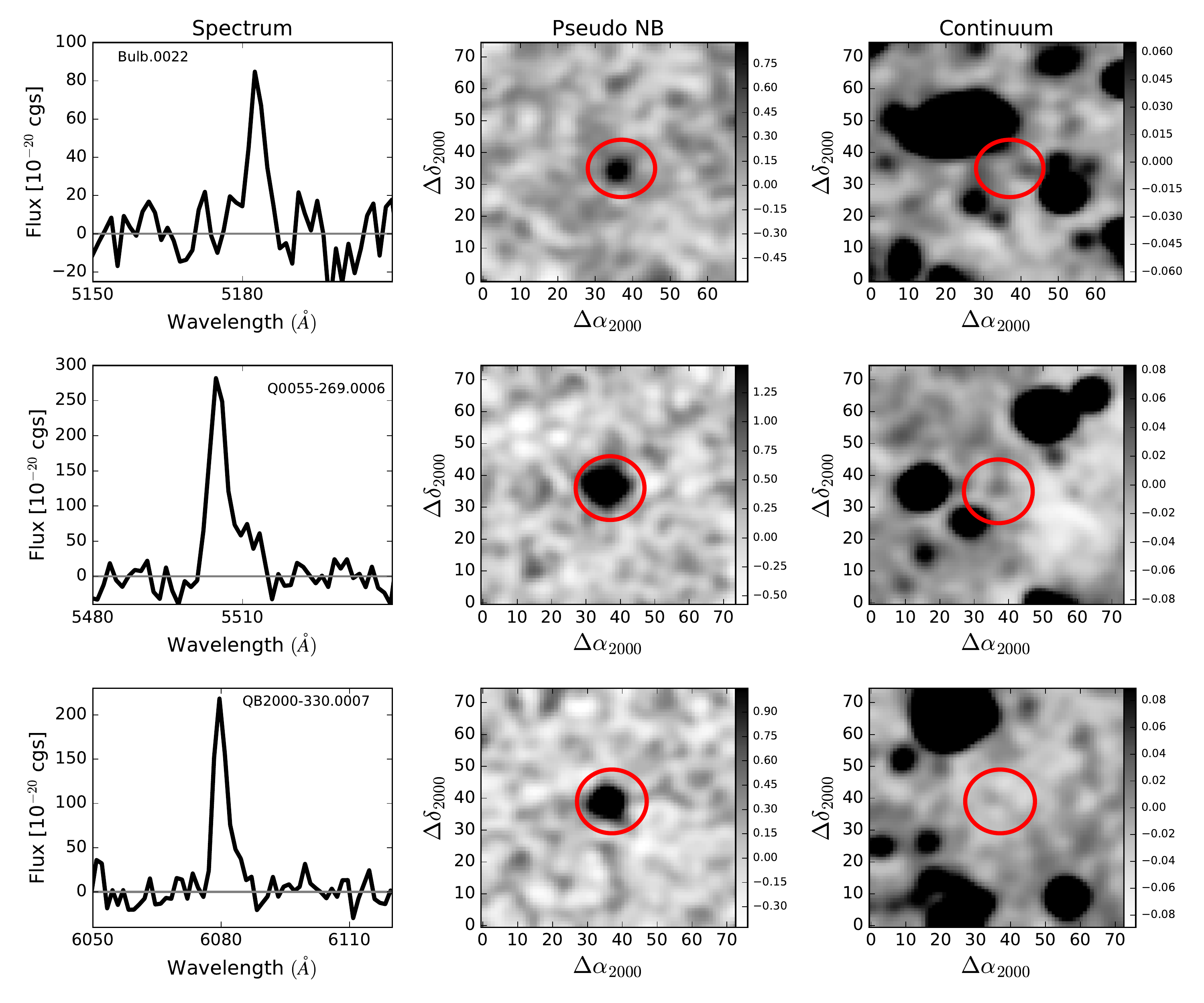}
\caption{High EW$_{0}$ objects detected in the control samples. Panels have the same meaning as in Fig.\ref{fig:DG2_low_z}. \label{fig:highEW}}
\end{figure}

\begin{figure}[!h]
\includegraphics[scale=0.85,width=\linewidth]{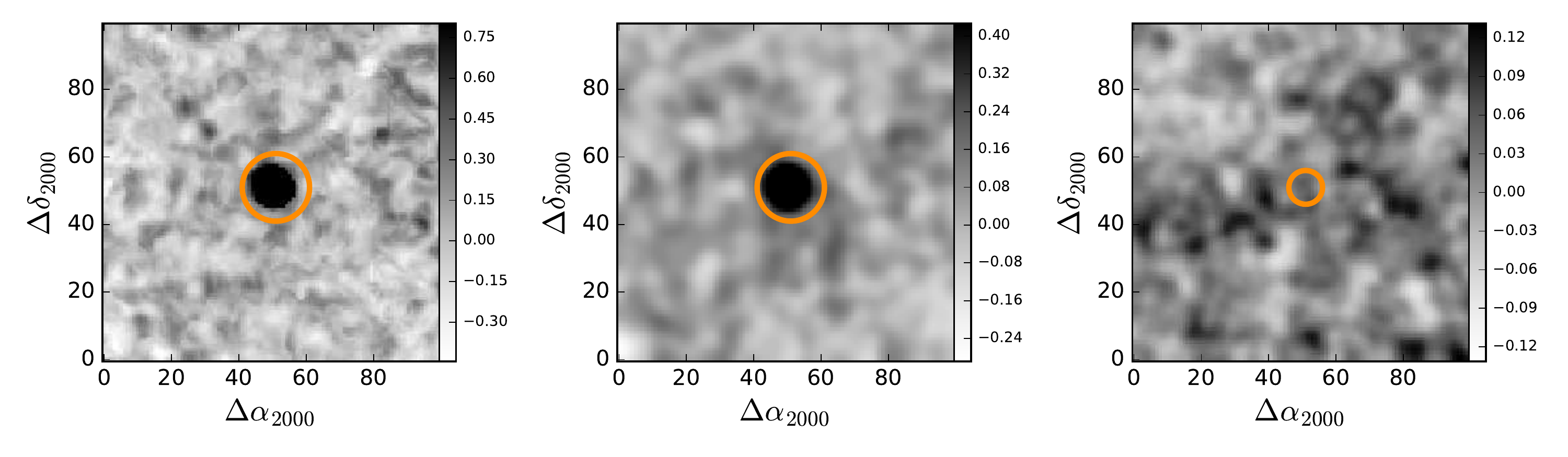}
\caption{Poststage images for the high$-$\textit{z} stacking. The \textit{optimally$-$extracted} image is shown in the first panel, while the classical pseudo$-$NB and the continuum images are shown in the second and third, respectively. The orange circles indicate the positions of the detected target at the center of the images. \label{fig:stack}}
\end{figure}






\begin{thebibliography}{}

\bibitem[Adam et al$.$(2015)]{2015A&A...576A..12A} Adam, R., Comis, B., Mac{\'{\i}}as-P{\'e}rez, J.-F., Adane, A., et al. 2015, \aap, 576, A12

\bibitem[Adelberger et al$.$(2006)]{2006ApJ...637...74A} Adelberger K.~L., Steidel C.~C., Kollmeier J.~A., Reddy N.~A., 2006, \apj, 637, 74

\bibitem[Astropy Collaboration et al$.$(2013)]{2013A&A...558A..33A} Astropy Collaboration, Robitaille, T.~P., Tollerud, E.~J., et al.\ 2013, \aap, 558, A33 

\bibitem[Bacon et al$.$(2010)]{2010SPIE.7735E..08B} Bacon, R., Accardo, M., Adjali, L., et al. 2010, \procspie, Vol. 7735, 773508

\bibitem[Bacon et al$.$(2015)]{2015A&A...575A..75B} Bacon, R., Brinchmann, J., Richard, J., et al. 2015, \aap, 575, A75

\bibitem[Bertin \& Arnouts (1996)]{1996A&AS..117..393B} Bertin, E., \& Arnouts, S., 1996, \aaps, 117, 393

\bibitem[Bina et al$.$(2016)]{2016A&A...590A..14B} Bina, D., Pell{\'o}, R., Richard, J., et al. 2016, \aap, 590, A14

\bibitem[Boera et al$.$(2014)]{2014MNRAS.441.1916B} Boera, E., Murphy, M.~T., Becker, G.~D., et al. 2014, \mnras, 441, 1916

\bibitem[Borisova et al$.$(2016a)]{2016ApJ...830..120B} Borisova, E., Lilly, S.~J., Cantalupo, S., Prochaska, J.~X., et al. 2016, \apj, 830, 120

\bibitem[Borisova et al$.$(2016b)]{2016ApJ...831...39B} Borisova, E., Cantalupo, S., Lilly, S.~J., Marino, R.~A., et al. 2016, \apj, 831, 39

\bibitem[Borisova (2016)]{ElenaThesis} Borisova, E. 2016, ETH Z$\ddot{u}$rich PhD Dissertation, 23603 

\bibitem[Bruns et al$.$(2012)]{2012MNRAS.421.2543B} Bruns, L.~R., Wyithe, J.~S.~B., Bland-Hawthorn, J. \& Dijkstra, M. 2012, \mnras, 421, 2543

\bibitem[Calverley et al$.$(2011)]{2011MNRAS.412.2543C} Calverley, A.~P., Becker, G.~D., Haehnelt, M.~G. \& Bolton, J.~S.. 2011, \mnras, 412, 2543

\bibitem[Cantalupo et al$.$(2005)]{2005ApJ...628...61C} Cantalupo, S., Porciani, C., Lilly, S.~J., \& Miniati, F. 2005, \apj, 628, 61

\bibitem[Cantalupo et al$.$(2007)]{2007ApJ...657..135C} Cantalupo, S., Lilly, S.~J., \& Porciani, C. 2007, \apj, 657, 135

\bibitem[Cantalupo et al$.$(2008)]{2008ApJ...672...48C} Cantalupo, S., Porciani, C., \& Lilly, S.~J. 2008, \apj, 672, 48

\bibitem[Cantalupo (2010)]{2010MNRAS.403L..16C} Cantalupo, S. 2010, \mnras, 403, L16

\bibitem[Cantalupo et al$.$(2012)]{2012MNRAS.425.1992C} Cantalupo S., Lilly S.~J., Haehnelt M.~G., 2012, \mnras, 425, 1992

\bibitem[Cantalupo et al$.$(2014)]{2014Natur.506...63C} Cantalupo S., Arrigoni-Battaia  F., Prochaska J.~X., Hennawi J.~F., Madau P., 2014, \nat, 506, 63

\bibitem[Cantalupo (2017)]{2017ASSL..430..195C} Cantalupo, S. 2017, Astrophysics and Space Science Library, 430, 195

\bibitem[Capak et al$.$(2012)]{2012AAS...21941001C} Capak, P.~L., Teplitz, H., Hanish, D., et al. 2012, American Astronomical Society Meeting Abstracts, 219, 410

\bibitem[Carswell et al$.$(1982)]{1982MNRAS.198...91C} Carswell, R.~F., Whelan, J.~A.~J., Smith, M.~G., Boksenberg, A. \& Tytler, D., 1982, \mnras, 198, 91

\bibitem[Charlot \& Fall (1993)]{1993ApJ...415..580C} Charlot, S., \& Fall, S.~M. 1993, \apj, 415, 580

\bibitem[Cimatti et al$.$(2002)]{2002A&A...392..395C} Cimatti, A., Mignoli, M., Daddi, E., et al. 2002, \aap, 392, 395

\bibitem[Cooksey et al$.$(2013)]{2013ApJ...763...37C} Cooksey, K.~L., Kao, M.~M., Simcoe, R.~A., et al. 2013, \apj, 763, 37

\bibitem[Dall'Aglio et al$.$(2008)]{2008A&A...491..465D} Dall'Aglio, A., Wisotzki, L. \& Worseck, G. 2008, \aap, 491, 465

\bibitem[Drake et al$.$(2016)]{2016arXiv160902920D} Drake, A.~B., Guiderdoni, B., Blaizot, J., et al. 2016, ArXiv e-prints, [arXiv:1609.02920]

\bibitem[Dekel et al$.$ (2009)]{2009Natur.457..451D} Dekel, A., Birnboim, Y., Engel, G., et al. 2009, \nat, 457, 451

\bibitem[Feldman \& Cousins (1998)]{1998PhRvD..57.3873F} Feldman, G.~J. \& Cousins, R.~D. 1998, \prd, 57, 3873

\bibitem[Francis \& Bland-Hawthorn (2004)]{2004MNRAS.353..301F} Francis, P.~ J., \& Bland-Hawthorn, J. 2004, \mnras, 353, 301

\bibitem[Francis \& McDonnell (2006)]{2006MNRAS.370.1372F} Francis P.~J., \& McDonnell S., 2006,  \mnras, 370, 1372

\bibitem[Fumagalli et al$.$(2011)]{2011Sci...334.1245F} Fumagalli, M., O\' Meara, J.~M., \& Prochaska, J.~X. 2011, Science, 334, 1245

\bibitem[Fumagalli et al$.$(2014)]{2014ApJ...780...74F} Fumagalli, M., Hennawi, J.~F., Prochaska, J.~X., et al. 2014, \apj, 780, 74

\bibitem[Fumagalli et al$.$(2016)]{2016MNRAS.462.1978F} Fumagalli, M., Cantalupo, S., Dekel, A., et al. 2016, \mnras, 462, 1978

\bibitem[Fumagalli et al$.$(2017)]{2017MNRAS.467.4802F} Fumagalli, M., Haardt, F., Theuns, T., et al. 2017, \mnras, 467, 4802


\bibitem[Fynbo et al$.$(2003)]{2003A&A...407..147F} Fynbo, J.~P.~U., Ledoux, C., M{\o}ller, P., Thomsen, B., \& Burud, I. 2003, \aap, 407, 147

\bibitem[Gavazzi et al$.$(2008)]{2008A&A...482...43G} Gavazzi, G., Giovanelli, R., Haynes, M.~P., et al. 2008, \aap, 482, 43

\bibitem[Giavalisco et al$.$(2011)]{2011ApJ...743...95G} Giavalisco, M., Vanzella, E., Salimbeni, S., et al. 2011, \apj, 743, 95

\bibitem[Giovanelli et al$.$(2005)]{2005AJ....130.2598G} Giovanelli, R., Haynes, M.~P., Kent, B.~R., Perillat, P., et al. 2005, \aj, 130, 2598

\bibitem[Gould \& Weinberg (1996)]{1996ApJ...468..462G} Gould, A. \& Weinberg, D.~H., 1996 \apj, 468, 462

\bibitem[Haiman \& Rees (2001)]{2001ApJ...556...87H} Haiman, Z., \& Rees, M.~J. 2001, \apj, 556, 87

\bibitem[Hennawi \& Prochaska (2013)]{2013ApJ...766...58H} Hennawi, J.~F. \& Prochaska, J.~X. 2013, \apj, 766, 58

\bibitem[Herenz et al$.$(2015)]{2015A&A...576A.115H} Herenz, E.~C., Wisotzki, L., Roth, M. \& Anders, F., 2015 \apj, 576, A115

\bibitem[Hinshaw et al$.$(2013)]{2013ApJS..208...19H} Hinshaw, G., Larson, D.,  Komatsu, E.,  Spergel, D.~N., et al. 2013, \apjs, 208, 19 

\bibitem[Hogan \& Weymann (1987)]{1987MNRAS.225P...1H} Hogan, C.~J. \& Weymann, R.~J. 1987, \mnras, 225, 1P

\bibitem[Husser et al$.$(2016)]{2016A&A...588A.148H} Husser, T.-O., Kamann, S., Dreizler, S., Wendt, M., et al. 2016, \aap, 588, A148

\bibitem[Johnson et al$.$(2015)]{2015MNRAS.452.2553J} Johnson, S.~D., Chen, H.-W. \& Mulchaey, J.~S., 2015, \mnras, 452, 2553

\bibitem[Kamann et al$.$(2016)]{2016A&A...588A.149K} Kamann, S., Husser, T.-O., Brinchmann, J., Emsellem, E., et al. 2016, \aap, 588, A149

\bibitem[Kashikawa et al$.$(2007)]{2007ApJ...663..765K} Kashikawa, N., Kitayama, T., Doi, M., et al. 2007, \apj, 663, 765

\bibitem[Kikuta et al$.$(2017)]{2017arXiv170504753K} Kikuta, S., Imanishi, M., Matsuoka, Y., et al. 2017, ArXiv e-prints, [arXiv:1705.04753]

\bibitem[Kim et al$.$(2009)]{2009ApJ...695..809K} Kim, S., Stiavelli, M., Trenti, M., Pavlovsky, C.~M., et al. 2009, \apj, 695, 809

\bibitem[Kobayashi et al$.$(2010)]{2010ApJ...708.1119K} Kobayashi, M.~A.~R., Totani, T. \& Nagashima, M. 2010, \apj, 708, 1119

\bibitem[Kollmeier et al$.$ (2010)]{2010ApJ...708.1048K} Kollmeier, J.~A., Zheng, Z., Dav{\'e}, R., et al. 2010, \apj, 708, 1048

\bibitem[Krumholz \& Dekel (2012)]{2012ApJ...753...16K} Krumholz, M. R., \& Dekel, A. 2012, \apj, 753, 16

\bibitem[Kuhlen et al$.$ (2012)]{2012ApJ...749...36K} Kuhlen, M., Krumholz, M. R., Madau, P., Smith, B. D., \& Wise, J. 2012, \apj, 749, 36

\bibitem[Kuhlen et al$.$ (2013)]{2013ApJ...776...34K} Kuhlen, M., Madau, P., \& Krumholz, M. R. 2012, \apj, 776, 34

\bibitem[Lee et al$.$ (2014)]{2014ApJ...795L..12L} Lee, K.-G., Hennawi, J.~F., Stark, C., Prochaska, J.~X., et al. 2014, \apjl, 795, L12

\bibitem[Lusso et al$.$ (2015)]{2015MNRAS.449.4204} Lusso, E., Worseck, G., Hennawi, J.~F., et al. 2015, \mnras, 449, 4204

\bibitem[Madau \& Dickinson (2014)]{2014ARA&A..52..415M} Madau, P., \& Dickinson, M., 2014, \araa, 52, 415

\bibitem[Malhotra \& Rhoads (2002)]{2002ApJ...565L..71M} Malhotra, S., \& Rhoads, J.~E. 2002, \apjl, 565, L71

\bibitem[Mazzucchelli et al$.$ (2017)]{2017ApJ...834...83M} Mazzucchelli, C., Ba{\~n}ados, E., Decarli, R., et al. 2017, \apj, 834, 83

\bibitem[Meiksin (2009)]{2009RvMP...81.1405M} Meiksin A.~A., 2009, Rev. Mod. Phys., 81, 1405

\bibitem[Meurer et al$.$ (1999)]{1999ApJ...521...64M} Meurer, G. R., Heckman, T. M., \& Calzetti, D. 1999, \apj, 521, 64

\bibitem[North et al$.$ (2017)]{2017arXiv170505728N} North, P.~L., Marino, R.~A., Gorgoni, C., et al. 2017, ArXiv e-prints, [arXiv:1705.05728]

\bibitem[Oke \& Gunn (1983)]{1983ApJ...266..713O} Oke J.~B., Gunn J.~E., 1983, \apj, 266, 713

\bibitem[O'Meara et al$.$ (2015)]{2015AJ....150..111O} O'Meara, J.~M., Lehner, N., Howk, J.~C., et al. 2015, \aj, 150, 111

\bibitem[Orsi et al$.$ (2012)]{2012MNRAS.425...87O}  Orsi, A., Lacey, C.~G., \& Baugh, C.~M. 2012,  \mnras, 425, 87

\bibitem[Ot{\'{\i}}-Floranes \& Mas-Hesse (2010)]{2010A&A...511A..61O} Ot{\'{\i}}-Floranes, H. \& Mas-Hesse, J.~M. 2010, \aap, 511, A61

\bibitem[Overzier (2016)]{2016A&ARv..24...14O} Overzier, R.~A., 2016, \aapr, 24, 14

\bibitem[P{\^a}ris et al$.$(2012)]{2012A&A...548A..66P} P{\^a}ris, I., Petitjean, P., Aubourg, {\'E}., et al. 2012, \aap, 548, A66

\bibitem[Prochaska et al$.$ (2013a)]{2013ApJ...762L..19P} Prochaska, J.~X., Hennawi, J.~F., \& Simcoe, R.~A. 2013, \apj, 762, L19

\bibitem[Prochaska et al$.$ (2013b)]{2013ApJ...776..136P} Prochaska, J.~X., Hennawi, J.~F., Lee, K.-G., et al. 2013, \apj, 776, 136

\bibitem[Raiter et al$.$(2010)]{2010A&A...523A..64R} Raiter, A., Schaerer, D., Fosbury, R.~A.~E. 2010, \aap, 523, A64

\bibitem[Rauch et al$.$ (2008)]{2008ApJ...681..856R} Rauch, M., Haehnelt, M., Bunker, A., et al. 2008, \apj, 681, 856

\bibitem[Schaerer (2002)]{2002A&A...382...28S} Schaerer, D., 2002, \aap, 382, 28

\bibitem[Schaerer (2003)]{2003A&A...397..527S} Schaerer, D., 2003, \aap, 397, 527

\bibitem[Schmidt et al$.$ (1987)]{1987ApJ...316L...1S} Schmidt, M., Schneider, D.~P. \& Gunn, J.~E. 1987, \apjl, 316, L1 

\bibitem[Seibert et al$.$ (2012)]{2012AAS...21934001S} Seibert, M., Wyder, T., Neill, J., Madore, B., et al. 2012, American Astronomical Society Meeting Abstracts, 219, 340

\bibitem[Shapiro \& Stockman (2001)]{Shastock} Shapiro, L.~G. \& Stockman, G.~C. 2001, Computer vision, Upper Saddle River, NJ: Prentice Hall

\bibitem[Schaye et al$.$ (2003)]{2003ApJ...596..768S} Schaye, J., Aguirre, A., Kim, T.-S., et al. 2003, \apj, 596, 768

\bibitem[Schneider et al$.$ (2010)]{2010AJ....139.2360S} Schneider, D.~P., Richards, G.~T., Hall, P.~B., et al. 2010, \aj, 139, 2360

\bibitem[Shen et al$.$ (2016)]{2016ApJ...831....7S} Shen, Y., Brandt, W.~N., Richards, G.~T., Denney, K.~D., et al. 2016, \apj, 831, 7 

\bibitem[Shibuya et al$.$ (2014)]{2014ApJ...785...64S} Shibuya, T., Ouchi, M., Nakajima, K., et al. 2014, \apj, 785, 64

\bibitem[Swinbank et al$.$(2015)]{2015MNRAS.449.1298S} Swinbank, A.~M., Vernet, J.~D.~R., Smail, I., et al. 2015, \mnras, 449, 1298

\bibitem[Trainor \& Steidel (2012)]{2012ApJ...752...39T} Trainor, R.~F., \& Steidel, C.~C. 2012, \apj, 752, 39

\bibitem[Trainor \& Steidel (2013)]{2013ApJ...775L...3T} Trainor, R.~F., \& Steidel, C.~C. 2013, \apj, 775, L3

\bibitem[Tolman (1930)]{1930PNAS...16..511T} Tolman, R.~C. 1930, Proceedings of the National Academy of Science, 16, 511

\bibitem[Tolman (1934)]{1934rtc..book.....T} Tolman, R.~C. 1934, Relativity, Thermodynamics, and Cosmology

\bibitem[Turner et al$.$(2014)]{2014MNRAS.445..794T} Turner, M.~L., Schaye, J., Steidel, C.~C., et al. 2014, \mnras, 445, 794

\bibitem[Uchiyama et al$.$ (2017)]{2017arXiv170406050U} Uchiyama, H., Toshikawa, J., Kashikawa, N., et al. 2017, ArXiv e-prints, [arXiv:1704.06050]

\bibitem[V{\'e}ron-Cetty \& V{\'e}ron (2010)]{2010A&A...518A..10V} V{\'e}ron-Cetty, M.~P., \& V{\'e}ron, P., 2010, \aap, 518, A10

\bibitem[Vogt et al$.$(1994)]{1994SPIE.2198..362V} Vogt, S.~S., Allen, S.~L., Bigelow, B.~C., et al. 1994,\procspie, 2198, 362

\bibitem[Weilbacher (2015)]{2015scop.confE..53W} Weilbacher, P.~M. 2015, Science Operations 2015: Science Data Management, [10.5281/zenodo.3465]

\bibitem[Wisotzki et al$.$(2016)]{2016A&A...587A..98W} Wisotzki, L., Bacon, R., Blaizot, J., et al. 2016, \aap, 587, A98

\bibitem[York et al$.$(2000)]{2000AJ....120.1579Y} York, D.~G., Adelman, J., Anderson, Jr., J.~E., et al. 2000, \aj, 120, 1579

\bibitem[Zafar et al$.$(2013)]{2013A&A...556A.141Z} Zafar, T., P{\'e}roux, C., Popping, A., et al. 2013, \aap, 556, A141

\end{thebibliography}
\end{document}